\documentclass[12pt]{article}
\usepackage{amsmath}
\usepackage{graphicx}
\usepackage{enumerate}
\usepackage{natbib}
\usepackage{doi}

\usepackage{url} 
\usepackage{nicematrix}
\usepackage{multibib}
\usepackage{hyperref}
\usepackage{centernot}

\newcommand{\dpath}{\centernot{\longleftrightarrow}}
\usepackage{tocloft}
\usepackage{etoolbox}
\usepackage{titletoc}
\newcites{supp}{Supplementary Material References}
\newcommand{\blind}{1}
\providecommand{\ExcludeFromLatexdiff}[1]{#1}

\def\spacingset#1{\renewcommand{\baselinestretch}%
	{#1}\small\normalsize} \spacingset{1}

\usepackage[mathscr]{eucal}

\newcommand{\Cross}{\mathbin{\tikz [x=2ex,y=2ex,line width=.2ex] \draw (0,0) -- (1,1) (0,1) -- (1,0);}}%
\usepackage{dlfltxbcodetips,bbm,mathtools,amsfonts,amsmath,amssymb,fancybox,graphicx,bm,latexsym,amsthm}

\newcommand{\Mcle}{\widehat{\bm\Theta}(\delta_N)}
\newcommand{\mcle}{\widehat{\bm\theta}}

\usepackage{tikz}
\usetikzlibrary{bayesnet}
\usetikzlibrary{fit,positioning}
\newcommand{\infnormvec}[1]{{\left|\!\left| #1 \right|\!\right|_\infty}}
\newcommand{\infnorm}[1]{{\left|\!\left|\!\left| #1 \right|\!\right|\!\right|_\infty}}

\newcommand{\spectralnorm}[1]{{|\!|\!| #1 |\!|\!|_2}}
\newcommand{\spectralnormsqrt}[1]{{|\!|\!| #1 |\!|\!|_2^2}}
\newcommand{\spectralnormvec}[1]{{|\!| #1 |\!|_2}}
\newcommand{\onenorm}[1]{{|\!|\!| #1 |\!|\!|_1}}

\newcommand{\bsuff}{\bm{b}}
\newcommand{\suff}{b}
\newcommand{\mP}{\mathscr{P}}

\usepackage[tikz]{bclogo}

\newcommand{\btheorem}{\begin{bclogo}[couleur={rgb:orange,0;yellow,0;white,1},arrondi=0.1,logo=\bcplume,ombre=true]{Theorem}}
	\newcommand{\ettheorem}{\end{bclogo}}

\newcommand{\bsh}{\begin{bclogo}[couleur={rgb:orange,0;yellow,0;white,1},arrondi=0.1,logo=\bcpanchant,ombre=true]}
	\newcommand{\esh}{\end{bclogo}}

\newcommand{\benum}{\begin{enumerate}}
	\newcommand{\eenum}{\end{enumerate}}

\newcommand{\bq}{\begin{quote}\em}
	\newcommand{\eq}{\end{quote}}
\newcommand{\bbq}{\begin{quote}\bf\em}
	\newcommand{\ebq}{\end{quote}}

\newcommand{\iid}{\msim\limits^{\mbox{\tiny iid}}}

\newcommand{\lte}{&\leq&}
\newcommand{\gte}{&\geq&}
\newcommand{\mR}{\mathbb{R}}
\newcommand{\mK}{\mathscr{K}}

\newcommand{\one}{\mathbbm{1}}

\newcommand{\mbE}{\mathbb{E}}
\newcommand{\mbC}{\mathbb{C}}

\newcommand{\mbP}{\mathbb{P}}
\newcommand{\mbQ}{\mathbb{Q}}

\newcommand{\hide}[1]{}
\newcommand{\ghost}[1]{}
\newcommand{\ba}{\begin{array}{llllllllll}}
	\newcommand{\ea}{\end{array}}
\newcommand{\bea}{\begin{equation}\begin{array}{llllllllll}}
		\newcommand{\eea}{\end{array}\end{equation}}
\newcommand{\nat}{\bm\theta}
\newcommand{\Nat}{\bm\Theta}
\newcommand{\truth}{\nat^\star}

\newcommand{\logit}{\mbox{logit}}
\newcommand{\beno}{\begin{equation}\begin{array}{llllllllll}\nonumber}
		\newcommand{\be}{\begin{equation}\begin{array}{llllllllll}}
				\newcommand{\ee}{\end{array}\end{equation}}
		\newcommand{\bcc}{\begin{equation}\begin{array}{cccccccccc}}
				\newcommand{\ecc}{\end{array}\end{equation}}
		\newcommand{\bi}{\begin{itemize}}
			\newcommand{\ei}{\end{itemize}}
		\newcommand{\ben}{\begin{enumerate}}
			\newcommand{\een}{\end{enumerate}}
\newcommand{\alert}[1]{\textcolor{red}{\bf{#1}}}

		\newcommand{\dsum}{\displaystyle\sum\limits}
		\newcommand{\dint}{\displaystyle\int\limits}
		\newcommand{\dprod}{\displaystyle\prod\limits}
		\newcommand{\dd}{\mathop{\mbox{d}}\nolimits}
		
		\newcommand{\mM}{\mbox{$\mathscr{M}$}}
		\newcommand{\mN}{\mathscr{N}}
		\newcommand{\mS}{\mathscr{S}}
		
		\newcommand{\mG}{\mathscr{G}}
		\newcommand{\mbG}{\mathscr{C}(\delta_N)}
		\newcommand{\mbH}{\mathscr{H}}
		\newcommand{\mA}{\mathscr{A}}
		\newcommand{\bH}{\bm{H}}
		\newcommand{\mV}{\mathscr{V}}
		\newcommand{\mE}{\mathscr{E}}
		\newcommand{\s}{\vspace{0.25cm}}
		\newcommand{\bc}{\bm{C}}
		\newcommand{\bx}{\bm{x}}
		
		\newcommand{\bw}{\bm{w}}

		\newcommand{\bA}{\bm{A}}
		\newcommand{\bB}{\bm{B}}

		\newcommand{\bX}{\bm{X}}
	    \newcommand{\mH}{\mathscr{H}}
	    \newcommand{\mW}{\mathscr{W}}
		\newcommand{\mX}{\mathscr{X}}
		\newcommand{\mT}{\mathscr{T}}
		
		\newcommand{\mY}{\mathscr{Y}}
		
		\newcommand{\mZ}{\mathscr{Z}}
		\newcommand{\mD}{\mathscr{D}}
		\newcommand{\mB}{\mathscr{B}}
		\newcommand{\bY}{\bm{Y}}
		\newcommand{\bC}{\bm{C}}
		
		\newcommand{\by}{\bm{y}}

		\newcommand{\bW}{\bm{W}}
		
		\newcommand{\bs}{\bm{s}}
		\newcommand{\bz}{\bm{z}}
		\newcommand{\bZ}{\bm{Z}}

		\newcommand{\bV}{\bm{V}}
		\newcommand{\mbV}{\mathbb{V}}

		\newcommand{\msim}{\mathop{\rm \sim}}

		\newcommand{\bI}{\bm{I}}

		\newcounter{comment}

		\newcounter{example}

		\newcounter{counterexample}

		\newcounter{definition}
		
		\newcounter{theorem}
		\newenvironment{theorem}[1][]{\refstepcounter{theorem}\par\medskip\indent%
			\textbf{Theorem~\thetheorem #1}. \rmfamily}{\medskip}
		
		\newcounter{proposition}
		\newenvironment{proposition}[1][]{\refstepcounter{proposition}\par\medskip\indent%
			\textbf{Proposition~\theproposition #1}. \rmfamily}{\medskip}
		
		\newcounter{result}

		\newcounter{tproof}
		\newcounter{corollary}
		\newenvironment{corollary}[1][]{\refstepcounter{corollary}\par\medskip\indent%
			\textbf{Corollary~\thecorollary #1}. \rmfamily}{\medskip}
		
		\newcounter{cproof}
		\newcounter{lemma}
		\newenvironment{lemma}[1][]{\refstepcounter{lemma}\par\medskip\noindent%
			\textbf{Lemma~\thelemma #1}. \rmfamily}{\medskip}
		
		\newcounter{com}

		\newcounter{lproof}
		\newcounter{assumption}
		\newenvironment{assumption}[1][]{\refstepcounter{assumption}\par\medskip\indent%
			\textbf{Condition~\theassumption #1}. \rmfamily}{\medskip}
		
		\newif\ifmydraft
		\mydraftfalse

		\ifmydraft
		
		\pagecolor{black!20}
		\else
		
		\fi

		\ifmydraft
		
		\pagecolor{black!20}
		\fi
		
		\usepackage{multicol}

		\newcommand{\bg}{g}

\newcommand{\norm}[1]{\left|\!\left|#1\right|\!\right|}

       \newcommand{\TV}[1]{\norm{#1}_{\text{TV}}}
		\newcommand{\normsup}[1]{\norm{#1}_\infty}

		\makeatletter
		\newcommand*{\deq}{\mathrel{\rlap{%
					\raisebox{0.3ex}{$\m@th\cdot$}}%
				\raisebox{-0.3ex}{$\m@th\cdot$}}=}
		\makeatother

		\newcommand{\mnorm}[1]{|\!|\!|#1|\!|\!|}

\begin{document}
	
\def\spacingset#1{\renewcommand{\baselinestretch} {#1}\small\normalsize} \spacingset{1}
\date{}

\spacingset{1.5}

\if1\blind
{
\title{\bf A Regression Framework for Studying Relationships among Attributes under Network Interference}
\author{Cornelius Fritz\\
Michael Schweinberger\thanks{Corresponding author.}\\
Subhankar Bhadra\\
David R.\ Hunter\\
Department of Statistics,
The Pennsylvania State University}
\maketitle
} \fi

\if0\blind
{
\title{\bf A Regression Framework for Studying Relationships among Attributes under Network Interference}
\maketitle
} \fi

\if0\blind
{
\vspace{-1.75cm}
} \fi

\spacingset{1.9}

\begin{abstract}
To understand how the interconnected and interdependent world of the twenty-first century operates and make model-based predictions,
joint probability models for networks and interdependent outcomes are needed.
We propose a comprehensive regression framework for networks and interdependent outcomes with multiple advantages,
including interpretability,
scalability,
and provable theoretical guarantees.
The regression framework can be used for studying relationships among attributes of connected units and captures complex dependencies among connections and attributes,
while retaining the virtues of linear regression,
logistic regression,
and other regression models by being interpretable and widely applicable.
On the computational side,
we show that the regression framework is amenable to scalable statistical computing
based on convex optimization of pseudo-likelihoods using minorization-maximization methods.
On the theoretical side,
we establish convergence rates for pseudo-likelihood estimators based on a single observation of dependent connections and attributes.
We demonstrate the regression framework using simulations and an application to hate speech on the social media platform X.

{\em Keywords:}
Dependent Data,
Generalized Linear Models,
Minorization-Maximization,
Pseudo-Likelihood
\end{abstract}

\section{Introduction}
\label{sec:introduction}

In the interconnected and interdependent world of the twenty-first century,
individual and collective outcomes---such as personal and public health, economic welfare,
or war and peace---are affected by relationships among individual,
corporate,
state, and non-state actors.
To understand how the world of the twenty-first century operates and make model-based predictions,
it is vital to study networks of relationships and gain insight into how the structure of networks affects individual and collective outcomes.

While the structure of networks has been widely studied \citep[see][and references therein]{Ko17},
the structure of networks is rarely of primary interest.
Instead,
we often wish to understand
how networks affect individual or collective outcomes.
For example,
social, economic, and financial relationships among individual and corporate actors can affect the welfare of people,
but the outcome of primary interest is the welfare of billions of people around the world.
Relationships among state and non-state actors can affect war and peace,
but the outcome of primary interest is the welfare of nations.
Contact networks mediate the spread of infectious diseases,
but the outcome of primary interest is public health.
A final example is causal inference under network interference:
If the outcomes of units are affected by the treatments or outcomes of other units,
the spillover effect of treatments on outcomes can be represented by an intervention network,
but the target of statistical inference is the direct and indirect causal effects of treatments on outcomes.

To learn how networks are wired and how the structure of networks affects outcomes of interest,
data on outcomes $\bY \coloneqq (Y_i)_{i=1}^N$ and connections $\bZ \coloneqq (Z_{i,j})_{i,j}^N$ among $N$ units are needed along with predictors $\bX \coloneqq (\bX_i)_{i=1}^N$.
Statistical work on joint probability models for dependent outcomes and connections $(\bY, \bZ) \mid \bX = \bx$ is scarce.
\citet{SnStSc05} and \citet{NiSn17} develop models for behavioral outcomes and connections using continuous-time Markov processes,
assuming that the behavioral outcomes and connections are observed at two or more time points.
\citet{WaFeHa24} combine Ising models for binary outcomes with exponential family models for binary connections,
with applications to causal inference \citep{clark_causal_2024}.
In a Bayesian framework,
\citet{FoHo15} unite models for continuous outcomes with latent variable models that capture dependencies among connections.
A common feature of these approaches is that the models and methods in these works may be useful in small populations with, say, hundreds of members, but may be less useful in large populations with, say, thousands or millions of members.
For example,
many of these models make dependence assumptions that are reasonable in small populations but are less reasonable in large populations.
In the special case of exponential-family models,
it is known that models that make unreasonable dependence assumptions can give rise to undesirable probabilistic and statistical behavior in large populations,
such as model near-degeneracy \citep{Ha03p,Sc09b,ChDi11}.
In addition,
these works rely on Monte Carlo and Markov chain Monte Carlo methods for moment- and likelihood-based inference,
which limits the scalability of the mentioned approaches.
Last,
but not least,
the theoretical properties of statistical procedures based on dependent outcomes and connections $(\bY, \bZ) \mid \bX = \bx$,
such as the convergence rates of estimators,
are unknown.

While statistical work on joint probability models for $(\bY, \bZ) \mid \bX = \bx$ is scarce,
recent progress has been made on conditional models for outcomes $\bY \mid (\bX, \bZ) = (\bx, \bz)$ and connections $\bZ \mid \bX = \bx$.
For example,
the literature on network-aware regression uses conditional models for outcomes $\bY \mid (\bX, \bZ) = (\bx, \bz)$:
\citet{lei2024} assume that the dependence among outcomes decays as a function of distance in the population network,
while \citet{li2019} and \citet{le2022} encourage outcomes of connected units to be similar.
A related branch of literature,
concerned with causal inference under network interference,
leverages conditional models for outcomes $\bY \mid (\bX, \bZ) = (\bx, \bz)$ given treatment assignments $\bX = \bx$ and connections $\bZ = \bz$.
Some of them consider fixed connections
\citep[e.g.,][]{
TTFuSh21,ogburn2024causal}, while others combine
conditional models for $\bY \mid (\bX, \bZ) = (\bx, \bz)$ with marginal models for $\bZ$, 
assuming that connections are independent \citep{LiWa22}.
Other works advance autoregressive network models for $\bY \mid (\bX, \bZ) = (\bx, \bz)$ \citep{
huang2019,huang2020,Zhetal20}.
Conditional models for connections $\bZ \mid \bX = \bx$ include stochastic block and exponential-family models with covariates \citep[e.g.,][]{Ha03p,
huang2024,wang2024,stein2025}.

All of the cited work is limited to special cases, 
such as real-valued outcomes or binary connections,
rather than presenting a comprehensive regression framework for studying relationships among attributes $(\bX, \bY)$ under network interference $\bZ$.
To fill the void left by existing work,
we propose a comprehensive regression framework for studying relationships among attributes $(\bX, \bY)$ under network interference $\bZ$ based on joint probability models for $(\bY, \bZ) \mid \bX = \bx$.
The proposed regression framework has important advantages over existing work,
including interpretability,
scalability,
and provable theoretical guarantees:
\begin{enumerate}
\item We show in Sections \ref{sec:glm} and \ref{sec:running_example} that the proposed regression framework can be viewed as a generalization of linear regression,
logistic regression,
and other regression models
for studying relationships among attributes under network interference,
adding a simple and widely applicable set of tools to the toolbox of data scientists.
We demonstrate the advantages of the regression framework with an application to hate speech on the social media platform X in Section~\ref{sec:applications}.
\item The proposed regression framework can be applied to small and large populations by leveraging additional structure to control the dependence among outcomes and connections,
facilitating the construction of models with complex dependencies among outcomes and connections in small and large populations.
\item We develop scalable methods using minorization-maximization algorithms for convex optimization of
pseudo-likelihoods in Section~\ref{sec:computing}.
To disseminate the regression framework and its scalable methods,
we provide an {\tt R} package.
\item We establish theoretical guarantees for pseudo-likelihood estimators in Section~\ref{sec:theory}.
To the best of our knowledge,
these are the first theoretical guarantees for joint probability models of $(\bY, \bZ) \mid \bX = \bx$ based on a single observation $(\by, \bz)$ of $(\bY, \bZ)$.
The simulation results in Section~\ref{sec:simulations} demonstrate that pseudo-likelihood estimators perform
well as the number of units $N$ and the number of parameters $p$ increases.
\end{enumerate}

In addition,
the regression framework has conceptual and statistical advantages:
\begin{enumerate}
\setcounter{enumi}{4}
\item Compared with conditional models for outcomes $\bY \mid (\bX,\, \bZ) = (\bx,\, \bz)$ and connections $\bZ \mid \bX = \bx$,
the proposed regression framework for $(\bY, \bZ) \mid \bX = \bx$ provides insight into outcome-connection dependencies,
in addition to outcome-outcome and connection-connection dependencies.
\item Compared with conditional models for outcomes $\bY \mid (\bX,\, \bZ) = (\bx,\, \bz)$,
conclusions based on the proposed regression framework are not limited to a specific population network $\bz$,
but can be extended to the superpopulation of all possible population networks.
In addition,
the proposed regression framework provides insight into the probability law governing the superpopulation of all possible population networks.
\item The proposed regression framework retains the advantages of two general approaches to building joint probability models for dependent data,
elucidated in the celebrated paper by \citet{Bj74}:
Specifying a joint probability distribution directly guarantees desirable mathematical properties,
while specifying it indirectly via conditional probability distributions helps build complex models from simple building blocks.
We show how to directly specify a joint probability model from simple building blocks.
The resulting regression framework possesses desirable mathematical properties
and induces conditional distributions that can be represented by regression models,
facilitating interpretation.
We showcase these advantages in Sections~\ref{sec:running_example} and \ref{sec:application.results}.
\end{enumerate}

We elaborate the proposed regression framework in the remainder of the article.
\hide{
with an emphasis on non-causal relationships among attributes under network interference.
While the regression framework can be used for studying causal relationships among attributes under network interference,
the question of how the regression framework can be used for causal inference an important topic in its own right,
which we leave to future research.
}

\section{Regression under Network Interference}
\label{sec:model}

Consider a population of $N \geq 2$ units $\mP_N \coloneqq \{1, \dots, N\}$,
where each unit $i \in \mP_N$ possesses
\bi
\item one or more binary,
count-,
or real-valued predictors $\bX_i \in \mX_i$,
which may include covariates and treatment assignments;
\item binary,
count-,
or real-valued outcomes or responses $Y_i \in \mY_i$;
\item binary,
count-,
or real-valued connections $Z_{i,j} \in \mZ_{i,j}$ to other units $j \in \mP_N \setminus\, \{i\}$,
which represent indicators of connections or weights of connections (e.g., the number of interactions between $i$ and $j$).
\ei
We first consider undirected connections, 
for which $Z_{i,j}$ equals $Z_{j,i}$,
and describe extensions to directed connections in Section~\ref{sec:applications}.
We write $\bX \coloneqq (\bX_i)_{1 \leq i \leq N}$,\,
$\bY \coloneqq (Y_i)_{1 \leq i \leq N}$,\,
$\bZ \coloneqq (Z_{i,j})_{1 \leq i < j \leq N}$,\,
$\mX \coloneqq \Cross_{i=1}^N\, \mX_i$,\,
$\mY \coloneqq \Cross_{i=1}^N\, \mY_i$,\, and
$\mZ \coloneqq \Cross_{i<j}^N\, \mZ_{i,j}$,
and refer to $\bY$ without $Y_i$ and $\bZ$ without $Z_{i,j}$ as $\bY_{-i} \in \mY_{-i}$ and $\bZ_{-\{i,j\}} \in \mZ_{-\{i,j\}}$,
respectively.
In line with Generalized Linear Models (GLMs),
we introduce a known scale parameter $\psi \in (0, +\infty)$ and define $Y_i^\star \coloneqq Y_i\, /\, \psi$ and $\bY_{-i}^\star \coloneqq \bY_{-i}\, /\, \psi$.
Throughout,
$\mathbb{I}(\cdot)$ is an indicator function,
which is $1$ if its argument is true and is $0$ otherwise.
We write $a_N = O(b_N)$ and $a_N=o(b_N)$ to indicate that $|a_N/b_N|$
remains bounded and $\lim_{N\to\infty} |a_N/b_N| =0$, respectively.


Following the bulk of the literature on regression models,
we condition on predictors $\bX = \bx$.
To construct joint probability models for dependent responses and connections $(\bY,\, \bZ) \mid \bX = \bx$,
we introduce a family of probability measures $\{\mbP_{\nat},\, \nat \in \Nat\}$ dominated by a $\sigma$-finite measure $\nu$,
with densities of the form
\be
\label{eq:ernm_model}
f_{\nat}(\by,\, \bz \mid \bx)
&=& \dfrac{1}{\varphi(\nat)} \left[\dprod_{i = 1}^N a_{\mY}(y_i)\, \exp\left(\nat_g^\top\, g_i(\bx_i,\, y_i^\star)\right)\right]\s
\\
&\times& \left[\dprod_{i=1}^{N-1}\, \dprod_{j=i+1}^N a_{\mZ}(z_{i,j})\, \exp\left(\nat_h^\top\, h_{i,j}(\bx,\, y_i^\star,\, y_j^\star,\, \bz)\right)\right]:
\ee
\begin{itemize}
\item $a_{\mY}: \mY_i \mapsto [0, +\infty)$ and $a_{\mZ}: \mZ_{i,j} \mapsto [0, +\infty)$ are known functions of responses $Y_i$ of units $i \in \mP_i$ and connections $Z_{i,j}$ of pairs of units $\{i, j\} \subset \mP_N$;
\item $g_i: \mX_i \times \mY_i  \mapsto \mathbb{R}^{q}$ are known functions describing the relationship of predictors $\bx_i$ and responses $Y_i$\, of\, units $i \in \mP_N$,
which can depend on $\psi$;
\item $h_{i,j}: \mX \times \mY_i \times \mY_j \times \mZ \mapsto \mathbb{R}^r$ are known functions specifying how the responses and connections of pairs of units $\{i, j\} \subset \mP_N$ depend on the predictors,
responses,
and connections to other units,
which can depend on  $\psi$;
\item $\nat \coloneqq (\nat_g,\, \nat_h) \in \Nat$ is a parameter vector of dimension $p \coloneqq q + r$,
where $\Nat \coloneqq \{\nat \in \mR^p: \varphi(\nat) < \infty\}$ and $\varphi: \Nat \mapsto (0, +\infty]$
ensures that $\int_{\mY \times \mZ}\, f_{\nat}(\by, \bz\, |\, \bx) \dd \nu(\by, \bz) = 1$,
with the dependence of $\varphi$ on $\bx$ suppressed;
\item $\nu$ is a $\sigma$-finite product measure of the form
\beno
\nu(\by, \bz)
&\coloneqq& \left[\dprod_{i=1}^N \nu_{\mY}(y_i)\right] \left[\dprod_{i=1}^{N-1}\, \dprod_{j=i+1}^N \nu_{\mZ}(z_{i,j})\right],
\ee
where $\nu_{\mY}$ and $\nu_{\mZ}$ are $\sigma$-finite measures that depend on the support sets of responses $Y_i$ and connections $Z_{i,j}$ (e.g.,
Lebesgue or counting measure).
\end{itemize}

{\bf Remark: Importance of Additional Structure.}
To respect real-world constraints and facilitate theoretical guarantees,
joint probability models for dependent responses and connections $(\bY, \bZ) \mid \bX = \bx$ should
leverage additional structure,
e.g.,
local dependence structure.
For one thing,
units in large populations may not be aware of most other units in the population, so
it is not credible that the responses and connections of units depend on the responses and
connections of all other units in the population.
In addition,
models permitting strong dependence among the responses and connections of all units in the
population may suffer from model near-degeneracy
\citep{Ha03p,Sc09b,ChDi11}.
By contrast,
\citet{StSc20} demonstrate that leveraging additional structure
to control dependence can lead to theoretical guarantees.
Motivated by these considerations,
we assume that each unit $i \in \mP_N$ has a known set of neighbors $\mN_i \subset \mP_N$,
which includes $i$ and is independent of connections $\bZ$,
and that the dependence among responses and connections $(\bY, \bZ) \mid \bX = \bx$ is local
in the sense that it is limited to overlapping neighborhoods.
We provide examples of joint probability models for $(\bY, \bZ) \mid \bX = \bx$ with local dependence in Sections \ref{sec:running_example} and \ref{sec:application.model}.

{\bf Remark: Fixed versus Random Design.}
In line with the bulk of the literature on regression models, 
we consider a fixed design:
We view predictors $\bX$ and neighborhoods $\mN_1, \ldots, \mN_N$ as exogenous and known,
and we do not make assumptions about the mechanism generating them.
Thus, 
Equation~\eqref{eq:ernm_model} specifies the joint probability density function of responses and connections $(\bY, \bZ)$ conditional on predictors $\bX = \bx$ and neighborhoods $\mN_1, \ldots, \mN_N$.
If $\bX$ and $\mN_1, \ldots, \mN_N$ were random, 
the conditional model for $(\bY, \bZ) \mid \bX = \bx,\, \mN_1, \ldots, \mN_N$ could be combined with marginal models for $\bX$ and $\mN_1, \ldots, \mN_N$.
In the social media application in Section \ref{sec:applications},
the neighborhoods are fixed and known:
The neighborhoods of users are the sets of followees,
because users choose whom to follow and hence who can influence them,
and these choices are observed.
If the neighborhoods were unobserved,
one could view them as unobserved constants (if neighborhoods were fixed) or unobserved variables (if neighborhoods were random) and learn them from attributes $(\bX, \bY)$ or connections $\bZ$.
The problem of how to learn neighborhoods is an open problem and constitutes a promising avenue for future research.

\subsection{GLM Representations}
\label{sec:glm}

The proposed joint probability models of $(\bY, \bZ) \mid \bX = \bx$ can be viewed as generalizations of Generalized Linear Models (GLMs) \citep{efron22}.
GLMs form a well-known, interpretable, and widely applicable statistical framework for univariate responses $Y_i \in \mathscr{Y}_i$ given predictors $\bx_i \in \mathbb{R}^d$ ($d \geq 1$),
including logistic regression ($Y_i \in \{0, 1\}$),
Poisson regression ($Y_i \in \{0, 1, \ldots \}$),
and linear regression ($Y_i \in \mR$).
GLMs are characterized by two properties:
\begin{enumerate}
\item {\em Conditional mean:}
The conditional mean $\mu_i(\eta_i) \coloneqq \mbE_{\eta_i}(Y_i \mid \bx_i)$ of response $Y_i \in \mY_i$,
conditional on predictors $\bx_i \in \mR^d$ with weights $\bm\beta \in \mathbb{R}^d$,
is a (possibly nonlinear) function of a linear predictor $\eta_i \coloneqq \bm\beta^\top \bx_i$.
\item {\em Conditional distribution:}
The conditional distribution of response $Y_i$ is an exponential family distribution with a known scale parameter $\psi \in (0,\, +\infty)$,
which admits a density with respect to a $\sigma$-finite measure $\nu_{\mY}$ of the form
\beno
f_{\eta_i}(y_i \mid \bx_i)
&\coloneqq& a_{\mathscr Y}(y_i)\, \exp\left(\dfrac{\eta_{i}\, y_i - b_{i}(\eta_{i})}{\psi}\right),
\ee
with cumulant-generating function
\beno
b_i(\eta_i)
&\coloneqq& \psi \, \log \dint_{\mY_i}\, a_{\mY}(y)\, \exp\left(\dfrac{\eta_i\, y}{\psi}\right) \dd \nu_{\mY}(y).
\ee
The conditional mean $\mu_i(\eta_i)$ can be obtained by differentiating $b_i(\eta_i)$:\break
$\mu_i(\eta_i) = \nabla_{\eta_i}\, b_i(\eta_i)$ \citep[Corollary 2.3,][pp.~35--36]{Br86}.
\end{enumerate}

The relationship to GLMs facilitates the interpretation and dissemination of results.
The following proposition clarifies the relationship to GLMs.

\begin{proposition}[: GLM Representation of Conditionals]
\label{prop_general_glm}
{\em
Consider any pair of units $\{i, j\} \subset \mP_N$ ($i < j$) and assume that $g_i$ and $h_{i,j}$ are affine functions of $y_i^\star$ for any given $(\bx,\, \by_{-i},\, \bz) \in \mX \times \mY_{-i} \times \mZ$,
in the sense that there exist known functions $g_{i,0}: \mX_i  \mapsto \mathbb{R}^q$,\,
$g_{i,1}: \mX_i \mapsto \mathbb{R}^q$,\,
$h_{i,j,0}: \mX \times \mY_j  \times \mZ \mapsto \mathbb{R}^r$,\,
and $h_{i,j,1}: \mX \times \mY_j \times \mZ \mapsto \mathbb{R}^r$ such that
\beno
g_i(\bx_i,\, y_i^\star)
\;\coloneqq\; g_{i,0}(\bx_i) + g_{i,1}(\bx_i)\; y_i^\star\s
\\
h_{i,j}(\bx,\, y_i^\star,\, y_j^\star,\, \bz)
\;\coloneqq\; h_{i,j,0}(\bx,\, y_j^\star,\, \bz) + h_{i,j,1}(\bx,\, y_j^\star,\, \bz)\; y_i^\star.
\ee
Then the conditional distribution of response\, $Y_i \mid (\bX,\, \bY_{-i},\, \bZ) = (\bx,\, \by_{-i},\, \bz)$ by unit $i$ can be represented by a GLM with linear predictor
\beno
\eta_i(\nat;\, \bx,\, \by_{-i}^\star,\, \bz)
&\coloneqq& \nat^\top \left(g_{i,1}(\bx_i),\; \dsum_{j\, \in\, \mP_N \setminus\, \{i\}}\, h_{i,j,1}(\bx,\, y_j^\star,\, \bz)\right)
\ee
and cumulant-generating function
\beno
b_i(\eta_i(\nat;\, \bx,\, \by_{-i}^\star,\, \bz))
&\coloneqq& \psi\, \log \dint_{\mY_i}\, a_{\mY}(y)\, \exp\left(\dfrac{\eta_i(\nat;\, \bx,\, \by_{-i}^\star,\, \bz)\; y}{\psi}\right) \dd \nu_{\mY}(y).
\ee
}
\end{proposition}
To ease the notation,
we henceforth write $\eta_i$ instead of\, $\eta_i(\nat;\, \bx,\, \by_{-i}^\star,\, \bz)$.

Proposition \ref{prop_general_glm} supplies a recipe for representing the conditional distribution of responses
$Y_i \mid (\bX,\, \bY_{-i},\, \bZ) = (\bx,\, \by_{-i},\, \bz)$ by a GLM:
\begin{enumerate}
\item {\em Conditional distribution:}
The conditional distribution of response $Y_i$ is an exponential family distribution,
which can be represented by a GLM with conditional mean $\mu_i(\eta_i)$, linear predictor $\eta_i$, and scale parameter $\psi$.
\item {\em Conditional mean:}
The conditional mean $\mu_i(\eta_i) \coloneqq \mbE_{\eta_i}(Y_i \mid \bx,\, \by_{-i},\, \bz)$ can be obtained by differentiating $b_i(\eta_i)$: 
$\mu_i(\eta_i) = \nabla_{\eta_i}\, b_i(\eta_i)$.
Since the map $\eta_i \mapsto \mu_i$ is one-to-one and invertible \citep[Theorem 3.6,][p.~74]{Br86},\,
$\eta_i$ can be obtained by inverting $\mu_i(\eta_i)$.
\end{enumerate}

Thus,
the proposed regression framework for dependent responses and connections $(\bY, \bZ) \mid \bX = \bx$ inherits the GLM advantages of being interpretable and widely applicable,
without assuming that responses or connections are independent.
As a result,
the proposed regression framework can be viewed as a generalization of GLMs.

\subsection{Example: Model Specification}
\label{sec:running_example}

We showcase how a joint probability model for dependent responses and connections $(\bY, \bZ) \mid \bX = \bx$ with local dependence can be constructed,
leveraging additional structure in the form of overlapping neighborhoods $\mN_1, \ldots, \mN_N$ to control the dependence among responses and connections in small and large populations.

We focus on units $i \in \mP_N$ with binary,
count-,
or real-valued predictors $x_i \in \mX_i$ and responses $Y_i \in \mY_i$ and binary connections $Z_{i,j} \in \{0, 1\}$.
Starting with $g_i$,
we capture the main effect of $Y_i^\star$ and the interaction effect of $x_i$ and $Y_i^\star$ by specifying $g_i$ as follows:
\be
\label{eq:gi}
\nat_g
\,\coloneqq\,
\left(
\begin{array}{ccc}
\alpha_{\mY}\\
\beta_{\mX,\mY}
\end{array}
\right)
\,\in\, \mR^2,
&
g_i
\,\coloneqq\,
\left(
\begin{array}{ccc}
y_i^\star\\
x_i\, y_i^\star
\end{array}
\right)
\,\in\, \mR^2.
\ee
Turning to $h_{i,j}$,
we define neighborhood-related terms
\be\label{eq:cstar}
c_{i,j} \coloneqq \one(\mN_i\, \cap\, \mN_j\, \neq \, \emptyset), \quad
d_{i,j}(\bz) \coloneqq \one(\exists\; k\, \in\, \mN_i\, \cap\, \mN_j\, :\, z_{i,k} = z_{k,j} = 1).
\ee
To capture unobserved heterogeneity in the propensities of units to form connections,
we introduce the
$N$-vector $\bm\alpha_{\mZ} \coloneqq (\alpha_{\mZ,1}, \ldots, \alpha_{\mZ,N}) \in \mR^N$.
In addition,
we penalize connections among units $i$ and $j$ with non-overlapping neighborhoods and capture transitive closure along with treatment and outcome spillover by specifying $h_{i,j}$ as follows:
\be
\label{eq:hij}
\nat_h
\coloneqq
\left(
\begin{array}{ccc}
\bm\alpha_{\mZ}\\
\lambda\\
\gamma_{\mZ,\mZ}\\
\gamma_{\mX,\mY,\mZ}\\
\gamma_{\mY,\mY,\mZ}
\end{array}
\right)
\in \mR^{N+4},
&
h_{i,j}
\coloneqq
\left(
\begin{array}{ccc}
\bm{e}_{i,j}\, z_{i,j}\\
-(1 - c_{i,j})\, z_{i,j} \log N\\
d_{i,j}(\bz)\, z_{i,j}\\
c_{i,j}\, (x_i\, y_j^\star + x_j\, y_i^\star)\, z_{i,j}\\
c_{i,j}\, y_i^\star\, y_{j}^\star\, z_{i,j}
\end{array}
\right)
\in \mR^{N+4},
\ee
where
$\bm{e}_{i,j}$ denotes the $N$-vector whose
whose $i$th and $j$th coordinates are $1$ and whose other coordinates are all $0$.
The parameters $\alpha_{\mZ,1}, \ldots, \alpha_{\mZ,N}$ can be interpreted as
the propensities of units $1, \dots, N$ to form connections;
$\lambda > 0$ discourages connections among units with non-overlapping neighborhoods;
$\gamma_{\mZ,\mZ}$ quantifies the tendency towards transitive closure among connections;
and $\gamma_{\mX,\mY,\mZ}$ and $\gamma_{\mY,\mY,\mZ}$ capture treatment and outcome spillover,
respectively.
Sections \ref{sec:glm_y} and \ref{subsec:GLMRepresentationOfZ} demonstrate that the
interpretation of these effects is facilitated by the fact that the conditional distributions of $Y_i$ and
$Z_{i,j}$ can be represented by GLMs.

\subsubsection{GLM Representation of Responses $Y_i$}
\label{sec:glm_y}

\begin{figure}[t!]
\centering
\includegraphics[width=0.45\linewidth, page = 1]{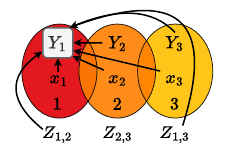}
\caption{
Given $N=3$ units $1, 2, 3$ with neighborhoods $\mN_1 \coloneqq \{1, 2\}$,\,
$\mN_2 \coloneqq  \{1, 2, 3\}$,\,
and $\mN_3 \coloneqq \{2, 3\}$,
the arrows indicate which predictors, responses,
and connections can affect the response $Y_1$ of unit $1$
according to the model specified by Equations~(\ref{eq:gi}) and~(\ref{eq:hij}).
}
\label{fig:overlap}
\end{figure}

To interpret the model specified by Equations~(\ref{eq:gi}) and~(\ref{eq:hij}),
we take advantage of the fact that the conditional distribution of response
$Y_i \mid (\bX,\, \bY_{-i},\, \bZ) = (\bx,\, \by_{-i},\, \bz)$ by unit $i$
can be represented by a GLM with linear predictor
\be
\label{eq:fc_x}
\eta_i
&=& \alpha_{\mY} + \beta_{\mX,\mY}\; x_{i} + \gamma_{\mX,\mY,\mZ}\, \dsum_{j:\, \mN_i\, \cap\, \mN_j\, \neq\, \emptyset}\, x_j\, z_{i,j}
+ \gamma_{\mY,\mY,\mZ}\, \dsum_{j:\, \mN_i\, \cap\, \mN_j\, \neq\, \emptyset}\, y_j^\star\, z_{i,j}.
\ee
Figure~\ref{fig:overlap} depicts the predictors, responses,
and connections that affect the conditional distribution of response $Y_i$.
We provide three specific examples,
depending on the support set of response $Y_i$.

{\bf Example 1: Real-valued Responses $Y_i \in \mR$.}
Let $\psi \in (0, +\infty)$ and
\beno
a_\mY(y_i)
&\coloneqq&  \dfrac{1}{\sqrt{2\, \pi\, \psi}} \, \exp \left(- \dfrac{y_i^2}{2\, \psi}\right)  \, \mathbb{I}(y_i \in \mR).
\ee
\begin{enumerate}
\item {\em Conditional distribution:}
The conditional distribution of response $Y_i$ is $N(\mu_i(\eta_i),\, \psi)$.
\item {\em Conditional mean:}
The conditional mean $\mu_i(\eta_i)$ can be obtained by differentiating $b_i(\eta_i) = \eta_i^2 /\, 2$ with respect to $\eta_i$,\,
giving $\mu_i(\eta_i) = \eta_i$:
\beno
\mu_i(\eta_i)
&=& \alpha_{\mY} + \beta_{\mX,\mY}\; x_{i} + \gamma_{\mX,\mY,\mZ}\, \dsum_{j:\, \mN_i\, \cap\, \mN_j\, \neq\, \emptyset}\, x_j\, z_{i,j}
+ \gamma_{\mY,\mY,\mZ}\, \dsum_{j:\, \mN_i\, \cap\, \mN_j\, \neq\, \emptyset}\, y_j^\star\, z_{i,j}.
\ee
Under certain restrictions on $\gamma_{\mY,\mY,\mZ}$,
the conditional distribution of $\bY \mid (\bX, \bZ) = (\bx, \bz)$ is $N$-variate Gaussian.
The restrictions on $\gamma_{\mY,\mY,\mZ}$ depend on the neighborhoods $\mN_i$ and $\mN_j$ and
connections $Z_{i,j}$ of pairs of units $\{i, j\} \subset \mP_N$;
see Proposition \ref{prop_gaussian_y_mid_z_2} in Section \ref{sec:proof_prop1} of the Supplementary Materials.
\end{enumerate}

{\bf Example 2: Count-valued Responses $Y_i \in \{0, 1, \dots\}$.}
Let $\psi \coloneqq 1$ and
\beno
a_\mY(y_i)
&\coloneqq& \dfrac{1}{y_i!}\; \mathbb{I}(y_i \in \{0, 1, \dots\}).
\ee
\begin{enumerate}
\item {\em Conditional distribution:}
The conditional distribution of response $Y_i$ is $\mbox{Poisson}(\mu_i(\eta_i))$.
\item {\em Conditional mean:}
The conditional mean $\mu_i(\eta_i)$ can be obtained by differentiating $b_i(\eta_i) = \exp(\eta_i)$ with respect to $\eta_i$,\,
giving $\mu_i(\eta_i) = \exp(\eta_i)$.
\end{enumerate}

{\bf Example 3: Binary Responses $Y_i \in \{0,\, 1\}$.}
Let $\psi \coloneqq 1$ and $a_{\mY}(y_i) \coloneqq \mathbb{I}(y_i \in \{0,1\})$.
\begin{enumerate}
\item {\em Conditional distribution:}
The conditional distribution of response $Y_i$ is $\mbox{Bernoulli}(\mu_i(\eta_i))$.
\item {\em Conditional mean:}
The conditional mean $\mu_i(\eta_i)$ can be obtained by differentiating $b_i(\eta_i) = \log(1 + \exp(\eta_i))$ with respect to $\eta_i$,\,
giving $\mu_i(\eta_i) = \logit^{-1}(\eta_i)$.
\end{enumerate}

{\bf Interpretation of Examples.}
According to Equations~(\ref{eq:gi}) and~(\ref{eq:hij}), regardless of the conditional distribution
of response $Y_i$,
$\alpha_{\mY}$ can be viewed as an intercept,
while $\beta_{\mX, \mY}$ captures the relationship between predictor $x_i$ and response $Y_i$.
The parameters $\gamma_{\mX,\mY,\mZ}$ and $\gamma_{\mY,\mY,\mZ}$ capture two distinct spillover effects:
\bi
\item {\em Treatment spillover:}
$\gamma_{\mX,\mY,\mZ} \neq 0$ allows the outcome $Y_{i}$ of unit $i$ to be affected by the treatments $x_j$ of its neighbors $j \in \mN_i$ and non-neighbors $j \not\in \mN_i$,\,
provided $\mN_i\, \cap\, \mN_j \neq \emptyset$ and $i$ and $j$ are connected (see Figure \ref{fig:overlap}).
\item {\em Outcome spillover:}
$\gamma_{\mY,\mY,\mZ} \neq 0$ allows the outcome $Y_{i}$ of unit $i$ to be affected by the outcomes $y_j$ of its neighbors $j \in \mN_i$ and non-neighbors $j \not\in \mN_i$,\,
provided $\mN_i\, \cap\, \mN_j \neq \emptyset$ and $i$ and $j$ are connected (see Figure \ref{fig:overlap}).
\ei
The proposed regression framework can be used for causal inference under network interference,
which studies treatment spillover.
That said,
the framework is considerably broader,
because it permits outcome spillover, 
and spillover need not be studied in a causal setting.

\subsubsection{GLM Representation of Connections $Z_{i,j}$}
\label{subsec:GLMRepresentationOfZ}

The conditional mean $\mu_{i,j}(\eta_{i,j}) \coloneqq \mbE_{\eta_{i,j}}(Z_{i,j} \mid \bx,\, \by,\, \bz_{-\{i,j\}}) = \logit^{-1}(\eta_{i,j})$ of connection $Z_{i,j} \in \{0, 1\}$ depends on the linear predictor
\beno
\label{eq:fc_z}
\eta_{i,j}
= \alpha_{\mZ,i} + \alpha_{\mZ,j}
- (1 - c_{i,j})\, \lambda\, \log N
+ c_{i,j} \left[\gamma_{\mZ,\mZ}\, \Delta_{i,j}(\bz)
+ \gamma_{\mX,\mY,\mZ}\, (x_i\, y_j^\star + x_j\, y_i^\star)
+ \gamma_{\mY,\mY,\mZ}\, y_i^\star\, y_j^\star \right],
\ee
where $\Delta_{i,j}: \mZ \mapsto \mR$ is the change in $\sum_{a<b}^N d_{a,b}(\bz)$ due to transforming $z_{i,j}$ from $0$ to $1$.
The logistic regression representation of $Z_{i,j} \mid (\bX, \bY, \bZ_{-\{i,j\}}) = (\bx, \by, \bz_{-\{i,j\}})$ facilitates interpretation:
e.g.,
$\alpha_{\mZ,i}$ captures heterogeneity among units $i$ in forming connections.
If $\lambda > 0$,
the sparsity-inducing term $- (1 - c_{i,j})\, \lambda\, \log N$
penalizes connections between pairs of units with non-overlapping neighborhoods,
where the $\log N$-term can be motivated in the special case of Bernoulli random graphs \citep{
KrCoHe23}:
If $Z_{i,j} \iid \mbox{Bernoulli}(\pi)$ and the expected degrees $\mbE \sum_{j=1}^N Z_{i,j}$ are bounded,
then $\pi = O(1/N)$ and $\logit(\pi) = O(\log N)$.
In addition,
the model captures three forms of dependencies.
First,
the model encourages $i$ and $j$ to be connected when $i$ and $j$ are both connected to some $k\, \in\, \mN_i\, \cap\, \mN_j$,
provided $\mN_i\, \cap\, \mN_j \neq \emptyset$ and $\gamma_{\mZ,\mZ} > 0$.
Second,
the model encourages $i$ and $j$ to be connected when $x_i \, y_j^\star > 0$ or $x_j \, y_i^\star > 0$,
provided $\mN_i\, \cap\, \mN_j \neq \emptyset$ and $\gamma_{\mX,\mY,\mZ} > 0$.
Third,
the model encourages $i$ and $j$ to be connected when $y_i^\star\; y_j^\star > 0$,
provided $\mN_i\, \cap\, \mN_j\, \neq\, \emptyset$ and~$\gamma_{\mY,\mY,\mZ}\, >\, 0$.

\section{Scalable Statistical Computing}
\label{sec:computing}

To learn the regression framework from a single observation $(\by, \bz)$ of dependent responses and connections $(\bY, \bZ) \mid \bX = \bx$, 
we develop scalable methods based on convex optimization of pseudo-likelihoods using minorization-maximization methods.

\subsection{Pseudo-Loglikelihood}
\label{sec:cl}

Let
\be
\label{eq:comlik}
\ell(\nat;\, \by,\, \bz)
&\coloneqq& \dsum_{i=1}^N \ell_i (\nat;\, \by,\, \bz) + \dsum_{i=1}^{N-1}\, \dsum_{j=i+1}^N \ell_{i,j}(\nat;\, \by,\, \bz),
\ee
where the dependence on predictors $\bx \in \mX$ is suppressed and $\ell_i$ and $\ell_{i,j}$ are defined by
\beno
\ell_{i}(\nat;\, \by,\, \bz)
&\coloneqq& \log f_{\nat}(y_i \mid \by_{-i},\, \bz)
&\mbox{and}&
\ell_{i,j}(\nat;\, \by,\, \bz)
&\coloneqq& \log f_{\nat}(z_{i,j} \mid \by,\, \bz_{-\{i,j\}}).
\ee
The pseudo-loglikelihood $\ell$ is based on full conditional densities of responses $Y_i$
and connections $Z_{i,j}$ and is hence tractable.
In addition,
$\ell$ is a sum of exponential family loglikelihood functions $\ell_{i}$ and $\ell_{i,j}$,
each of which is
concave and twice differentiable on the convex set $\Nat$ \citep[Theorem 1.13, p.~19 and Lemma 5.3, p.~146]{Br86},
proving Lemma \ref{concave}:
\begin{lemma}[: Convexity and Smoothness]
\label{concave}
{\em
The set $\Nat$ is convex
and the pseudo-loglikelihood function $\ell: \Nat \mapsto \mR$,
considered as a function of\, $\nat$ for fixed $(\by, \bz) \in \mY \times \mZ$,
is twice differentiable with a negative semidefinite Hessian matrix on $\Nat$.}
\end{lemma}

In light of the tractability and concavity of $\ell$,
it makes sense to base statistical learning on pseudo-likelihood estimators of the form
\be
\label{delta}
\widehat\Nat(\delta_N)
&\coloneqq& \left\{\nat\, \in\, \Nat:\; |\!|\nabla_{\nat}~\ell(\nat;\, \by,\, \bz)|\!|_\infty\; \leq\; \delta_N\right\},
\ee
where $\nabla_{\nat}$ denotes the gradient with respect to $\nat$
while $|\!|\bm{v}|\!|_\infty \,\coloneqq\, \max_{1 \leq k \leq p}\, |v_k|$ denotes the $\ell_\infty$-norm of vectors $\bm{v} \in \mR^p$.
The quantity $\delta_N \in [0, +\infty)$ can be viewed as a convergence criterion of a root-finding algorithm and can depend on $N$.
The set $\widehat\Nat(\delta_N)$ consists of maximizers of $\ell$ when $\delta_N = 0$,
and maximizers and near-maximizers when $\delta_N > 0$.

\subsection{Minorization-Maximization (MM)}
\label{sec:two.stage}

While pseudo-likelihood estimators $\mcle \in \Mcle$ can be obtained by standard root-finding algorithms,
inverting the $p \times p$ negative Hessian
of\, $\ell$ at each iteration is time-consuming,
because inversions require $O(p^3)$ operations
and $p$ can increase with $N$.
We thus divide the task of estimating $p$ parameters into two subtasks using MM methods \citep{HunterLange2004}.
In the example model specified by Equations~(\ref{eq:gi}) and~(\ref{eq:hij}) for
binary,
count-,
or real-valued predictors and responses $(X_i, Y_i)$ and binary connections $Z_{i,j}$,
we partition $\nat \in \mR^{N + 6}$ into $N$ nuisance parameters,\,
$\nat_1 \coloneqq (\alpha_{\mZ,1},\, \dots,\, \alpha_{\mZ,N}) \in \mR^N$,
and $6$ parameters of primary interest,\,
$\nat_2 \coloneqq (\lambda,\, \alpha_{\mY},\, \beta_{\mX, \mY},\, \gamma_{\mZ, \mZ},\,
\gamma_{\mX, \mY, \mZ},\, \gamma_{\mY, \mY, \mZ}) \in \mR^6$.
We then partition the negative Hessian of $\ell$ accordingly:
\be \label{Hessian}
-\nabla_{\nat}^2~\ell(\nat;\, \by,\, \bz)
&\coloneqq&
\begin{pmatrix}
\mathbf{A}(\nat)  & \mathbf{B}(\nat)\\
\mathbf{B}(\nat)^\top & \mathbf{C}(\nat) &
\end{pmatrix},
\ee
where $\mathbf{A}(\nat) \in \mR^{N \times N}$,\,
$\mathbf{B}(\nat) \in \mR^{N \times 6}$,\,
and $\mathbf{C}(\nat) \in \mR^{6\times 6}$.
We suppress the dependence of $\mathbf{A}(\nat)$,
$\mathbf{B}(\nat)$,
and $\mathbf{C}(\nat)$ on $(\by,\, \bz)$
and henceforth write $\ell(\nat_1, \nat_2;\, \by,\, \bz)$ instead of $\ell(\nat;\, \by,\, \bz)$.

Iteration $t + 1$ then consists of two steps:
\bi
\item[] {\bf Step 1:}
Find $\nat_1^{(t+1)}$ satisfying $\ell(\nat_1^{(t+1)},\, \nat_2^{(t)};\, \by,\, \bz)\, \ge\, \ell(\nat_1^{(t)},\, \nat_2^{(t)};\, \by,\, \bz)$.
\item[] {\bf Step 2:}
Find $\nat_2^{(t+1)}$ satisfying $\ell(\nat_1^{(t+1)},\, \nat_2^{(t+1)};\, \by,\, \bz)\, \ge\, \ell(\nat_1^{(t+1)},\, \nat_2^{(t)};\, \by,\, \bz)$.
\ei
In Step 1,
it is inconvenient to invert the high-dimensional $N \times N$ matrix
\be
\label{A}
\bA(\nat^{(t)})
&\coloneqq& -\dsum_{i<j}^N\, \nabla_{\nat_1}^2\, \ell_{i,j}(\nat_1,\, \nat_2^{(t)};\, \by,\, \bz) \Big|_{\nat_1=\nat_1^{(t)}}
&=&  \dsum_{i<j}^N \pi_{i,j}^{(t)}\, (1-\pi_{i,j}^{(t)})\, \bm{e}_{i,j}\, \bm{e}_{i,j}^\top,
\ee
where $\pi_{i,j}^{(t)} \coloneqq \mbP_{\nat^{(t)}}(Z_{i,j} = 1 \mid \by,\, \bz_{-\{i,j\}})$.
We thus increase $\ell$ by maximizing a minorizer of $\ell$,
replacing $\bA(\nat^{(t)})$ by a constant matrix $\bA^\star$ that
only needs to be inverted once.

\begin{lemma}[: Minorizer]
\label{lemma.minorizer}
{\em
Define
\beno
\bA^\star
&\coloneqq& \dfrac14\,\dsum_{i<j}^N \bm{e}_{i,j}\, \bm{e}_{i,j}^\top
\ =\ \dfrac14\, \left[(N-2)\, \bI + \bm{1} \bm{1}^\top\right]
\ =\  \left[ \dfrac{4}{N-2}\, \left(\bI - \dfrac{1}{2\, N-2}\, \bm{1} \bm{1}^\top\right) \right]^{-1},
\ee
where $\bI$ is the $N \times N$ identity matrix and $\bm{1}$ is the $N$-vector of ones.
Then the function
\beno
m(\nat_1;\, \nat_1^{(t)},\, \nat_2^{(t)},\, \by,\, \bz)
&\coloneqq& \ell(\nat_1^{(t)},\, \nat_2^{(t)};\, \by,\, \bz)\s
\\
&+& \left(\nabla_{\nat_1}\, \ell(\nat_1,\, \nat_2^{(t)};\, \by,\, \bz)\Big|_{\nat_1=\nat_1^{(t)}}\right)^\top (\nat_1 - \nat_1^{(t)})\s
\\
&+& \dfrac12\, (\nat_1 - \nat_1^{(t)})^\top\, (-\bA^\star)\, (\nat_1 - \nat_1^{(t)})
\ee
is a minorizer of $\ell$
at $\nat_1^{(t)}$ for fixed $\nat_2^{(t)}$,
in the sense that
\beno
m(\nat_1;\, \nat_1^{(t)},\, \nat_2^{(t)},\, \by,\, \bz)
&\leq& \ell(\nat_1,\, \nat_2^{(t)};\, \by,\, \bz)\; \mbox{ for all }\; \nat_1 \in \mR^N\s,
\\
m(\nat_1^{(t)};\, \nat_1^{(t)},\, \nat_2^{(t)},\, \by,\, \bz)
&=& \ell(\nat_1^{(t)},\, \nat_2^{(t)};\, \by,\, \bz).
\ee
}
\end{lemma}

\vspace{-.6cm}
\noindent
Lemma \ref{lemma.minorizer} is proved in Section \ref{sec:proof_lemma2} of the Supplementary Materials.
Step 1 may be implemented by an MM algorithm,
as the closed-form maximizer of the minorizer $m$ is
\be
\label{Step1Update}
\nat_1^{(t+1)}
&\coloneqq& \nat_1^{(t)} +
\left(\mathbf{A}^\star \right)^{-1}
\left(\nabla_{\nat_1}\, \ell(\nat_1, \nat_2^{(t)};\, \by,\, \bz)\Big|_{\nat_1=\nat_1^{(t)}}\right).
\ee
We accelerate the MM step in Equation~\eqref{Step1Update} with quasi-Newton methods.
Details can be found in Section~\ref{sec:quasi.newton} of the Supplementary Materials.
Compared to a Newton-Raphson algorithm,
the accelerated MM-step reduces the per-iteration computational complexity from $O(N^3)$ to $O(N^2)$.
Step 2 updates the low-dimensional parameter vector of interest $\nat_2^{(t+1)} \in \mR^6$ given the high-dimensional nuisance parameter vector $\nat_1^{(t+1)} \in \mR^N$ by a Newton-Raphson step.
The concavity of $\ell$,
established in Lemma~\ref{concave},
guarantees that
\beno
\ell(\nat_1^{(t)},\, \nat_2^{(t)};\, \by,\, \bz)
\lte \ell(\nat_1^{(t+1)},\, \nat_2^{(t)};\, \by,\, \bz)
\lte \ell(\nat_1^{(t+1)},\, \nat_2^{(t+1)};\, \by,\, \bz).
\ee

\hide{
We accelerate the algorithm by replacing the MM step in Equation~\eqref{Step1Update} with a quasi-Newton step.
Details can be found in Section~\ref{sec:quasi.newton} of the Supplementary Materials.
Compared to a Newton-Raphson algorithm,
the accelerated MM-step reduces the per-iteration computational complexity from $O(N^3)$ to $O(N^2)$.
Step 2 updates $\nat_2^{(t+1)}$ given $\nat_1^{(t+1)}$ by a Newton-Raphson step.
The concavity of $\ell$,
established in Lemma~\ref{concave},
guarantees that
\beno
\ell(\nat_1^{(t+1)},\, \nat_2^{(t+1)};\, \by,\, \bz)
&\geq& \ell(\nat_1^{(t+1)},\, \nat_2^{(t)};\, \by,\, \bz).
\ee 

{\bf Remarks.}
The MM algorithm for binary connections can be extended to non-binary connections using the de Pierro method \citep{becker_em_1997}.
If the scale parameter $\psi$ is unknown,
it can be estimated by iterative methods.

}

\subsection{Quantifying Uncertainty}
\label{sec:standard_errors}

The uncertainty about the maximum pseudo-likelihood estimator $\widehat\nat$ of the data-generating parameter vector $\truth$ can be quantified based on the covariance matrix of the sampling distribution of $\widehat\nat$, which we derive as follows:
The mean-value theorem for vector-valued functions \citep[][Equations (2) and (3), pp.~68--69]{OrRh00}
implies that there exist real numbers $t_1, \ldots, t_p \in (0,\, 1)$ such that
\be\label{eq:MeanValue}
\nabla_{\nat}\; \ell(\nat;\, \by,\, \bz)\Big|_{\nat=\widehat\nat}
- \nabla_{\nat}\; \ell(\nat;\, \by,\, \bz)\Big|_{\nat=\truth}
&=& \bm{H}(\widehat\nat,\, \truth;\, \by,\, \bz)\; (\widehat\nat - \truth),
\ee
where
\beno
\bm{H}(\widehat\nat,\, \truth;\, \by,\, \bz)
&\coloneqq&
\left(
\begin{array}{cccccc}
g_1^\prime(\truth + t_1\, (\widehat\nat - \truth);\, \by,\, \bz)
\\
\vdots
\\
g_p^\prime(\truth + t_p\, (\widehat\nat - \truth);\, \by,\, \bz)
\end{array}
\right).
\ee
Here,
$g_k(\nat;\, \by,\, \bz)$ is the $k$th coordinate of $\nabla_{\nat}\,\, \ell(\nat;\, \by, \bz)$ and $g_k^\prime(\nat;\, \by,\, \bz)$ is the row vector of partial derivatives of $g_k(\nat;\, \by,\, \bz)$ with respect to $\nat$ ($k = 1, \ldots, p$).
Leveraging Equation~(\ref{eq:MeanValue}) along with $\nabla_{\nat}\,\, \ell(\nat;\, \by,\, \bz)|_{\nat=\widehat\nat} = \bm{0}$ gives the exact covariance matrix $\widehat\nat$:
\beno
\mbV_{\truth}(\widehat\nat)
&=& \mbV_{\truth}\left[-\bm{H}(\widehat\nat,\, \truth;\, \bY, \bZ)^{-1}\; \nabla_{\nat}\; \ell(\nat;\, \bY,\, \bZ)\Big|_{\nat=\truth}\right].
\ee
If $N$ is large,
$\truth$ can be replaced by $\widehat\nat$,
because $|\!|\widehat\nat - \truth|\!|_\infty$ is small with high probability according to Theorem \ref{thm:mple_consistency} in Section \ref{sec:theory}.
The resulting approximation of $\mbV_{\truth}(\widehat\nat)$ is
\be
\label{approximate.covariance}
\mbV_{\widehat\nat}(\widehat\nat)
&=& \mbV_{\widehat\nat}\left[-\nabla_{\nat}^2\; \ell(\nat;\, \bY,\, \bZ)\Big|_{\nat=\widehat\nat}^{-1}\;\, \nabla_{\nat}\; \ell(\nat;\, \bY,\, \bZ)\Big|_{\nat=\widehat\nat}\right],
\ee
which can be estimated by simulating responses and connections $(\bY, \bZ) \mid \bX = \bx$
from $\mbP_{\widehat\nat}$ using Markov chain Monte Carlo methods.

{\bf Remark: Asymptotic Distribution.}
Establishing asymptotic normality for pseudo-likelihood estimators based on a single observation of
dependent responses and connections $(\bY, \bZ) \mid \bX = \bx$ in scenarios with $p \to \infty$
parameters is an open problem.
Asymptotic normality results in the most closely related literature---the literature in applied probability concerned with Ising models, Gibbs measures, and Markov random fields in single-observation scenarios \citep[e.g.,][]{JeKu94,CoJa98}---assume the presence of lattice structure and a fixed number of parameters $p$,
in addition to other assumptions motivated by applications in physics.
In the current setting,
none of these assumptions holds,
although simulation results
in Section \ref{sec:simulations} suggest that the sampling distribution of
$\widehat\nat$ is approximately normal and that normal-based confidence
intervals based on Equation~(\ref{approximate.covariance}) achieve close-to-nominal coverage
probabilities.

\vspace{-.5cm}

\section{Theoretical Guarantees}
\label{sec:theory}

We establish convergence rates for pseudo-likelihood estimators $\widehat\Nat(\delta_N)$ based on a single observation of dependent responses and connections $(\bY, \bZ) \mid \bX = \bx$.
To cover a wide range of models for binary, count-, and real-valued responses and connections,
we introduce a general theoretical framework and showcase convergence rates in a specific example.

We denote by $\truth \in \Nat \subseteq \mR^p$ the data-generating parameter vector and by $\mB_{\infty}(\truth,\, \rho)\, \coloneqq\, \{\nat \in \mathbb{R}^p: \infnormvec{\nat-\truth} < \rho\}$
a hypercube with center $\truth \in \Nat$ and width $2\, \rho \in (0, +\infty)$.
Let
\[
\mathscr{I}(S) \,\coloneqq\,
\left\{(\by,\, \bz) \in \mY \times \mZ:\; -\nabla_{\nat}^2\; \ell(\nat;\, \by,\, \bz) \mbox{ is invertible for all $\nat \in S$}\right\}
\]
and,
for some $\epsilon^\star \in (0, +\infty)$ and $\mbH \,\subseteq\, \mathscr{I}\left(\mB_{\infty}(\truth,\, \epsilon^\star)\right)$, let
\[
\Lambda_{N}(\truth)
\,\coloneqq\, \sup\limits_{(\by,\, \bz)\, \in\, \mbH}\; \sup\limits_{\nat\, \in\, \mB_{\infty}(\truth,\, \epsilon^\star)}\, \mnorm{(-\nabla_{\nat}^2\; \ell(\nat;\, \by,\, \bz))^{-1}}_{\infty},
\]
where $\mnorm{.}_\infty$ is the $\ell_\infty$-induced matrix norm.
The set $\mbH$ can be a proper subset of\, $\mathscr{I}\left(\mB_{\infty}(\truth,\, \epsilon^\star)\right)$,
provided $\mbH$ is a high probability subset of $\mY \times \mZ$.
The definition of $\mbH$ is motivated by the fact that characterizing the set of all $(\by,\, \bz) \in \mY \times \mZ$\, for which the Hessian is invertible can be challenging,
but finding a sufficient condition for invertibility is often possible.

\begin{theorem}[: Convergence Rate]
\label{thm:mple_consistency}
{\em Consider a single observation of
$(\bY, \bZ) \in \mY \times \mZ$ generated by model \eqref{eq:ernm_model} with parameter vector\, $\truth \in \Nat \subseteq \mR^p$,
where $\mY \times \mZ$ is a finite,
countably infinite,
or uncountable set.
Assume that there exists a sequence $\rho_1, \rho_2, \ldots \in [0, +\infty)$ satisfying $\rho_N = o(1)$ so that the events $|\!|\nabla_{\nat}\; \ell(\nat;\, \bY, \bZ)|_{\nat = \truth} - \mbE\; \nabla_{\nat}\; \ell(\nat;\, \bY, \bZ)|_{\nat = \truth}|\!|_{\infty} < \delta_N$ and $(\bY, \bZ) \in \mbH$ occur with probability $1 - o(1)$,
where $\delta_N \coloneqq \rho_N / (2\, \Lambda_N(\truth))$.
Then there exists a positive integer $N_0$ such that,
for all $N > N_0$,
the random set $\widehat\Nat(\delta_N)$ is non-empty and,
with probability $1 - o(1)$,
satisfies
\beno
\widehat\Nat(\delta_N)
&\subseteq& \mB_{\infty}(\truth,\, \rho_N).
\ee
}
\end{theorem}

\vspace{-1cm}

Theorem \ref{thm:mple_consistency} is proved in Section \ref{sec:proof.theorem1}
of the Supplementary Materials.
The requirement $\delta_N \coloneqq \rho_N / (2\, \Lambda_N(\truth))$ implies $\rho_N \propto \delta_N\, \Lambda_N(\truth)$,
so the convergence rate depends on
\bi
\item the strength of concentration of the gradient $\nabla_{\nat}\,\, \ell(\nat;\, \bY, \bZ)|_{\nat = \truth}$ around its expectation $\mbE\, \nabla_{\nat}\,\, \ell(\nat;\, \bY, \bZ)|_{\nat = \truth}$ via $\delta_N$;
\item the inverse negative Hessian $(-\nabla_{\nat}^2~\ell(\nat;\, \by, \bz))^{-1}$ in a neighborhood $\mB_{\infty}(\truth,\, \epsilon^\star)$ of $\truth \in \Nat$ and a high probability subset $(\by, \bz) \in \mH$ of $\mY \times \mZ$ via $\Lambda_N(\truth)$.
\ei
The strength of concentration of $\nabla_{\nat}\,\, \ell(\nat;\, \bY, \bZ)|_{\nat = \truth}$ can be quantified by concentration inequalities for dependent random variables.
In general,
the strength of concentration depends on the sample space and the tails of the distribution,
the smoothness of the functions $g_i$ and $h_{i,j}$,
and the dependence induced by model \eqref{eq:ernm_model}.
To control the dependence among responses and connections $(\bY, \bZ) \mid \bX = \bx$,
one can take advantage of additional structure (e.g., one or more neighborhood structures, non-overlapping or overlapping subpopulations, or a metric space in which units are embedded).
For example,
each unit can have one or more neighborhoods (e.g., geographical neighbors and colleagues in the workplace),
and the responses and connections of the unit can be affected by any geographical neighbor and any colleague.
Theoretical guarantees can be obtained as long as the neighborhoods are not too large and do not overlap too much.

Specific convergence rates depend on the model.
To demonstrate,
consider predictors $x_i \in \mR$,
responses $Y_i \in \{0, 1\}$,
and connections $Z_{i,j} \in \{0, 1\}$ generated by a model capturing heterogeneity in the
propensities $\alpha_{\mZ,1},\, \dots, \alpha_{\mZ,N}$ of units $1, \dots, N$
to form connections,
transitive closure among connections with weight $\gamma_{\mZ,\mZ}$,
and treatment spillover with weight $\gamma_{\mX,\mY,\mZ}$;
compare Equations~(\ref{eq:gi}) and~(\ref{eq:hij}) in Section~\ref{sec:running_example}.
Since $Y_i \in \{0, 1\}$ and $Z_{i,j} \in \{0, 1\}$,
it is reasonable to specify $a_{\mY}(y_i) \coloneqq \mathbb{I}(y_i \in \{0,1\})$ and $a_{\mZ}(z_{i,j}) \coloneqq \mathbb{I}(z_{i,j} \in \{0,1\})$.
Convergence rates can be obtained under the following conditions.
\ExcludeFromLatexdiff{
\begin{assumption}[: Predictors]
\label{as:1_main}
{\em There exist finite constants $0 < c < C$ such that,
for each $i \in \mP_N$,
$x_i \in [0,\, C]$ and there exists $j \in \mP_N \setminus\, \{i\}$ such that $\mN_i\, \cap\, \mN_j \neq \emptyset$ and $x_j \in [c,\, C]$.
}
\end{assumption}
\vspace{-1cm}
\begin{assumption}[: Parameters]
\label{as:theta_main}
{\em The parameter space is $\Nat = \mR^{N+2}$ and there exists a constant $A \in (0, +\infty)$,
not depending on $N$, such that $\normsup{\truth} < A$.
}
\end{assumption}
\begin{assumption}[: Dependence]
\label{as:neigh0}
{\em The population $\mP$ consists of overlapping subpopulations $\mA_1, \mA_2, \ldots$,
which can be represented as vertices of a subpopulation graph $\mG_{\mA}$ with an edge
connecting $\mA_k$ and $\mA_l$ if $\mA_k\, \cap\, \mA_l \neq \emptyset$ ($k<l$).
For each $\mA_k$,
the number of subpopulations at geodesic distance $K$ in $\mG_{\mA}$ is $O(\log K)$.
For each $i \in \mP_i$,
the neighborhood is
\beno
\mN_i
&\coloneqq& \{j\, \in\, \mP_N: \mbox{ there exists } k \in \{1, 2, \ldots\} \text{ such that } i \in \mA_k \text{ and } j \in \mA_k\}.
\ee
There exists a constant $B \in (0,\, +\infty)$ such that $\max_{1 \leq i \leq N} |\mN_i| < B$.
}
\end{assumption}
}

Condition \ref{as:1_main} imposes restrictions on $\bx \in \mR^N$.
Condition \ref{as:theta_main} requires that
the data-generating parameter vector $\truth$ be contained in a compact
subset of\, $\Nat = \mR^{N+2}$.
The set of estimators $\widehat\Nat(\delta_N)$ is not restricted by Condition \ref{as:theta_main}
and consists of all $\nat \in \mR^{N+2}$ such that $|\!|\nabla_{\nat}~\ell(\nat;\, \bY, \bZ)|\!|_\infty\, \leq\, \delta_N$.
Condition \ref{as:theta_main} can be weakened in special cases,
allowing $|\!|\truth|\!|_\infty = O(\log N)$;
see Section \ref{sec:cor_nonoverlapping} of the Supplementary Materials.
Condition \ref{as:neigh0} controls the dependence among responses and connections $(\bY, \bZ) \mid \bX = \bx$
and can be weakened to $\max_{1 \leq i \leq N} |\mN_i| = O(\log N)$, 
as demonstrated by \citet{StSc20} in the special case of connections $\bZ$ (without predictors $\bX$ and responses $\bY$).

\begin{corollary}[: Example of Convergence Rate]
\label{theorem.model}
{\em
Consider a single observation of dependent responses and connections $(\bY, \bZ)$ generated by the model with parameter vector $\truth \coloneqq (\alpha_{\mZ,1}^\star,\, \dots, \alpha_{\mZ,N}^\star,\, \gamma_{\mZ,\mZ}^\star,\, \gamma_{\mX, \mY, \mZ}^\star) \in \mR^{N+2}$.
If Conditions \ref{as:1_main}--\ref{as:neigh0} hold,
there exist constants $K \in (0, +\infty)$ and $0 < L \leq U < +\infty$ along with an integer $N_0 \in \{3, 4, \dots\}$ such that,
for all $N > N_0$,
the quantity $\delta_N$ satisfies
\beno
L\, \sqrt{N \log N}
&\leq& \delta_N
&\leq& U\, \sqrt{N \log N},
\ee
and the random set $\widehat\Nat(\delta_N)$ is non-empty and satisfies
\beno
\widehat\Nat(\delta_N)
&\subseteq& \mB_{\infty}\left(\truth,\; K\,  \sqrt{\dfrac{\log N}{N}}
\right)
\ee
with probability at least $1 - 6\, / N^2$.
}
\end{corollary}

Corollary~\ref{theorem.model} is proved in Section~\ref{sec:proof.corollary1}
of the Supplementary Materials.
The same method of proof can be used to establish convergence rates for pseudo-likelihood estimators $\widehat\Nat(\delta_N)$ based on other models for dependent responses and connections $(\bY, \bZ) \mid \bX = \bx$,
provided there is additional structure to control the dependence among responses and connections.

\section{Simulation Results}
\label{sec:simulations}

To evaluate the performance of pseudo-likelihood estimators
$\widehat\nat \in \widehat\Nat(\delta_N)$ and the accompanying uncertainty quantification,
we simulate data from the example model specified by Equations~(\ref{eq:gi}) and~(\ref{eq:hij}).
The coordinates of the nuisance parameter vector,
$\nat_1^\star \coloneqq (\alpha_{\mZ,1}^\star$, $\dots$, $\alpha_{\mZ,N}^\star) \in \mR^N$,
are independent Gaussian draws with mean $-3/2$ and standard deviation $3/10$.
The parameter vector of primary interest,
$\nat_2^\star \coloneqq (\lambda^\star$,
$\alpha_{\mY}^\star$,
$\beta_{\mX,\mY}^\star$,
$\gamma_{\mZ,\mZ}^\star$,
$\gamma_{\mX,\mY,\mZ}^\star$,
$\gamma_{\mY,\mY,\mZ}^\star) \in \mR^6$,
is specified as $(1/5,\, -1,\, 3,\, 4/5,\, 1/2,\, -1/2)$.
The sparsity parameter $\lambda^\star = 1 / 5$
ensures that each unit has on average approximately 30 connections, regardless of the
value of $N$.
The neighborhood structure is based on $L=(N-25)/25$ intersecting
subpopulations $\mathscr{A}_1, \ldots, \mathscr{A}_L$,
where $\mathscr{A}_l$ consists of the 50 units $1+25\, (l-1), \ldots, 25\, (l+1)$ ($l = 1, \dots, L-1$).
For each $i \in \mP_N$, we define
the neighborhood $\mN_i \subset \mP_N$ to be the 50- or 75-unit union of all subpopulations
$\mathscr{A}_l$ containing $i$.
\hide{
Since the bounds on the quantity $\delta_N$ in Corollary \ref{theorem.model} depend on unknown constants,
it is impossible to determine whether $|\!|\nabla_{\nat}~\ell(\nat)|_{\nat=\nat^{(t+1)}}|\!|_\infty\; \leq\; \delta_N$.
We thus declare convergence when both $\spectralnormvec{\widehat{\nat}^{(t+1)} - \widehat{\nat}^{(t)}}$ and $|(\ell(\widehat{\nat}^{(t+1)}) - \ell(\widehat{\nat}^{(t)}))\, /\, \ell(\widehat{\nat}^{(t)})|$ are less than $10^{-6}$.
}

\begin{figure}[t]
\centering
\includegraphics[width = .42\linewidth, keepaspectratio]{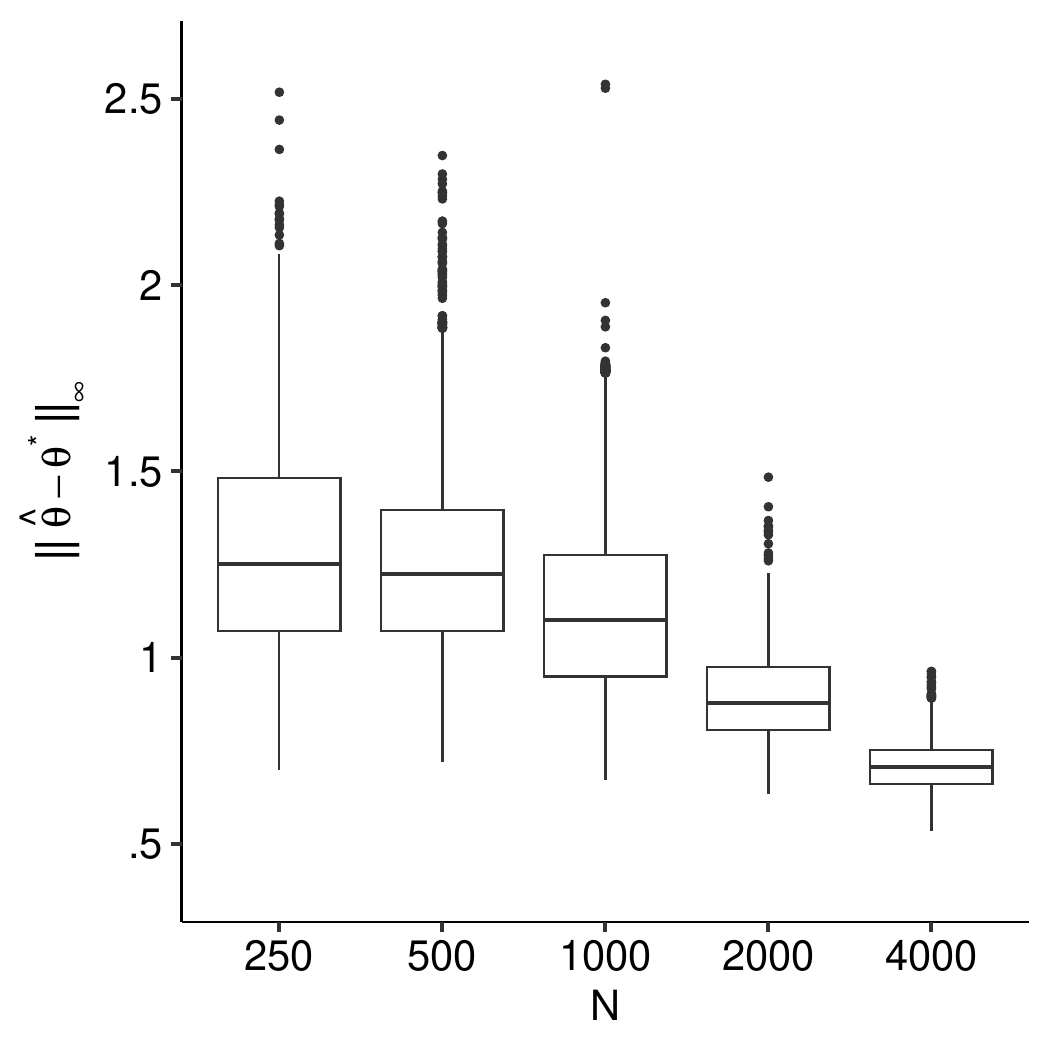}
\includegraphics[width = .42\linewidth, keepaspectratio]{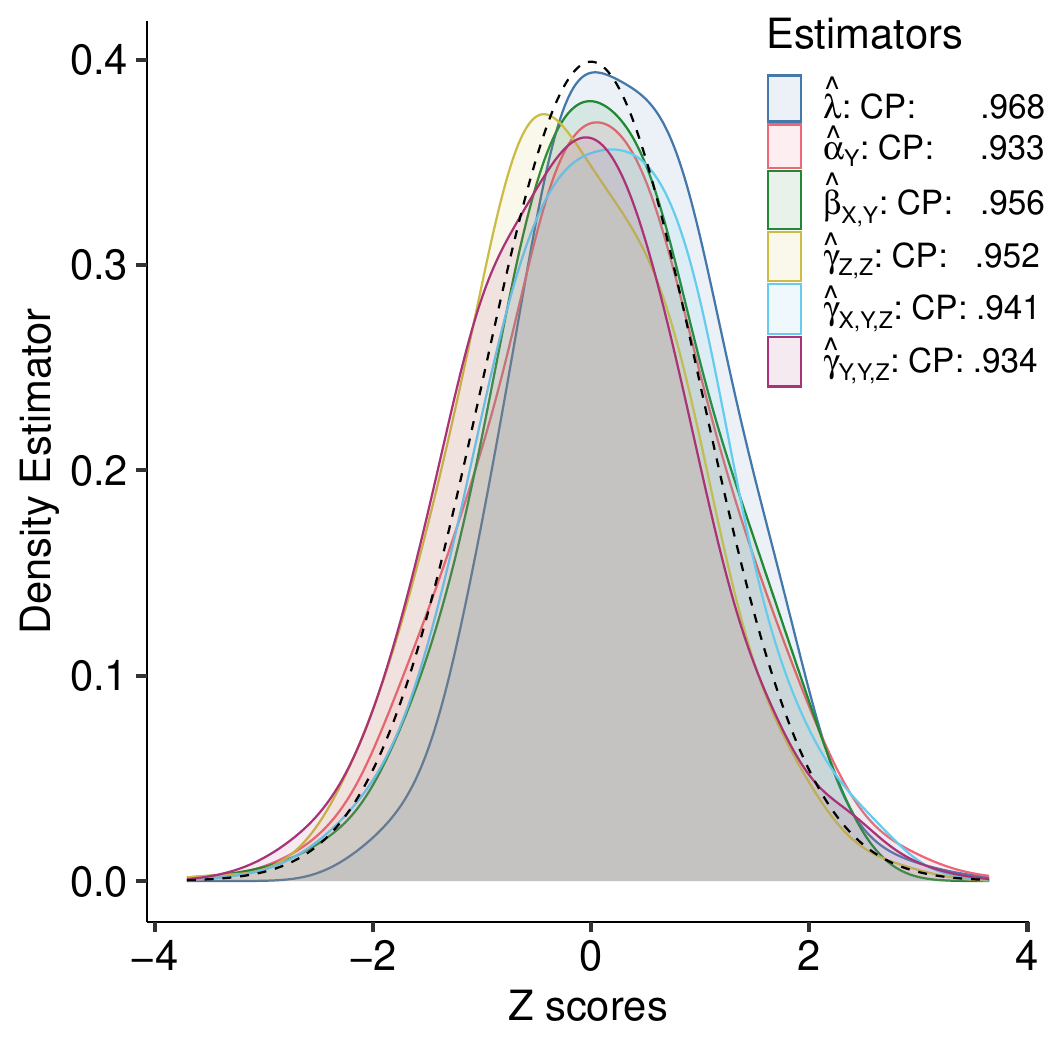}
\caption{
Simulation results based on 1,000 replications.
Left:
Statistical error $|\!|\widehat\nat - \truth|\!|_\infty$ of maximum pseudo-likelihood estimator $\widehat\nat \in \mR^{N+6}$ as a function of $N$.
Right:
Kernel density estimators of $Z$-scores $(\widehat\lambda - \lambda^\star)/\text{S.E.}(\widehat \lambda)$,\,
$(\widehat\alpha_{\mY} - \alpha_{\mY}^\star)/\text{S.E.}(\widehat\alpha_{\mY})$,\,
$(\widehat\beta_{\mX,\mY} - \beta_{\mX,\mY}^\star)/\text{S.E.}(\widehat\beta_{\mX,\mY})$,\,
$(\widehat\gamma_{\mZ,\mZ} - \gamma_{\mZ,\mZ}^\star)/\text{S.E.}(\widehat\gamma_{\mZ,\mZ})$,\,
$(\widehat\gamma_{\mX,\mY,\mZ} - \gamma_{\mX,\mY,\mZ}^\star)/ \text{S.E.}(\widehat\gamma_{\mX,\mY,\mZ})$,\,
and $(\widehat\gamma_{\mY,\mY,\mZ} - \gamma_{\mY,\mY,\mZ}^\star)/\text{S.E.}(\widehat\gamma_{\mY,\mY,\mZ})$ based on $N = 500$ units,
where the dashed line corresponds to the standard normal density and
CP is the coverage probability of interval estimators with nominal coverage probability $.95$.
\vspace{.1cm}
}
\label{fig:simulation_N_increasing}
\end{figure}

In Figure~\ref{fig:simulation_N_increasing}, the left panel shows that
$|\!|\widehat\nat - \truth|\!|_\infty$ decreases as $N$ increases.
The right panel depicts the empirical distributions of the standardized univariate estimators
and the empirical coverage probabilities, 
demonstrating that the covariance estimator in Equation \eqref{approximate.covariance}
appears accurate and normal-based inference seems reasonable.
As discussed in Section \ref{sec:introduction},
comparisons to other estimation approaches are infeasible due to their lack of scalability.

\section{Hate Speech on X}
\label{sec:applications}

We analyze posts of U.S.\ state legislators on the social media platform X in the six months preceding the insurrection at the U.S. Capitol on January 6, 2021 \citep{kim_attention_2022},
with a view to studying how hate speech depends on the attributes of legislators and connections among them.
Using Large Language Models (LLMs),
we classify the contents of 109,974 posts by $N = $ 2,191 legislators as ``non-hate speech'' or ``hate speech,''
as explained in Section \ref{sec:app.add} of the Supplementary Materials.
The response $Y_i$ of legislator $i$ indicates whether $i$ released at least one post classified as hate speech.
We use four covariates:
$x_{i,1}$ indicates that legislator $i$'s party affiliation is Republican,
$x_{i,2}$ indicates that legislator $i$ is female,
$x_{i,3}$ indicates that legislator $i$ is white,
and $x_{i,4}$ is the state legislature that legislator $i$ is a member of (e.g., New York).
The directed connections $Z_{i,j}$ are based on the mentions and reposts exchanged between
January 6, 2020 and January 6, 2021:
$Z_{i,j} = 1$ if legislator $i$ mentioned or reposted posts by legislator $j$ in a post.
To construct the neighborhoods $\mN_i$ of legislators $i$,
we exploit the fact that users of X choose whom to follow and that these choices are known,
so $\mN_i$ is defined as the union of $\{i\}$ and the set of users followed by $i$.

\subsection{Model Specification}
\label{sec:application.model}

To accommodate binary responses $Y_i \in \{0, 1\}$ and connections $Z_{i,j} \in \{0, 1\}$ that are directed, 
i.e.,
$Z_{i,j}$ may not be equal to $Z_{j,i}$,
we consider a model of the form
\be
\label{joint.pdf.directed}
f_{\nat}(\by,\, \bz \mid \bx)
&\propto& \left[\dprod_{i = 1}^N a_{\mY}(y_i)\, \exp(\nat_g^\top\, g_i (\bx_i,\, y_i^\star))\right]\s
\\
&\times& \left[\dprod_{i=1}^N\, \dprod_{j=1,\, j \neq i}^N a_{\mZ}(z_{i,j})\, \exp(\nat_h^\top\, h_{i,j}(\bx,\, y_i^\star,\, y_j^\star,\, \bz))\right],
\ee
where $y_i^\star \coloneqq y_i / \psi = y_i$ because $\psi \coloneqq 1$ when $Y_i \in \{0, 1\}$;
see Example 3 in Section \ref{sec:glm_y}.
Since $y_i^\star = y_i$,
we henceforth write $y_i$ instead of $y_i^\star$.

Using the definitions of $c_{i,j}$ and $d_{i,j}$ in Equation~(\ref{eq:cstar}),
we specify $g_i$ and $h_{i,j}$ as follows:
\be\label{eq:gi_directed}
\nat_g
\,\coloneqq\,
\left(
\begin{array}{ccc}
\alpha_{\mY}\\
\beta_{\mX,\mY,m},\; m = 1, 2, 3
\end{array}
\right),
&
g_i
\,\coloneqq\,
\left(
\begin{array}{ccc}
y_i\\
x_{i,m}\, y_i,\; m = 1, 2, 3
\end{array}
\right),
\ee
\be\label{eq:hij_directed}
\nat_h
\coloneqq
\left(
\begin{array}{ccc}
\bm\alpha_{\mZ,O}\\
\bm\alpha_{\mZ,I}\\
\lambda\\
\gamma_{\mX,\mZ,1}\\
\gamma_{\mX,\mZ,m},\; m = 2, 3, 4\\
\gamma_{\mY,\mZ}\\
\gamma_{\mZ,\mZ,1}\\
\gamma_{\mZ,\mZ,2}\\
\gamma_{\mX,\mY,\mZ}
\end{array}
\right),
&
h_{i,j}
\coloneqq
\left(
\begin{array}{ccc}
\bm{e}_i\, z_{i,j}\\
\bm{e}_j\, z_{i,j}\\
-(1 - c_{i,j})\, z_{i,j} \log N\\
c_{i,j}\, x_{i,1}\, z_{i,j}\\
c_{i,j}\, \mathbb{I}(x_{i,m} = x_{j,m})\, z_{i,j},\; m = 2, 3, 4\\
c_{i,j}\, y_j\, z_{i,j}\\
\dfrac12\, z_{i,j}\, z_{j,i}\\
d_{i,j}(\bz)\, z_{i,j}\\
c_{i,j}\, x_{i,1}\, y_j\, z_{i,j}
\end{array}
\right),
\ee
\hide{
\alert{
This is just an alternative way to write it needing less space. Delete if not wanted.
\be\label{eq:gi_directed}
g_i(\bx_i,\, y_i)^\top
\,\coloneqq\,
\left(
y_i,
x_{i,1}\, y_i,
x_{i,2}\, y_i,
x_{i,3}\, y_i
\right)
\mbox{ and }
\nat_g^\top
\,\coloneqq\,
\left(
\alpha_{\mY},
\beta_{\mX,\mY,1},
\beta_{\mX,\mY,2},
\beta_{\mX,\mY,3}
\right)
\,\in\, \mR^4
\ee
\be\label{eq:hij_directed}
h_{i,j}(\bx,\, y_i,\, y_j,\, \bz)^\top
&\coloneqq&
(\bm{e}_i\, z_{i,j},
\bm{e}_j\, z_{i,j},
-(1 - c_{i,j})\, z_{i,j} \log N,
c_{i,j}\, x_{i,1}\, z_{i,j}, \\
&~&
c_{i,j}\, \mathbb{I}(x_{i,2} = x_{j,2})\, z_{i,j},
c_{i,j}\, \mathbb{I}(x_{i,3} = x_{j,3})\, z_{i,j},\\
&~&
c_{i,j}\, \mathbb{I}(x_{i,4} = x_{j,4})\, z_{i,j},
c_{i,j}\, y_j\, z_{i,j},
z_{i,j}\, z_{j,i}/2,
d_{i,j}(\bz)\, z_{i,j},\\
&~&
c_{i,j}\, x_i\, y_j\, z_{i,j}
)
\ee
and
\beno
\nat_h^\top
\coloneqq
\left(
\bm\alpha_{\mZ,I},
\bm\alpha_{\mZ,O},
\lambda,
\gamma_{\mX,\mZ,1},
\gamma_{\mX,\mZ,2},
\gamma_{\mX,\mZ,3},
\gamma_{\mX,\mZ,4},
\gamma_{\mY,\mZ},
\gamma_{\mZ,\mZ,1},
\gamma_{\mZ,\mZ,2},
\gamma_{\mX,\mY,\mZ}
\right)
\in \mR^{2\, N + 9}
\ee
}
}
where the $i$th coordinate of $N$-vector $\bm{e}_i \in \{0, 1\}^N$ is $1$ and all other coordinates are $0$.
Here,
$\bm\alpha_{\mZ,O} \coloneqq (\alpha_{\mZ,O,1}, \ldots, \alpha_{\mZ,O,N}) \in \mR^N$ quantifies
the activity of legislators $1, \dots, N$, i.e., their tendency to mention or repost posts of other
legislators;
$\bm\alpha_{\mZ,I} \coloneqq (\alpha_{\mZ,I,1}, \ldots, \alpha_{\mZ,I,N}) \in \mR^N$ quantifies the
attractiveness of legislators $1, \dots, N$, i.e., the tendency for other legislators to mention or repost
posts by them;
$\lambda > 0$ discourages connections between legislators with non-overlapping neighborhoods;
$\gamma_{\mX,\mZ,1}, \dots, \gamma_{\mX,\mZ,4} \in \mR$ capture the effects of covariates $x_{i,1}, \dots, x_{i,4}$ on connections $Z_{i,j}$;
$\gamma_{\mY,\mZ} \in \mR$ is the weight of the interaction of $Y_j$ and $Z_{i,j}$;\,
$\gamma_{\mZ,\mZ,1} \in \mR$ quantifies the tendency to reciprocate connections;
$\gamma_{\mZ,\mZ,2} \in \mR$ quantifies the tendency to form transitive connections;
and $\gamma_{\mX,\mY,\mZ}$ captures spillover from covariate $x_{i,1}$ on response $Y_j$ through connection $Z_{i,j}$;
note that the spillover effect should not be interpreted as a causal effect,
because the party affiliations $x_{i,1}$ of legislators $i$ are not under the control of investigators \citep{kim_attention_2022}.
Since $\sum_{i=1}^N Z_{i,j} = \sum_{j=1}^N Z_{i,j}$ with probability $1$,
we set $\alpha_{\mZ,I,N} \coloneqq 0$
to address the identifiability problem that would result if all $\alpha_{\mZ,O,i}$ and $\alpha_{\mZ,I,j}$ were allowed to vary freely.
These model terms were chosen based on domain knowledge,
because model selection is an open problem:
For instance, 
the statistic $c_{i,j}\, x_{i,1}\, y_j\, z_{i,j}$ with weight $\gamma_{\mX,\mY,\mZ}$ is included to assess whether the party affiliation $x_{i,1}$ of state legislators $i$ affects posts $y_j$ of state legislators $j$ who are connected ($z_{i,j} = 1$) and whose neighborhoods overlap ($c_{i,j} = 1$).
In practice,
data scientists can consult domain experts to make informed choices regarding model specifications.

The specified model is estimated by an extension of the algorithm in Section~\ref{sec:two.stage} to directed connections;
see Section \ref{sec:mm_directed} of the Supplementary Materials.

\subsection{Results}
\label{sec:application.results}

\begin{table}[!tbp]
\caption{Maximum pseudo-likelihood estimates and standard errors based on the model specified by Equations \eqref{eq:gi_directed} and \eqref{eq:hij_directed}.\label{tbl:res}} 
\begin{center}
\begin{tabular}{crrcrr}
\hline
\multicolumn{1}{c}{Weight}&\multicolumn{1}{c}{Estimate}&\multicolumn{1}{c}{Standard Error}&\multicolumn{1}{c}{Weight}&\multicolumn{1}{c}{Estimate}&\multicolumn{1}{c}{Standard Error}\tabularnewline
\hline
$\alpha_{\mathcal Y}$&$-.893$&.134&$\gamma_{\mathcal Z, \mathcal Z, 1}$&$2.57$&.033\tabularnewline
$\beta_{\mathcal X,\mathcal Y,1}$&$-.257$&.105&$\gamma_{\mathcal Z, \mathcal Z, 2}$&$.604$&.037\tabularnewline
$\beta_{\mathcal X,\mathcal Y,2}$&$.069$&.094&$\gamma_{\mathcal X, \mathcal Z,1}$&$-.007$&.07\tabularnewline
$\beta_{\mathcal X,\mathcal Y,3}$&$-.034$&.127&$\gamma_{\mathcal X, \mathcal Z,2}$&$.236$&.016\tabularnewline
$\gamma_{\mathcal Y, \mathcal Z}$&$.035$&.005&$\gamma_{\mathcal X, \mathcal Z,3}$&$.756$&.025\tabularnewline
$\gamma_{\mathcal X, \mathcal Y, \mathcal Z}$&$.038$&.013&$\gamma_{\mathcal X, \mathcal Z,4}$&$4.729$&.049\tabularnewline
&&&$\lambda$&$.184$&.006\tabularnewline
\hline
\end{tabular}\end{center}
\end{table}

To interpret the results,
we exploit the fact that the conditional distributions of responses $Y_i$ and connections $Z_{i,j}$ can be represented by logistic regression models,
with log odds
\beno
\label{log.odds.y}
\log\dfrac{\mathbb{P}_{\nat}(Y_i = 1 \mid \text{others}) }{1-\mathbb{P}_{\nat}(Y_i = 1 \mid \text{others})}
&=& \alpha_{\mathscr{Y}} + \dsum_{m = 1}^{3}\beta_{\mX,\mY,m}\, x_{i,m}
+ \dsum_{j:\, \mathscr N_i\, \cap\, \mathscr N_j \neq \emptyset} (\gamma_{\mathscr{Y}, \mathscr{Z}} + \gamma_{\mathscr{X}, \mathscr{Y}, \mathscr{Z}}\, x_{j,1})\, z_{j,i}
\ee
and
\beno
&& \log\dfrac{\mathbb{P}_{\nat} (Z_{i,j} = 1 \mid \text{others}) }{1-\mathbb{P}_{\nat}(Z_{i,j} = 1 \mid\text{others})}
\;=\; \alpha_{\mZ,O,i} + \alpha_{\mZ,I,j} + \dfrac{1}{2}\, \gamma_{\mZ,\mZ,1}\, z_{j,i}
- (1 - c_{i,j})\, \lambda\, \log N\s
\\
&+& c_{i,j} \left(\gamma_{\mZ,\mZ,2}\; \Delta_{i,j}(\bz) + \gamma_{\mX,\mZ,1} \, x_{i,1}
+ \dsum_{m = 2}^{4} \gamma_{\mX, \mZ,m}\, \mathbb{I}(x_{i,m} = x_{j,m})
+ \gamma_{\mY,\mZ}\, y_{j}
+ \gamma_{\mX,\mY,\mZ}\, x_{i,1}\, y_j\right).
\ee
For instance,
the positive sign of $\widehat\gamma_{\mX, \mY, \mZ} = .038$ suggests that the more Republicans interact with legislator $i$,
the higher is the conditional probability that legislator $i$ uses offensive text in a post,
holding everything else constant.
Alternatively,
one can interpret $\widehat\gamma_{\mX, \mY, \mZ}$ in terms of the conditional probability of observing a connection:
The positive sign of $\widehat\gamma_{\mX, \mY, \mZ} = .038$ indicates that Republican legislators are more likely to interact with legislators who post harmful language.
Other estimates align with expectations.
For example,
serving for the same state is the strongest predictor for reposting and mentioning activities ($\widehat\gamma_{\mX, \mZ, 4} = 4.729$),
while matching gender ($\widehat\gamma_{\mX, \mZ, 2} = .236$) and race ($\widehat\gamma_{\mX, \mZ, 3} = .756$) likewise increase the conditional probability to interact.
At the same time,
connections affect other connections:
For example,
forming a connection that leads to a transitive connection is observed more often than expected under the model with $\gamma_{\mZ,\mZ,2} = 0$,
holding everything else constant.

\s

\subsection{Model Assessment}
\label{sec:model.assessment}

\begin{figure}[t!]
\centering
\includegraphics[width=0.7\linewidth]{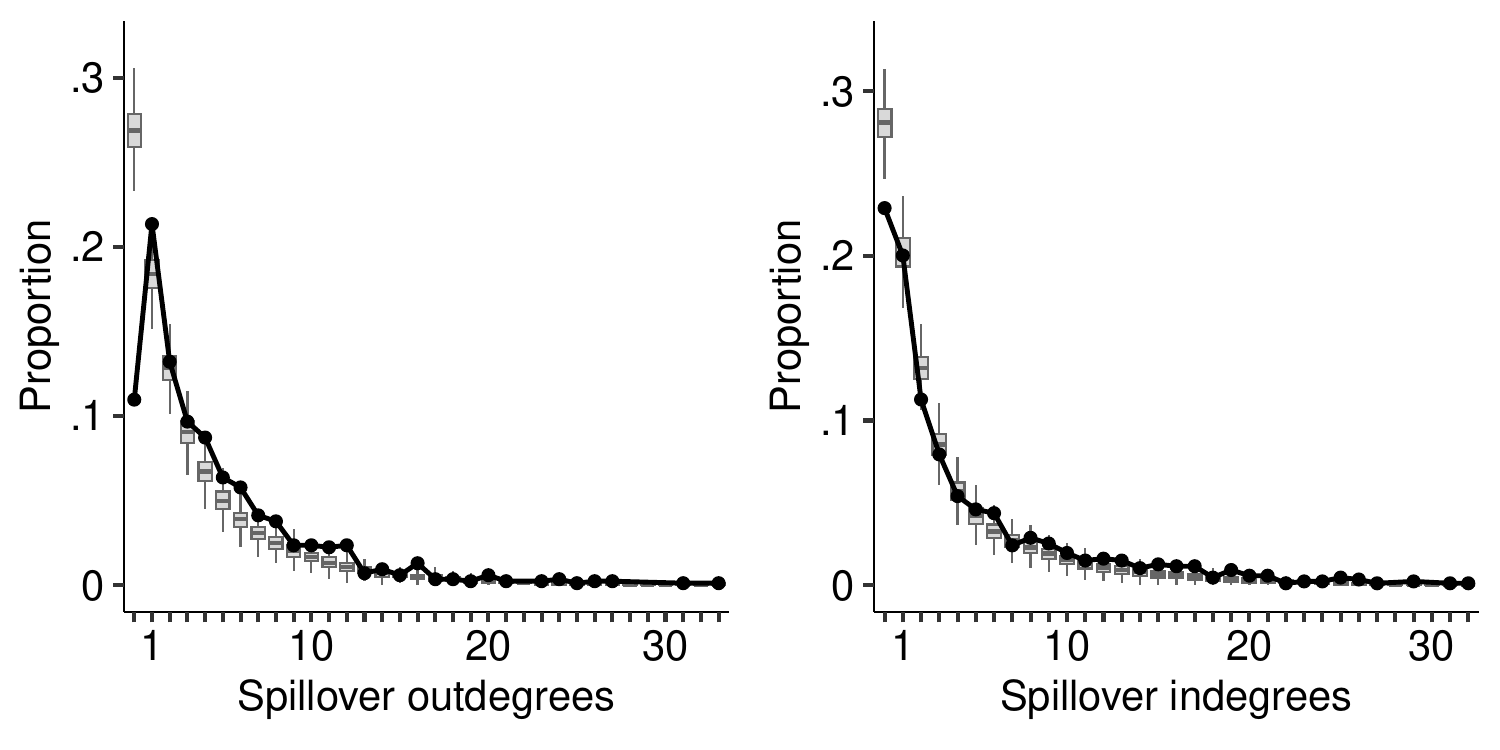}
\caption{Model-based predictions of spillover in- and out-degrees of U.S.\ state legislators in the subnetwork
with $i$ being Republican,
$j$ using offensive language,
and the neighborhoods of $i$ and $j$ overlapping,
i.e.,
$x_{i,1} = Y_j = 1$ and $\mathscr{N}_i\, \cap\, \mN_i\, \neq\, \emptyset$.
By construction, the
possible connections in the subnetwork act as potential channels of spillover.
The spillover in- and out-degrees are defined as the respective degree of a unit in the subnetwork with $x_{i,1} = Y_j = 1$ and $\mathscr{N}_i\, \cap\, \mN_i\, \neq\, \emptyset$.
The observed spillover in- and out-degrees are shown by solid lines with overlaid points.
\s
}
\label{fig:plotdegree}
\end{figure}

\begin{figure}[t!]
\centering
\includegraphics[width=0.7\linewidth]{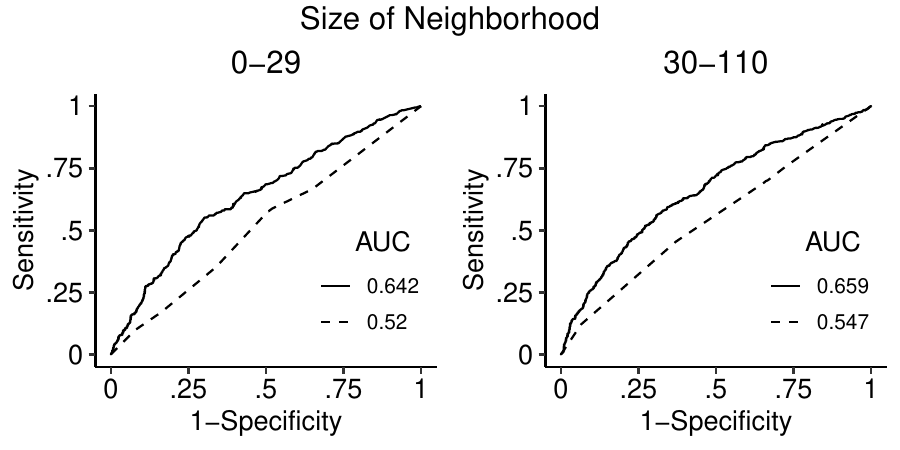}
\caption{
Comparing model-based predictions of $Y_i \mid \bX_i = \bx_i$ without network interference (solid) to model-based predictions of $Y_i \mid (\bX, \bY_{-i}, \bZ) = (\bx, \by_{-i}, \bz)$ with network interference (dashed) based on the area under the curve (AUC).
Sensitivity is the true positive rate, the fraction among legislators predicted to make offensive posts who actually do so.
Specificity is the true negative rate, the fraction among legislators predicted not to make offensive posts who actually do not do so.
\label{fig:roc}
}
\end{figure}
\hide{
\begin{table}[!t]
\caption{Comparison of maximum pseudo-likelihood estimates based on the logistic regression model for $Y_i \mid \bX_i = \bx_i$ without network interference and the joint probability model for $(\bY, \bZ) \mid \bX = \bx$ with network interference.\label{tbl:comp}} 
\begin{center}
\begin{tabular}{crrrr}
\hline
\multicolumn{1}{c}{Weight}&\multicolumn{1}{c}{$\alpha_{\mathcal Y}$}&\multicolumn{1}{c}{$\beta_{\mathcal X,\mathcal Y,1}$}&\multicolumn{1}{c}{$\beta_{\mathcal X,\mathcal Y,2}$}&\multicolumn{1}{c}{$\beta_{\mathcal X,\mathcal Y,3}$}\tabularnewline
\hline
Without interference: estimate (S.E.) &$-.101$ (.103)&$-.235$ (.097)&$.032$ (.093)&$-.169$ (.113)\tabularnewline
With interference: estimate (S.E.) &$-.893$ (.134)&$-.257$ (.105)&$.069$ (.094)&$-.034$ (.127)\tabularnewline
\hline
\end{tabular}\end{center}
\end{table}
}
\begin{table}[!t]
\caption{Comparison of maximum pseudo-likelihood estimates based on the logistic regression model for $Y_i \mid \bX_i = \bx_i$ without network interference and the joint probability model for $(\bY, \bZ) \mid \bX = \bx$ with network interference.\label{tbl:comp}} 
\begin{center}
\begin{tabular}{lrrrr}
\hline
\multicolumn{1}{c}{}&\multicolumn{1}{c}{$\alpha_{\mathcal Y}$}&\multicolumn{1}{c}{$\beta_{\mathcal X,\mathcal Y,1}$}&\multicolumn{1}{c}{$\beta_{\mathcal X,\mathcal Y,2}$}&\multicolumn{1}{c}{$\beta_{\mathcal X,\mathcal Y,3}$}\tabularnewline
\hline
Without network interference&$-.101$ (.103)&$-.235$ (.097)&$.032$ (.093)&$-.169$ (.113)\tabularnewline
With network interference&$-.893$ (.134)&$-.257$ (.105)&$.069$ (.094)&$-.034$ (.127)\tabularnewline
\hline
\end{tabular}\end{center}
\end{table}

We assess the model using model-based predictions of $(\bY, \bZ) \mid \bX = \bx$.
First,
we focus on the subnetwork of all pairs of legislators $\{i, j\} \subset \mP_N$ with $x_{i,1} = Y_j = 1$ and $\mN_i\, \cap\, \mN_j\, \neq\, \emptyset$,
with a view to assessing how well the interplay of $x_{i,1}$, $Y_j$, and $Z_{i,j}$ can be represented by the model.
Figure \ref{fig:plotdegree} shows that the model captures the effect of $x_{i,1}$ on $Y_j$ among pairs of legislators $\{i, j\} \subset \mP_N$ with $\mN_i\, \cap\, \mN_j\, \neq\, \emptyset$.
Second,
we compare predictions of responses $Y_i$ based on models with and without network interference.
Predictions without network interference are based on the logistic regression model for $Y_i \mid \bX_i = \bx_i$ with weights $\alpha_{\mY}$, $\beta_{\mX,\mY,1}$, $\beta_{\mX,\mY,2}$, and $\beta_{\mX,\mY,3}$,
which assumes that the posts $Y_i$ of legislators $i$ are independent and do not depend on the connections $\bZ$ among the legislators.
By contrast,
predictions with network interference are based on the joint probability model for $(\bY, \bZ) \mid \bX = \bx$ specified by Equation \eqref{joint.pdf.directed},
which allows the posts $Y_i$ of state legislators $i$ to be affected by the posts $Y_j$ of connected legislators $j$ whose sets of followees overlap.
Figure~\ref{fig:roc} demonstrates that predictions based on models with network interference outperform those without network interference,
suggesting that posts of connected state legislators with overlapping sets of followees are interdependent.
Table~\ref{tbl:comp} compares estimates of $\alpha_{\mathcal Y}$,
$\beta_{\mathcal X,\mathcal Y,1}$, 
$\beta_{\mathcal X,\mathcal Y,2}$,
and $\beta_{\mathcal X,\mathcal Y,3}$ based on models with and without network interference.
While both models agree on the signs of parameter estimates,
the estimates of $\beta_{\mathcal X,\mathcal Y,2}$ and $\beta_{\mathcal X,\mathcal Y,3}$ differ by a factor of 2 and 5,
respectively,
suggesting that network interference affects parameter estimates.
Third,
we demonstrate in Section \ref{sec:app.plots} of the Supplementary Materials that the model preserves salient features of connections $\bZ$.

\section{Discussion}
\label{sec:discussion}

The proposed regression framework is flexible,
allowing data scientists to specify a wide range of models for dependent responses and
connections $(\bY, \bZ) \mid \bX = \bx$.

The large set of possible models raises the question of how data scientists can select a model from a
set of candidate models.
While model selection for independent responses $Y_i \mid \bX_i = \bx_i$ is well-established
and model selection for independent connections $\bZ \mid \bX = \bx$ is an active area of
research \citep[e.g.,][]{Wa15,wang2024,stein2025},
model selection for dependent responses and connections $(\bY, \bZ) \mid \bX = \bx$
with $p \to \infty$ parameters is an open problem.
Two model selection ideas that hold promise are those of \citet{RaWaLa10} for
high-dimensional graphical models for dependent
responses $\bY$  and those of \citet{ChenChen12} for
high-dimensional generalized linear models for
independent responses $Y_i \mid \bX_i = \bx_i$ based on the extended BIC.

Likewise,
open questions remain in the
realm of uncertainty quantification, as discussed in
Section~\ref{sec:standard_errors}.
For example,
a proof of asymptotic normality based on dependent data
remains elusive.
A related avenue for future research is Godambe information:
If $-\nabla_{\nat}^2\,\, \ell(\nat;\, \bY, \bZ)|_{\nat=\widehat\nat}^{-1}$\, were constant,
it could be pulled out of the approximate covariance matrix in Equation
\eqref{approximate.covariance},
giving rise to Godambe information.
Simulations suggest that uncertainty quantification
based on Godambe information achieves comparable  accuracy to the method
reported here,
while avoiding multiple matrix inversions.

The question of neighborhood recovery
is another important direction for future research,
as discussed in Section~\ref{sec:model}.

\label{last_page}

\begin{center}
{\large\bf Supplementary Materials}
\end{center}

\noindent
The supplementary materials contain proofs of all theoretical results.

\begin{center}
{\large\bf Acknowledgements}
\end{center}

\noindent
The authors are indebted to the constructive comments and suggestions of an anonymous associate editor and two referees,
which have led to numerous improvements.

\begin{center}
{\large\bf Disclosure Statement}
\end{center}

\noindent
The authors report there are no competing interests to declare.

\if1\blind
\begin{center}
	{\large\bf Funding}
\end{center}

\noindent
The authors acknowledge support by DFG award FR 4768/1-1 (CF) and ARO award W911NF-21-1-0335 (CF, MS, SB).
\fi

\vspace{-.295cm}

\bibliographystyle{chicago}
\bibliography{base}
\label{last_page_ref}

\newpage

\begin{center}
{\LARGE\textbf{Supplementary Materials:\\ 
A Regression Framework for Studying Relationships among Attributes under Network Interference}}
\end{center}

\setcounter{section}{0}

\setcounter{page}{1}

\appendix
\numberwithin{equation}{section}
\renewcommand{\cftsecfont}{\mdseries}
\renewcommand{\cftsecpagefont}{\mdseries}
\setlength{\cftbeforesecskip}{0 pt} 
\renewcommand{\cftsecleader}{\cftdotfill{\cftdotsep}} 
\startcontents
\printcontents{ }{1}{}

\section{Proofs of Propositions \ref{prop_general_glm} and \ref{prop_gaussian_y_mid_z_2}}
\label{sec:proof_prop1}

\textsc{Proof of Proposition \ref{prop_general_glm}.}
The joint probability density function of $(\bY,\, \bZ) \mid \bX = \bx$ stated in Equation \eqref{eq:ernm_model} in Section \ref{sec:model} implies that the conditional probability density function of $Y_i \mid (\bX,\, \bY_{-i},\, \bZ) = (\bx,\, \by_{-i},\, \bz)$ can be written as
\beno
&& f_{\nat}(y_i \mid \bx,\, \by_{-i},\, \bz)
\;=\; \dfrac{f_{\nat}(y_i,\, \by_{-i},\, \bz \mid \bx)}{\dint_{\mY_i} f_{\nat}(y,\, \by_{-i},\, \bz \mid \bx) \dd \nu_{\mY}(y)}\s\s
\\ 
&=& \dfrac{a_{\mY}(y_i)\, \exp\left(\nat_g^\top\, g_{i,1}(\bx_i)\, y_i^\star + \left(\dsum_{j\, \in\, \mP_N \setminus\, \{i\}}\; \nat_h^\top\, h_{i,j,1}(\bx,\, y_j^\star,\, \bz)\right) y_i^\star \right) }{\dint_{\mY_i} a_{\mY}(y)\, \exp \left(\nat_g^\top\, g_{i,1} (\bx_i)\, y^\star + \left(\dsum_{j\, \in\, \mP_N \setminus\, \{i\}}\; \nat_h^\top\, h_{i,j,1}(\bx,\, y_j^\star,\, \bz)\right) y^\star\right) \dd \nu_{\mY}(y)}\s\s
\\
&=& a_{\mY}(y_i) \, \exp \left(\dfrac{\eta_{i}(\nat;\, \bx,\, \by_{-i}^\star,\, \bz)\, y_i - b_{i}(\eta_{i}(\nat;\, \bx,\, \by_{-i}^\star,\, \bz))}{\psi} \right),
\ee
where $y^\star \coloneqq y / \psi$,\, 
$y_i^\star \coloneqq y_i / \psi$,\,
and $\by_{-i}^\star \coloneqq \by_{-i} / \psi$,
while 
\beno
\eta_i(\nat;\, \bx,\, \by_{-i}^\star,\, \bz)
\;\coloneqq\; \nat^\top \left(g_{i,1}(\bx_i),\; \dsum_{j\, \in\, \mP_N \setminus\, \{i\}}\, h_{i,j,1}(\bx,\,  y_j^\star,\, \bz)\right)\s
\\
b_i(\eta_i(\nat;\, \bx,\, \by_{-i}^\star,\, \bz))
\;\coloneqq\; \psi \, \log \dint_{\mY_i}\, a_{\mY}(y) \exp\left(\dfrac{\eta_i(\nat;\, \bx,\, \by_{-i}^\star,\, \bz)\, y}{\psi}\right) \dd \nu_{\mY}(y).
\ee

\s

\begin{proposition} \label{prop_gaussian_y_mid_z_2}
{\em 
Consider Example 1 in Section \ref{sec:glm_y}.
Let $\bm{U} \in \{0, 1\}^{N \times N}$ be the $N \times N$ matrix with elements
\be
\label{udef}
u_{i,j} 
&\coloneqq& c_{i,j}\, z_{i,j}
&=& \one(\mN_i\, \cap\, \mN_j\, \neq \, \emptyset)\, z_{i,j},
\ee
and let $\bm{v} \in \mR^N$ be the $N$-vector with coordinates
\be
\label{vdef}
v_i 
&\coloneqq& \alpha_{\mY} + \beta_{\mX,\mY}\; x_{i} + \gamma_{\mX,\mY,\mZ}\, \dsum_{j\, \in\, \mP_N \setminus\, \{i\}}\, u_{i,j}\; x_j.
\ee
Denote by $\bI$ the $N \times N$ identity matrix and define $\xi_{\mY,\mY,\mZ} \,\coloneqq\, \gamma_{\mY,\mY,\mZ}/\psi$.
If $(\bI - \xi_{\mY,\mY,\mZ}\; \bm{U})$ is positive definite,
the conditional distribution of $\bY \mid (\bX,\, \bZ) = (\bx,\, \bz)$ is $N$-variate Gaussian with mean vector $(\bI -\xi_{\mY,\mY,\mZ}\; \bm{U})^{-1}\, \bm{v}$ and covariance matrix $\psi\, (\bI - \xi_{\mY,\mY,\mZ}\; \bm{U})^{-1}$. 
}
\end{proposition}

{\bf Remark.}
The requirement that $(\bI - \xi_{\mY,\mY,\mZ}\; \bm{U})$ be positive definite imposes restrictions on $\gamma_{\mY,\mY,\mZ}$.
The restrictions on $\gamma_{\mY,\mY,\mZ}$ depend on the neighborhoods $\mN_i$ of units $i \in \mP_N$ and connections $Z_{i,j}$ among pairs of units $\{i, j\} \subset \mP_N$.

\s

\textsc{Proof of Proposition \ref{prop_gaussian_y_mid_z_2}.} 
Example 1 in Section \ref{sec:glm_y} demonstrates that the conditional distribution of $Y_i \mid (\bX,\, \bY_{-i},\, \bZ) = (\bx,\,\by_{-i},\, \bz)$ is Gaussian with conditional mean
\begin{align}
\label{cond_mean}
\mbE(Y_i\mid \bx,\,\by_{-i},\, \bz)
~&= ~\alpha_{\mY} + \beta_{\mX,\mY}\, x_{i} + \gamma_{\mX,\mY,\mZ}\, \dsum_{j\, \in\, \mP_N \setminus\, \{i\}}\, u_{i,j}\; x_j
+ \gamma_{\mY,\mY,\mZ}\, \dsum_{j\, \in\, \mP_N \setminus\, \{i\}}\, u_{i,j}\; y_j^\star\s
\notag\\
&=~ v_i + \xi_{\mY,\mY,\mZ}\, \dsum_{j\, \in\, \mP_N \setminus\, \{i\}}\, u_{i,j}\; y_j, 	
\end{align}
where
\beno 
v_i &\coloneqq& \alpha_{\mY} + \beta_{\mX,\mY}\, x_{i} + \gamma_{\mX,\mY,\mZ}\, \dsum_{j\, \in\, \mP_N \setminus\, \{i\}}\, u_{i,j}\; x_j 
\ee
and
\beno  
\xi_{\mY,\mY,\mZ} &\coloneqq& \dfrac{\gamma_{\mY,\mY,\mZ}}{\psi}.
\ee
The conditional variance of $Y_i \mid (\bX,\, \bY_{-i}, \bZ) = (\bx,\, \by_{-i},\, \bz)$ is
\be
\label{cond_var}
\mathbb{V}(Y_i \mid \bx,\,\by_{-i},\, \bz)
&=& \psi.
\ee
Let $\bm{m} \coloneqq (m_i) \in \mR^N$ be the conditional mean of $\bY \mid (\bX,\, \bZ) = (\bx,\, \bz)$.
Upon taking expectation on both sides of \eqref{cond_mean} conditional on $(\bX,\, \bZ) = (\bx,\, \bz)$, 
we obtain
\be
\label{marg_mean}
m_i
&=& v_i + \xi_{\mY,\mY,\mZ}\, \dsum_{j\, \in\, \mP_N \setminus\, \{i\}}\, u_{i,j}\; m_j,
\ee
which implies that
\beno
\label{v-exp}
v_i
&=& m_i - \xi_{\mY,\mY,\mZ}\, \dsum_{j\, \in\, \mP_N \setminus\, \{i\}}\, u_{i,j}\; m_j
\ee
and hence
\begin{align}
\label{cond_mean2}
\mbE(Y_i \mid \bx,\,\by_{-i},\, \bz) 
&= ~v_i + \xi_{\mY,\mY,\mZ}\, \dsum_{j\, \in\, \mP_N \setminus\, \{i\}}\, u_{i,j}\; y_j\s
\notag\\
&=~ m_i - \xi_{\mY,\mY,\mZ}\, \dsum_{j\, \in\, \mP_N \setminus\, \{i\}}\, u_{i,j}\; m_j + \xi_{\mY,\mY,\mZ}\, \dsum_{j\, \in\, \mP_N \setminus\, \{i\}}\, u_{i,j}\; y_j\s
\notag\\
&=~ m_i + \xi_{\mY,\mY,\mZ}\, \dsum_{j\, \in\, \mP_N \setminus\, \{i\}}\, u_{i,j}\; (y_j - m_j)\s
\notag\\
&=~ m_i - \dsum_{j\, \in\, \mP_N \setminus\, \{i\}}\, b_{i,j}\; (y_j - m_j),
\end{align}
where
\beno
b_{i,j} 
&\coloneqq& -\, \xi_{\mY,\mY,\mZ}\; u_{i,j}.
\ee
By comparing Equations \eqref{cond_var} and \eqref{cond_mean2} to Equations (2.17) and (2.18) of \citetsupp{rue2005gaussian} and invoking Theorem 2.6 of \citetsupp{rue2005gaussian},
we conclude that the conditional distribution of $\bY \mid (\bX,\, \bZ) = (\bx,\, \bz)$ is $N$-variate Gaussian with mean vector $\bm{m} \in \mR^N$ and precision matrix $\bm{P} \in \mR^{N \times N}$ with elements
\beno
\label{prec_matrix}
p_{i,j}
&\coloneqq& 
\begin{cases}
\dfrac{1}{\psi} & \mbox{if } i = j\s
\\
\dfrac{b_{i,j}}{\psi} & \mbox{if } i \neq j,
\end{cases}
\ee
provided $u_{i,j} = u_{j,i}$ for all $i \neq j$ and $\bm{P}$ is positive definite;
note that $u_{i,j} = u_{j,i}$ is satisfied in undirected networks with $z_{i,j} = z_{j,i}$.

To state these results in matrix form,
note that \eqref{marg_mean} can be expressed as
\beno 
\bm{m} 
&=& \bm{v} + \xi_{\mY,\mY,\mZ}\; \bm{U}\, \bm{m},
\ee
implying
\beno
\bm{m}
&=& (\bI - \xi_{\mY,\mY,\mZ}\; \bm{U})^{-1}\; \bm{v},
\ee
while $\bm{P}$ can be expressed as
\beno
\bm{P}
&=& \dfrac{1}{\psi}\, (\bI - \xi_{\mY,\mY,\mZ}\; \bm{U}),
\ee
implying
\beno
\bm{P}^{-1} 
&=& \psi\, (\bI -\xi_{\mY,\mY,\mZ}\; \bm{U})^{-1}.
\ee

To conclude,
the conditional distribution of $\bY \mid (\bX,\, \bZ) = (\bx,\, \bz)$ is $N$-variate Gaussian with mean vector $(\bI - \xi_{\mY,\mY,\mZ}\; \bm{U})^{-1}\, \bm{v}$ and covariance matrix $\psi\, (\bI - \xi_{\mY,\mY,\mZ}\; \bm{U})^{-1}$,
provided $(\bI - \xi_{\mY,\mY,\mZ}\; \bm{U})$ is positive definite.

\section{Proofs of Lemmas \ref{concave} and \ref{lemma.minorizer}}
\label{sec:proof_lemma2}

\mbox{}

\vspace{-1cm}

\textsc{Proof of Lemma \ref{concave}.} 
Lemma \ref{concave} is proved in the sentence preceeding the statement of Lemma \ref{concave} in Section \ref{sec:cl}.

\textsc{Proof of Lemma \ref{lemma.minorizer}.} 
Letting $\Nat_1$ denote the parameter space of $\nat_1$, 
suppose that
$v: \Nat_1 \mapsto \mR$ is any twice differentiable function and that 
$\nabla^2\, v(\nat) - \bm{M}$ is non-negative definite for all $\nat\in \Nat_1$
for some constant matrix $\bm{M} \in \mathbb{R}^{d \times d}$ ($d \geq 1$).
Then the function $u: \Nat_1 \mapsto \mR$ given by
\beno
u(\nat_1)
&\coloneqq& v(\nat_0) + (\nat_1-\nat_0)^\top\, \nabla\, v(\nat_0) + \dfrac12 (\nat_1 - \nat_0)^\top \bm{M} (\nat_1-\nat_0),
&& \nat_0 \in \Nat_1
\ee
satisfies $u(\nat_1) \le v(\nat_1)$ for all $\nat_1 \in \Nat_1$,
because Taylor's theorem (Theorem 6.11, \citealpsupp[p.~124]{magnus_matrix_2019}) gives
\beno
u(\nat_1)-v(\nat_1) 
\;=\; \dfrac12\, (\nat_1-\nat_0)^\top \left[\nabla^2\, v(\dot\nat) -\bm{M}\right] (\nat_1-\nat_0),
\ee
where $\dot\nat \coloneqq \phi\, \nat_0 + (1-\phi)\, \nat_1 \in \Nat_1$ ($\phi \in [0,\, 1]$).
The inequality $1/4\, \ge\, \pi_{i,j} \, (1-\pi_{i,j})$ implies that
\beno
-[\bA(\nat_1) - \bA^\star] 
&=& \dsum_{i=1}^N\, \dsum_{j=i+1}^N \left[\dfrac14 - \pi_{i,j}^{(t)}\, (1-\pi_{i,j}^{(t)})\right]\, \bm{e}_{i,j}\, \bm{e}_{i,j}^\top
\ee
is non-negative definite.  
Lemma~\ref{concave} proves that $\nat_1$ is concave and that
the restriction of $\ell(\nat)$ to $\nat_1$ has the properties of $v(\nat_1)$ stated above,
proving Lemma \ref{lemma.minorizer}.

\section{Proof of Theorem \ref{thm:mple_consistency}}
\label{sec:proof.theorem1}

Theorem \ref{thm:mple_consistency} is a generalization of Theorem 2 of \citetsupp[][abbreviated as S25]{StSc20} from exponential family models for binary connections $\bZ$ to exponential family models for binary,
count-,
and real-valued responses and connections $(\bY, \bZ) \mid \bX = \bx$.
We henceforth suppress predictors $\bx \in \mX$.

\s

\textsc{Proof of Theorem \ref{thm:mple_consistency}.}
Let $\bs(\nat;\, \by, \bz) \coloneqq \nabla_{\nat}\; \ell(\nat;\, \by, \bz)$ and consider events
\[
\begin{array}{cll}
\mbG
&\coloneqq& \left\{(\by,\, \bz) \in \mY \times \mZ:\, \norm{\bs(\truth;\, \by,\, \bz)}_{\infty}\; \leq\; \delta_N\right\}\s
\\
\mbH
&\subseteq& \left\{(\by,\, \bz) \in \mY \times \mZ:\; -\nabla_{\nat}^2\; \ell(\nat;\, \by,\, \bz) \mbox{ is invertible for all $\nat \in \mB_{\infty}(\truth,\, \epsilon^\star)$}\right\}.
\end{array}
\]
Define
\[
\begin{array}{cll}
\Lambda_{N,\by, \bz}(\truth)
&\coloneqq& \sup\limits_{\nat\, \in\, \mB_{\infty}(\truth,\, \epsilon^\star)}\, \mnorm{(-\nabla_{\nat}^2\; \ell(\nat;\, \by,\, \bz))^{-1}}_{\infty},\;\;\; (\by,\, \bz) \in \mbH\s
\\
\Lambda_{N}(\truth)
&\coloneqq& \sup\limits_{(\by,\, \bz)\, \in\, \mbH}\, \Lambda_{N,\by, \bz}(\truth).
\end{array}
\]
It follows from results of S25 that $s(\nat;\, \by, \bz)$,
considered as a function of $\nat \in\Nat$ for fixed $(\by, \bz) \in \mbH$,
is a homeomorphism and is continuously differentiable.

\s

{\bf In the event $(\bY, \bZ) \in \mbG$,
the set ${\bm\Theta}(\delta_N)$ is non-empty.}
By construction of the sets $\mbG$ and $\widehat\Nat(\delta_N)$,
the set $\widehat\Nat(\delta_N)$ is non-empty for all $(\by,\, \bz) \in \mbG$, 
because $\widehat\Nat(\delta_N)$ contains the data-generating parameter vector $\truth \in \Nat$ provided $(\by,\, \bz) \in \mbG$:
\beno
\truth 
&\in& \widehat\Nat(\delta_N)
&\coloneqq& \left\{\nat \in \Nat:\; \norm{\bs(\nat;\, \by,\, \bz)}_{\infty}\, \leq\, \delta_N\right\}.
\ee

\s

{\bf In the event $(\bY, \bZ) \in \mbG\, \cap\,\, \mbH$,\,
the set $\widehat\Nat(\delta_N)$ satisfies $\widehat\Nat(\delta_N) \subseteq \mB_{\infty}(\truth,\, \rho_N)$ provided $N > N_0$.}
By assumption,
there exists a sequence $\rho_1, \rho_2, \dots \in [0, +\infty)$ such that $\rho_N = o(1)$.
Therefore, 
there exists an integer $N_0 \in \{1, 2, \dots\}$ such that $\rho_N < \epsilon^\star$ for all $N > N_0$.
Consider any $N > N_0$ and any $(\by,\, \bz) \in \mbG\, \cap\, \mbH$.
Since $\bs^{-1}(\, \cdot\, ;\, \by,\, \bz)$ is continuous on $\Nat$,
there exists, 
for each $(\by,\, \bz) \in \mbH$, 
a real number $\epsilon_N(\rho_N) \in (0,\, +\infty)$ (which depends on $(\by,\, \bz) \in \mbH$) such that 
\be
\label{continuity}
\norm{\bs(\nat;\, \by,\, \bz) - \bs(\truth;\, \by,\, \bz)}_{\infty}
&\leq& \epsilon_N(\rho_N)
&\mbox{implies}&
\norm{\nat - \truth}_{\infty} 
&\leq& \rho_N. 
\ee
As $\bs(\nat;\, \by, \bz)$ is a homeomorphism and continuously differentiable,
we can invoke Lemma 1 of S25 to conclude that $\epsilon_N(\rho_N)$ is related to $\rho_N$ by the following inequality: 
\be
\label{eq:lower_relation}
\dfrac{\rho_N}{\Lambda_{N,\by, \bz}(\truth)}
&\leq& \epsilon_N(\rho_N).
\ee
To take advantage of \eqref{eq:lower_relation},
observe that,
for all $\nat \in \widehat\Nat(\delta_N)$ and all $(\by,\, \bz) \in \mbG\, \cap\, \mbH$,
\begin{equation}
\begin{array}{llllllll}
\label{bound000}
\norm{\bs(\nat;\, \by, \bz) - \bs(\truth;\, \by, \bz)}_{\infty}
\,\leq\, \norm{\bs(\nat;\, \by, \bz)}_{\infty} + \norm{\bs(\truth;\, \by, \bz)}_{\infty}
\,\leq\, 2\, \delta_N
= \dfrac{\rho_N}{\Lambda_{N}(\truth)},
\end{array}
\end{equation}
because $\norm{\bs(\nat;\, \by,\, \bz)}_\infty \leq \delta_N$ for all $\nat \in \widehat\Nat(\delta_N)$,\,
$\norm{\bs(\truth;\, \by,\, \bz)}_{\infty} \leq \delta_N$ for all $(\by,\, \bz) \in \mbG\, \cap\, \mbH$,\,
and $\delta_N \coloneqq \rho_N\, /\, (2\, \Lambda_N(\truth))$.
Using \eqref{bound000} along with the definition of $\Lambda_{N}(\truth) \coloneqq \sup_{(\by,\, \bz) \in \mbH}\, \Lambda_{N,\by, \bz}(\truth) > 0$,
we obtain
\be
\label{bound00}
\norm{\bs(\nat;\, \by,\, \bz) - \bs(\truth;\, \by,\, \bz)}_{\infty}
&\leq& \dfrac{\rho_N}{\Lambda_{N}(\truth)}
&\leq& \dfrac{\rho_N}{\Lambda_{N,\by, \bz}(\truth)},
\ee
and,
using \eqref{eq:lower_relation},
\be
\label{delta.x}
\norm{\bs(\nat;\, \by,\, \bz) - \bs(\truth;\, \by,\, \bz)}_{\infty}
&\leq& \dfrac{\rho_N}{\Lambda_{N,\by, \bz}(\truth)}
&\leq& \epsilon_N(\rho_N).
\ee
In light of the fact that
\beno
\norm{\bs(\nat;\, \by,\, \bz) - \bs(\truth;\, \by,\, \bz)}_{\infty}
&\leq& \epsilon_N(\rho_N)
&\mbox{implies}&
|\!|\nat - \truth|\!|_\infty 
&\leq& \rho_N,
\ee
the set $\widehat\Nat(\delta_N)$ is non-empty and satisfies
\be
\label{keyresult}
\widehat\Nat(\delta_N) 
&\subseteq& \mB_{\infty}(\truth,\, \rho_N)
\ee
in the event $(\bY, \bZ) \in \mbG \cap\, \mbH$,
provided $N > N_0$.

\s

{\bf The event $(\bY, \bZ)\, \in\, \mbG\, \cap\, \mbH$ occurs with probability $1 - o(1)$.} 
The probability of event $(\bY, \bZ)\, \in\, \mbG\, \cap\, \mbH$ is bounded below by
\beno
\label{union.bound}
\mbP\left((\bY, \bZ) \in \mbG \cap\, \mbH\right)
\gte 1 - \mbP\left((\bY, \bZ) \not\in \mbG\right) - \mbP\left((\bY, \bZ) \not\in \mbH\right)
&=& 1 - o(1).
\ee
The above inequality stems from a union bound,
while the identity follows from the assumption that the probabilities of the events $(\bY, \bZ) \not\in \mbG$ and $(\bY, \bZ) \not\in \mbH$ satisfy
\beno
\mbP\left((\bY, \bZ) \not\in \mbG\right)
&=& \mbP\left(\norm{\bs(\truth;\, \bY, \bZ) - \mbE\,\, \bs(\truth;\, \bY, \bZ)}_{\infty} \geq \delta_N\right)
&=& o(1)\s
\\
\mbP\left((\bY, \bZ) \not\in \mbH\right)
&=& o(1),
\ee
where the first result leverages the fact that $\mbE\,\, \bs(\truth;\, \bY, \bZ) = \bm{0}$ by Lemma 7 of S25. 

\s

{\bf Conclusion.} 
Combining \eqref{keyresult} with \eqref{union.bound} establishes that,
for all $N > N_0$,
the random set $\widehat\Nat(\delta_N)$ is non-empty and,
with probability $1 - o(1)$,
satisfies
\beno
\widehat\Nat(\delta_N) 
&\subseteq& \mB_{\infty}(\truth,\, \rho_N).
\ee

\section{Corollaries \ref{theorem.model} and \ref{theorem.model_nonoverlapping}}
\label{sec:proof.corollary1}

To state and prove Corollaries \ref{theorem.model} and \ref{theorem.model_nonoverlapping},
we first introduce notation along with background on conditional independence graphs \citepsupp{graphical.models} and couplings \citepsupp{Li02}.

\subsection{Notation and Background}
\label{sec:setup}

We consider the model of Corollary \ref{theorem.model},
with joint probability mass function
\be
\label{eq:joint}
\mbP_{\nat}\left((\bY,\, \bZ) = (\by,\, \bz) \mid \bX = \bx \right) 
&\propto& \exp\left(\nat^\top\, \bsuff(\bx,\,\by,\, \bz)\right).
\ee 
The parameter vector is $\nat \coloneqq (\alpha_{\mZ,1}, \ldots, \alpha_{\mZ,N}, \gamma_{\mZ, \mZ},\, \gamma_{\mX, \mY, \mZ})\; \in\; \mR^{N+2}$ and the vector of sufficient statistics is $\bsuff(\bx,\,\by,\, \bz) \in \mathbb{R}^{N+2}$,
with coordinates 
\begin{itemize}
\item $\suff_i(\bx,\,\by,\, \bz) \coloneqq \sum_{j \in \mP_N \setminus\, \{i\}}\, z_{i,j}$ ($i = 1, \ldots, N$),
\item $\suff_{N+1}(\bx,\,\by,\, \bz) \coloneqq \sum_{i=1}^N \sum_{j=i+1}^N d_{i,j}(\bz)\, z_{i,j}$,
\item $\suff_{N+2}(\bx,\,\by,\, \bz) \coloneqq \sum_{i=1}^N \sum_{j=i+1}^N\, c_{i,j}\, (x_i\, y_j + x_j\, y_i)\, z_{i,j}$,
\end{itemize}
where the terms $c_{i,j}$ and $d_{i,j}(\bz)$ are defined as follows:
\be\label{eq:cstar_sm}
c_{i,j} \coloneqq \one(\mN_i\, \cap\, \mN_j\, \neq \, \emptyset)
\\
d_{i,j}(\bz) \coloneqq \one(\exists\; k\, \in\, \mN_i\, \cap\, \mN_j\, :\, z_{i,k} = z_{k,j} = 1).
\ee

In light of $\psi \coloneqq 1$,
we do not distinguish between $\by$ and $\by^\star$ or $y_i$ and $y_i^\star$.
To ease the presentation,
we write $ Y_i \mid \bx,\, \, \by_{-i},\, \bz$ rather than $Y_i \mid (\bX,\, \, \bY_{-i}, \,\bZ) = (\bx,\, \, \by_{-i},\, \bz)$,
and $Z_{i,j} \mid \bx,\, \by,\, \bz_{-\{i,j\}}$ rather than $Z_{i,j} \mid (\bX,\, \bY,\, \bZ_{-\{i,j\}}) = (\bx,\, \by,\, \bz_{-\{i,j\}})$.
Expectations,
variances,
and covariances with respect to the conditional distributions of $Y_i \mid \bx,\, \by_{-i},\, \bz$ and $Z_{i,j} \mid \bx,\, \by,\, \bz_{-\{i,j\}}$ are denoted by $\mbE_{\mY,i}$, $\mbV_{\mY,i}$, $\mbC_{\mY,i}$ and $\mbE_{\mZ,i,j}$, $\mbV_{\mZ,i,j}$, $\mbC_{\mZ,i,j}$,
respectively.

{\bf Conditional independence graph.} 
Let $M \coloneqq N + \binom{N}{2}$ be the total number of responses and connections and  
\be
\label{eq:def_W}
\bW 
\,\coloneqq\, (W_1,\, \ldots,\, W_M) 
\,\coloneqq\, (Y_1,\, \ldots,\, Y_N,\, Z_{1,2},\, \ldots,\, Z_{N-1,N}) 
\,\in\, \mW
\,\coloneqq\, \{0,\, 1\}^{N + \binom{N}{2}} 
\ee
be the vector consisting of responses and connections. 
The conditional independence structure of the model can be represented by a conditional independence graph $\mG \coloneqq (\mV,\, \mE)$ with a set of vertices $\mV \coloneqq \{W_1, \ldots, W_M\}$ and a set of undirected edges $\mE$.
We refer to elements of $\mV$ and $\mE$ as vertices and edges of $\mG$.
There are two distinct subsets of vertices in $\mG$: 
\bi
\item the subset $\mV_{\mY}\, \coloneqq\, \{W_1, \dots, W_N\}$ corresponding to responses $Y_1, \ldots, Y_N$;
\item the subset $\mV_{\mZ}\, \coloneqq\, \{W_{N+1}, \dots, W_M\}$ corresponding to connections $Z_{1,2}, \ldots, Z_{N-1,N}$.
\ei
An undirected edge between two vertices in $\mG$ represents dependence of the two corresponding random variables conditional on all other random variables. 
The vertices in $\mG$ are connected to the following subsets of vertices (neighborhoods):
\begin{itemize}
\item The neighborhood of $Y_i$ in $\mG$ consists of all $Y_j$ and all $Z_{i,j}$ such that $j\, \in\, \mP_N \setminus\, \{i\}$ and $\mN_i \, \cap \,  \mN_j \neq \emptyset$.
\item The neighborhood of $Z_{i,j}$ in $\mG$ consists of
\begin{enumerate}
\item $Y_i$ and $Y_j$;
\item all $Z_{i,h}$ and $Z_{j,h}$ such that $h \in \mP_N \setminus\, \{i,\, j\}$ and $h \in \mN_i \, \cap \,  \mN_j$;
\item all $Z_{i,h}$ and $Z_{j,h}$ such that $h \in \mP_N \setminus\, \{i,\, j\}$ and $h \not\in \mN_i\, \cap\, \mN_j$ provided that either $j \in \mN_i \,\cap\, \mN_h$ holds or $i \in \mN_j\, \cap\, \mN_h$ holds.
\end{enumerate}    
\end{itemize}
Let $d_{\mG}(i,j)$ be the length of the shortest path from vertex $W_i \in  \mV$ to vertex $W_j \in  \mV$ in $\mG$ and let $\mS_{\mG, i, k}$ be the set of vertices with distance $k\in \left\{1, 2, \ldots\right\}$ to the $i$th vertex $W_i$ in $\mG$:
\beno
\label{def:s_set}
\mS_{\mG, i, k} 
&\coloneqq& \left\{W_j \in \mV\setminus \{W_i\}: d_{\mG}(i,j) = k \right\}.
\ee
We define the maximum degree of vertices relating to connections in $\mG$ as follows: 
\be
\label{eq:def_d}
D_N 
&\coloneqq& \underset{1\, \leq\, i\, \leq\, M}{\text{max}}\; |\mS_{\mG, i, 1}|.
\ee
\hide{ 

In principle,
the quantity $D_N$ can depend on $N$:
e.g.,
S25 demonstrate that $D_N$ can grow as fast as $D_N = O(\log N)$ in special cases.
That said,
we consider $D_N$ to be constant,
which simplifies results.
We henceforth assume that the constant $D_N$ does not depend on $N$ and satisfies $D_N \coloneqq D \in \{2, 3, \ldots\}$.

}

{\bf Coupling matrix.} 
Let $\bW_{a:b} \coloneqq (W_a,\, \ldots,\, W_b) \in \mW_{a:b}$ be the subvector consisting of responses and connections with indices $1 \leq a \leq  b \leq M$.
The set of random variables excluding the random variable $W_v \in \mV$ with $v \in \{1, \ldots, M\}$ is denoted by  $\bw_{-v} \in \mW_{-v}$.      
Consider any $\bm{a} \in \{0,1\}^{M-i}$ and define 
\beno 
\mbP_{\truth, \bw_{1:(i-1)},w_i}(\bW_{(i+1):M} = \bm{a})  
&\coloneqq& \mbP_{\truth}(\bW_{(i+1):M} = \bm{a} \mid (\bW_{1:(i-1)}, W_i) =  (\bw_{1:(i-1)}, w_i)).
\ee
We use the total variation distance between the conditional distributions $\mbP_{\truth, \bw_{1:(i-1)},0}$ and $\mbP_{\truth, \bw_{1:(i-1)},1}$ for quantifying the amount of dependence induced by the model,
where $\truth \in \Nat$ is the data-generating parameter vector.
The total variation distance between $\mbP_{\truth, \bw_{1:(i-1)},0}$ and $\mbP_{\truth, \bw_{1:(i-1)},1}$ can be bounded from above by using coupling methods \citepsupp{Li02}. 
A coupling of $\mbP_{\truth, \bw_{1:(i-1)},0}$ and $\mbP_{\truth, \bw_{1:(i-1)},1}$ is a joint probability distribution $\mbQ_{\truth, i, \bw_{1:(i-1)}}$ for a pair of random vectors $(\bW_{(i+1):M}^\star,\, \bW_{(i+1):M}^{\star\star}) \in \{0,1\}^{M-i} \times \{0,1\}^{M-i}$ with marginals $\mbP_{\truth, \bw_{1:(i-1)},0}$ and $\mbP_{\truth, \bw_{1:(i-1)},1}$. 
For convenience, 
we define $(\bW^\star, {\bW}^{\star\star}) \in \{0,\, 1\}^M \times \{0,\, 1\}^M$, 
where the first $i$ elements are given by $\bW^\star_{1:i} = (\bw_{1:(i-1)},\, 0)$ and ${\bW}^{\star\star}_{1:i} = (\bw_{1:(i-1)},\, 1)$,
respectively.
The basic coupling inequality \citepsupp[][Theorem 5.2, p.\ 19]{Li02} shows that any coupling satisfies
\be
\label{eq:tv_to_c}
\TV{\mbP_{\truth, \bw_{1:(i-1)},0}-\mbP_{\truth, \bw_{1:(i-1)},1}} &\leq& \mbQ_{\truth, i, \bw_{1:(i-1)}}(\bW_{(i+1):M}^{\star} \neq \bW_{(i+1):M}^{\star\star}),
\ee
where $\TV{.}$ denotes the total variance distance between probability measures. 
If the two sides in Equation \eqref{eq:tv_to_c} are equal,
the coupling is called optimal.
An optimal coupling is guaranteed to exist,
but may not be unique \citepsupp[][pp.\ 99--107]{Li02}.
To prove Corollary \ref{theorem.model}, 
we need an upper bound on the spectral norm $\spectralnorm{\mD(\truth)}$ of the coupling matrix $\mD(\truth)$,
so we construct a coupling that is convenient but may not be optimal. 

A coupling $\mbQ_{\truth, i, \bw_{1:(i-1)}}$ of $\mbP_{\truth, \bw_{1:(i-1)},0}$ and $\mbP_{\truth, \bw_{1:(i-1)},1}$ can be constructed as follows:
\begin{enumerate}
\item[] \textbf{Step 1:} 
Set $\mathscr{U} = \{1, \ldots, i\}$ and $\mK = \{1, \ldots, M\}$. 
\item[] \textbf{Step 2:} 
Set $\mathscr{A} = \{j \in \mK\, \setminus\, \mathscr{U}:\; (W_i,\, W_j) \in \mE \text{ with } i \in \mathscr{U} \text{ and $j \in \mK\, \setminus\, \mathscr{U}$ such that } W_j^\star \neq W_j^{\star \star}\}$.
\begin{itemize}
\item[(a)] If $\mathscr{A} \neq \emptyset$, 
pick the smallest element $j \in \mathscr{A}$ and let $(W_j^\star,\, W_j^{\star \star})$ be distributed according to an optimal coupling of $\mbP_{\truth}(W_j = \cdot \mid \bW_{\mathscr{U}} = \bw_{\mathscr{U}}^{\star})$ and \linebreak 
$\mbP_{\truth}(W_j = \cdot \mid \bW_{\mathscr{U}} = \bw_{\mathscr{U}}^{\star \star})$. 
\item[(b)] If $\mathscr{A} = \emptyset$, pick the smallest element $j \in \mK \,\setminus \, \mathscr{U}$ and let $(W_j^\star, W_j^{\star \star})$ be distributed according to an optimal coupling of $\mbP_{\truth}(W_j = \cdot \mid \bW_{\mathscr{U}} = \bw_{\mathscr{U}}^{\star})$ and\linebreak
$\mbP_{\truth}(W_j = \cdot \mid \bW_{\mathscr{U}} = \bw_{\mathscr{U}}^{\star \star})$. 
\end{itemize}
\item[] \textbf{Step 3:} 
Replace $\mathscr{U}$ by $\mathscr{U} \, \cup\, \{j\}$ and repeat Step 2 until $\mK \,\setminus \, \mathscr{U} = \emptyset$.
\end{enumerate}
Based on $\mbQ_{\truth, i, \bw_{1:(i-1)}}$,
we construct a coupling matrix $\mD(\truth) \in \mathbb{R}^{M \times M}$ with elements 
\beno 
\mD_{i,j}(\truth) 
&\coloneqq& 
\begin{cases}
0& \mbox{if } i<j\\ 
1& \mbox{if } i=j\\ 
\max\limits_{\bw_{1:(i-1)}\, \in\, \mW_{1:i-1}} \mbQ_{\truth, i, \bw_{1:(i-1)}}(W_{j}^{\star} \neq W_{j}^{\star\star})& \mbox{if } i>j.
\end{cases}
\ee

\s

{\bf Overlapping subpopulations.}  
To obtain convergence rates based on a single observation of dependent random variables $\bW$,
we need to control the dependence of $\bW$ in the form of $\spectralnorm{\mD(\truth)}$.
In line with the simulation setting in Section \ref{sec:simulations}, 
we therefore assume that overlapping subpopulations $\mA_1, \mA_2, \ldots$ characterize the neighborhoods.
The neighborhood $\mN_i$ of unit $i \in \mP_N$ is then defined as
\be
\label{neigh.def} 
\mN_i 
&\coloneqq& \{j\, \in\, \mP_N: \mbox{ there exists } k \in \{1, 2, \ldots\} \text{ such that } i \in \mA_k \text{ and } j \in \mA_k\}.
\ee
Let $\mG_\mA$ be a subpopulation graph with a set of vertices $\mV_\mA \coloneqq \left\{\mA_1,  \mA_2 , \ldots\right\}$ and a set of edges connecting distinct subpopulations $\mA_k$ and $\mA_l$ with $\mA_k\, \cap\, \mA_l \neq \emptyset$. 
Define
\be
\label{s.def}
\mS_{\mG_\mA,\, i,k}  
&\coloneqq& \left\{\mA_j \in \mV_\mA\setminus \{\mA_i\}:\; d_{\mG_\mA}(i,j) = k \right\}.
\ee
Using the background introduced above,
we restate Condition \ref{as:neigh0} more formally.


\begin{assumption}[: Dependence] 
\label{as:neigh}
{\em The population $\mP$ consists of intersecting subpopulations $\mA_1, \mA_2, \ldots$,
whose intersections are represented by subpopulation graph $\mG_{\mA}$.
Let $D_N \in \{2, 3, \ldots\}$ be defined by \eqref{eq:def_d} and $\mS_{\mG_\mA,\, i,k}$ be defined by \eqref{s.def},
and assume that
\beno 
\max\limits_{k\, \in\, \{1, 2, \ldots\}}~|\mS_{\mG_\mA, k,l}| 
&\leq& \omega_1 + \dfrac{\omega_2}{2\, D_N^3} \log(l + 1),
&& l = 1, 2, \ldots,
\ee
where $\omega_1 \, \geq\, 0$ and  $0 \, \leq \, \omega_2\, \leq \, \min\limits\{\omega_1, \, 1/((\omega_1+1)\, |\log(1-U)|)\}$ with $U \coloneqq (1+\exp(-A))^{-1} > 0$. 
The constant $A > 0 $ is identical to the constant $A$ in Condition \ref{as:theta_main}. 
In addition,
for each unit $i \in \mP_i$,
the neighborhood $\mN_i$ is defined by \eqref{neigh.def},
and there exists a constant $B \in (0,\, +\infty)$ such that $\max_{1 \leq i \leq N} |\mN_i| < B$.
}
\end{assumption}

\hide{
Let $D \in \{2, 3, \ldots\}$ be the constant defined in \eqref{eq:def_d} and assume that
\beno 
\max\limits_{k \in \{1, 2, \ldots\}}~|\mS_{\mG_\mA, k,l} | 
&\leq& \omega_1 + \dfrac{\omega_2}{2\, D^3} \log(l + 1),
&& l = 1, 2, \ldots,
\ee
where $\omega_1 \, \geq\,  0$ and  $0 \, \leq \, \omega_2\, \leq \, \min\limits\{\omega_1, \, 1/((\omega_1+1)\, |\log(1-U)|)\}$
with $U \coloneqq (1+\exp(-A))^{-1} > 0 $. 
The constant $A > 0 $ is identical to the constant $A$ in Condition \ref{as:theta_main}. 
}

The assumption $\max_{1 \leq i \leq N} |\mN_i| < B$ implies that $D_N$ is bounded above by a constant $D \in \{2, 3, \ldots\}$.
\hide{
Condition \ref{as:neigh0} can be relaxed in special cases:
e.g.,
it is possible to replace $\max_{1 \leq i \leq N} |\mN_i| < B$ by $\max_{1 \leq i \leq N} |\mN_i| = O(\log N)$ in special cases (as demonstrated by S25 for connections $\bZ$, without responses $\bY$).
That said,
in the absence of replications,
it may not be possible to allow the neighborhood sizes to grow faster than $\log N$.
}

\subsection{Proof of Corollary \ref{theorem.model}}
\label{sec:theorem.model}

To prove Corollary \ref{theorem.model},
define
\be
\label{def.h}
\mH 
\;\coloneqq\;
\mH_1\, \cap\, \mH_2\s
\\
\mH_1
\;\coloneqq\;
\left\{\bw \in \mW:\; \dsum_{i=1}^N \infnormvec{\bH_{i,1}(\bw)} \geq \dfrac{N}{2\, (1 + \chi(\truth))^2}\right\}\s
\\
\mH_2
\;\coloneqq\;
\left\{\bw \in \mW:\; \dsum_{i=1}^N \infnormvec{\bH_{i,2}(\bw)} \geq \dfrac{c^2\, N}{2\, (1 + \chi(\truth))}\right\}\s
\\ 
\bH_{i,1}(\bw)
\;\coloneqq\;
(d_{i,1}(\bz),\, \ldots,\, d_{i,i-1}(\bz),\, d_{i,i+1}(\bz),\, \ldots,\, d_{i,N}(\bz))\s
\\
\bH_{i,2}(\bw) 
\;\coloneqq\;
(c_{i,1}\, x_1^2\, z_{i,1},\, \ldots,\, c_{i,i-1}\, x_{i-1}^2\, z_{i,i-1},\, c_{i,i+1}\, x_{i+1}^2\, z_{i,i+1},\, \ldots,\,
c_{i,N}\, x_N^2\, z_{i,N})
\ee
and
\be
\label{def.chi}
\chi(\truth)
&\coloneqq& \exp(C\, D^2\, (|\!|\truth|\!|_\infty + \epsilon^\star)),
\ee
where the constants $0 < c < C < \infty$ and $D \in \{2, 3, \ldots\}$ are identical to the corresponding constants defined in Condition \ref{as:1_main} and Equation \eqref{eq:def_d}, 
respectively.\s

\textsc{Proof of Corollary \ref{theorem.model}.} 
We prove Corollary \ref{theorem.model} using Theorem \ref{thm:mple_consistency} in five steps:
\bi
\item[] {\bf Step 1:} 
We bound
\beno
\mbP\left(\infnormvec{\nabla_{\nat}\; \ell(\nat; \bW)|_{\nat = \truth} - \mbE\, \nabla_{\nat}\; \ell(\nat; \bW)|_{\nat = \truth}} < \dfrac{\rho_N}{2\, \Lambda_N(\truth)}\right)
&\geq& 1 - \tau\left(\dfrac{\rho_N}{2\, \Lambda_N(\truth)}\right),
\ee
and choose $\rho_N$ so that $1 - \tau(\rho_N / (2\, \Lambda_N(\truth)))\, \geq\, 1 - 2\, / \max\{N,\, p\}^2$.
\item[] {\bf Step 2:} 
We show that $-\nabla_{\nat}^2~\ell(\nat;\, \bw)$ is invertible for all $\nat \in \mB_{\infty}(\truth,\, \epsilon^\star)$ and all $\bw \in \mH$.
\item[] {\bf Step 3:}
We prove that the event $\bW \in \mH$ occurs with probability at least\break 
$1 - \upsilon(\rho_N / (2\, \Lambda_N(\truth)))\, \geq\,  1 - 4\, / \max\{N,\, p\}^2$.
\item[] {\bf Step 4:} 
We bound $\delta_N$.
\item[] {\bf Step 5:} 
We bound $\rho_N$.
\ei
The proof of Corollary \ref{theorem.model} leverages auxiliary results supplied by Lemmas \ref{lemma.bounds.lambda}, 
\ref{lemma.bounds.psi}, 
and \ref{lemma.bounds.d},
which show that there exists an integer $N_1 \in \{3, 4, \dots\}$ such that,
for all $N > N_1$,
\beno
\Lambda_N(\truth) 
&\leq& C_1\; \dfrac{\chi(\truth)^9}{N} & \text{by Lemma \ref{lemma.bounds.lambda}}\s
\\ 
\sqrt{N/2} 
&\leq& \Psi_N  ~\leq~ C_2\, \sqrt{N} & \text{by Lemma \ref{lemma.bounds.psi}}\s
\\ 
\spectralnorm{\mD(\truth)} 
&\leq& C_3 & \text{by Lemma  \ref{lemma.bounds.d}},
\ee 
where $C_1 > 0 $,\,
$C_2 > 0 $,\, and
$C_3\geq 1$ are constants.

{\bf Step 1:}
Since $\bW \in \{0, 1\}^{M \times M}$,\,
Lemma 6 of S25 establishes 
\beno
\mbP\left(\infnormvec{\nabla_{\nat}\; \ell(\nat; \bW)|_{\nat = \truth} - \mbE\, \nabla_{\nat}\; \ell(\nat; \bW)|_{\nat = \truth}} < \dfrac{\rho_N}
{2\, \Lambda_N(\truth)}\right)
\gte 1 - \tau\left(\dfrac{\rho_N}{2\, \Lambda_N(\truth)}\right),
\ee
where
\beno 
\tau\left(\dfrac{\rho_N}{2\, \Lambda_N(\truth)}\right)
&\coloneqq& 2\, \exp\left( - \dfrac{\rho_N^2}{32\, \Lambda_N(\truth)^2\; (1 + D)^2\; \mnorm{\mD_N(\truth)}_2^2\; \Psi_N^2} + \log \, p \right),
\ee 
with $D \in \{2, 3, \ldots\}$ defined in \eqref{eq:def_d}.
Choosing
\be
\label{eq:rho_n}
\rho_N
&\coloneqq& \sqrt{96}\; \Lambda_N(\truth)\, (1+D)\, \spectralnorm{\mD(\truth)}\, \Psi_N \sqrt{\log\, \max\{N,\, p\}}
\ee
implies that the event
\beno
\infnormvec{\nabla_{\nat}\; \ell(\nat; \bW)|_{\nat = \truth} - \mbE\, \nabla_{\nat}\; \ell(\nat; \bW)|_{\nat = \truth}}
&<& \dfrac{\rho_N}{2\, \Lambda_N(\truth)}
\ee
occurs with probability at least
\beno
1 - \tau\left(\dfrac{\rho_N}{2\, \Lambda_N(\truth)}\right)
&\geq& 1 - \dfrac{2}{\max\{N,\, p\}^2}.
\ee

{\bf Step 2:} 
Let $\mH$ be defined in \eqref{def.h}.
Lemma \ref{lemma.bounds.lambda} establishes that $-\nabla_{\nat}^2\; \ell(\nat;\, \bw)$ is invertible for all $\nat \in \mB_{\infty}(\truth,\, \epsilon^\star)$ and all $\bw \in \mH$.

{\bf Step 3:}
Lemma \ref{lemma.h} shows that there exists an integer $N_2 \in \{3, 4, \dots\}$ such that,
for all $N > N_2$,
the event $\bW \in \mH$ occurs with probability at least
\beno
1 - \upsilon(\delta_N)
&=& 1 - \dfrac{4}{\max\{N,\, p\}^2}.
\ee

{\bf Step 4:}
The quantity
\beno
\delta_N
&\coloneqq& \dfrac{\rho_N}{2\, \Lambda_N(\truth)} 
&=& \sqrt{24}\;\, (1+D)\, \spectralnorm{\mD(\truth)}\, \Psi_N\, \sqrt{\log\, \max\{N,\, p\}}
\ee
is bounded below by
\beno
\delta_N
&\geq& \sqrt{24}\;\, D\, \sqrt{N/2}\; \sqrt{\log N}
&=& \sqrt{12}\;\, D\, \sqrt{N \log N}
\ee
and is bounded above by
\beno
\delta_N
&\leq& \sqrt{24}\;\, C_2\; C_3\, (2\, D)\, \sqrt{N}\, \sqrt{2\, \log N}
&=& \sqrt{192}\;\, C_2\; C_3\, D\, \sqrt{N \log N},
\ee
using $D \in \{2, 3, \ldots\}$, $1 \leq \spectralnorm{\mD(\truth)} \leq C_3$,\,
$\sqrt{N / 2} \leq \Psi_N \leq C_2\, \sqrt{N}$,\,
and $\max\{N,\, p\} = p = N+2$.
Since $C_2 > 0 $, $C_3\geq 1$, and $D \in \{2, 3, \ldots\}$ defined in \eqref{eq:def_d} are constants,
there exist constants $0 \,< \,L \,\leq\, U \, <\,\infty $ such that
\beno
L\; \sqrt{N \log N}
&\leq& \delta_N 
&\leq& U\; \sqrt{N \log N}.
\ee

{\bf Step 5:}
Substituting the bounds on $\Lambda_N(\truth)$,\,
$\Psi_N$,\,
and $\spectralnorm{\mD(\truth)}$ supplied by Lemmas \ref{lemma.bounds.lambda}, 
\ref{lemma.bounds.psi}, 
and \ref{lemma.bounds.d}
into \eqref{eq:rho_n} reveals that
\be
\label{eq:rho_n_approx}
\rho_N 
&\coloneqq& \sqrt{96}\; \Lambda_N(\truth)\, (1+D)\, \spectralnorm{\mD(\truth)}\, \Psi_N \sqrt{\log\, \max\{N,\, p\}}\s
\\
&\leq& \sqrt{96}\;\, C_1\; C_2\; C_3\; (2\, D)\; \dfrac{\chi(\truth)^9}{N}\; \sqrt{N \log(N+2)}\s
\\
&\leq& \sqrt{768}\;\, C_1\; C_2\; C_3\; D\, \chi(\truth)^9\; \sqrt{\dfrac{\log N}{N}},
\ee
using $\max\{N,\, p\} = p = N+2$ and $\log(N+2) \leq \log(2\, N) \leq 2\, \log N$ ($N\geq 2$).
To bound $\chi(\truth)$,
we invoke Condition \ref{as:theta_main}:
\beno
\chi(\truth)^9
\;\coloneqq\; \exp(C\, D^2\, (|\!|\truth|\!|_\infty + \epsilon^\star))^9
\;\leq\; \exp(C\, D^2\, (A + \epsilon^\star))^9
\;=\; \exp(9\; C\, D^2\, (A + \epsilon^\star)).
\ee
Define
\beno
K
&\coloneqq& \sqrt{768}\;\, C_1\; C_2\; C_3\; D\, \exp(9\; C\, D^2\, (A + \epsilon^\star))
&>& 0.
\ee
Since $A$, $C$, $C_1$, $C_2$, $C_3$, $D$, and $\epsilon^\star$ are independent of $N$,
so is $K$.
We conclude that
\beno
\rho_N
\lte K\; \sqrt{\dfrac{\log N}{N}} 
&\to& 0 
&\mbox{as}& N \to \infty.
\ee

{\bf Conclusion.}
Theorem \ref{thm:mple_consistency} implies that,
for all $N > N_0 \coloneqq \max\{N_1, N_2\}$,
the random set $\widehat\Nat(\delta_N)$ is non-empty and satisfies
\beno
\widehat\Nat(\delta_N) 
&\subseteq& \mB_{\infty}\left(\truth,\; K\, \sqrt{\dfrac{\log N}{N}} 
\right)	
\ee
with probability at least 
\beno
1 - \tau(\delta_N) - \upsilon(\delta_N)
&\geq& 1 - \dfrac{6}{\max\{N,\, p\}^2}
&\geq& 1 - \dfrac{6}{N^2},
\ee
using $\max\{N,\,  p\}^2 = p^2 = (N + 2)^2 \geq N^2$.

\hide{
Old version: 

\textsc{Proof of Corollary \ref{theorem.model}.} 
We prove Corollary \ref{theorem.model} using Theorem \ref{thm:mple_consistency} in three steps:
\bi
\item[1.] We bound 
\beno
\mbP\left(\infnormvec{\nabla_{\nat}\; \ell(\nat; \bW)|_{\nat = \truth} - \mbE\, \nabla_{\nat}\; \ell(\nat; \bW)|_{\nat = \truth}} < \dfrac{\rho_N}{2\, \Lambda_N(\truth)}\right)
&\geq& 1 - \tau\left(\dfrac{\rho_N}{2\, \Lambda_N(\truth)}\right)
\ee
and choose $\rho_N$ so that $1 - \tau(\rho_N / (2\, \Lambda_N(\truth)))\, \geq\, 1 - 2\, / \max\{N, p\}^2$.
\item[2.] We establish that $-\nabla_{\nat}^2~\ell(\nat;\, \bw)$ is invertible for all $\mB_{\infty}(\truth,\, \epsilon^\star)$ and all $\bw \in \mH$.
In addition,
we prove that the event $\bW \in \mH$ occurs with probability at least $1 - \upsilon(\rho_N / (2\, \Lambda_N(\truth)))\, \geq\,  1 - 4\, / \max\{N, p\}^2$.
\item[3.] We bound $\rho_N$.
\ei
The proof of Corollary \ref{theorem.model} leverages auxiliary results supplied by Lemmas \ref{lemma.bounds.lambda}, 
\ref{lemma.bounds.psi}, 
and \ref{lemma.bounds.d},
which show that there exists an integer $N_1 \in \{3, 4, \dots\}$ such that,
for all $N > N_1$,
\beno
\Lambda_N(\truth) 
&\leq& C_1\; \dfrac{D^7\, \chi(\truth)^9}{N} & \text{by Lemma \ref{lemma.bounds.lambda}}\s
\\ 
\sqrt{N/2} 
&\leq& \Psi_N  ~\leq~ C_2\, \sqrt{N} & \text{by Lemma \ref{lemma.bounds.psi}}\s
\\ 
\spectralnorm{\mD(\truth)} 
&\leq& C_3 & \text{by Lemma  \ref{lemma.bounds.d}},
\ee 
where $C_1 > 0 $,\,
$C_2 > 0 $,\,
$C_3\geq 1$,
and $D \in \{2, 3, \ldots\}$ are constants,
with $D$ defined in \eqref{eq:def_d}.

{\bf 1. Concentration of $\nabla_{\nat}\; \ell(\nat; \bW)|_{\nat = \truth}$.}
Since $\bW \in \{0, 1\}^{M \times M}$,
Theorem 1 of \citetsupp[][p.\ 207]{Chetal07} can be invoked along with a union bound to establish that
\beno
\mbP\left(\infnormvec{\nabla_{\nat}\; \ell(\nat; \bW)|_{\nat = \truth} - \mbE\, \nabla_{\nat}\; \ell(\nat; \bW)|_{\nat = \truth}} < \dfrac{\rho_N}
{2\, \Lambda_N(\truth)}\right)
\gte 1 - \tau\left(\dfrac{\rho_N}{2\, \Lambda_N(\truth)}\right),
\ee
where
\beno 
\tau\left(\dfrac{\rho_N}{2\, \Lambda_N(\truth)}\right)
&\coloneqq& 2\, \exp\left( - \dfrac{\rho_N^2}{32\, \Lambda_N(\truth)^2\; (1 + D)^2\; \mnorm{\mD_N(\truth)}_2^2\; \Psi_N^2} + \log \, p \right).
\ee 
Choosing
\be
\label{eq:rho_n}
\rho_N
&\coloneqq& \sqrt{96}\; \Lambda_N(\truth)\, (1+D)\, \spectralnorm{\mD(\truth)}\, \Psi_N \sqrt{\log\, \max\{N,\, p\}}
\ee
implies that the event
\beno
\infnormvec{\nabla_{\nat}\; \ell(\nat; \bW)|_{\nat = \truth} - \mbE\, \nabla_{\nat}\; \ell(\nat; \bW)|_{\nat = \truth}}
&<& \delta_N
\ee
occurs with probability at least
\beno
1 - \tau(\delta_N)
&\geq& 1 - \dfrac{2}{\max\{N,\, p\}^2}.
\ee
The quantity $\delta_N$ is defined as
\beno
\delta_N
&\coloneqq& \dfrac{\rho_N}{2\, \Lambda_N(\truth))}
&=& \sqrt{24}\;\, (1+D)\, \spectralnorm{\mD(\truth)} \, \Psi_N \sqrt{\log\, \max\{N,\, p\}},
\ee
which is bounded below by
\beno
\delta_N
&\geq& \sqrt{24}\;\, D\, \sqrt{N/2}\; \sqrt{\log N}
&=& \sqrt{12}\;\, D\, \sqrt{N \log N}
\ee
and is bounded above by
\beno
\delta_N
&\leq& \sqrt{24}\;\, (2\, D)\, C_2\, C_3\, \sqrt{N}\, \sqrt{2\, \log N}
&=& \sqrt{192}\;\, C_2\, C_3\, D\, \sqrt{N \log N},
\ee
using $D \in \{2, 3, \ldots\}$,\,
$1 \leq \spectralnorm{\mD(\truth)} \leq C_3$ (Lemma \ref{lemma.bounds.d}),\,
$\sqrt{N / 2} \leq \Psi_N \leq C_2\, \sqrt{N}$ (Lemma \ref{lemma.bounds.psi}),\,
and $\max\{N,\, p\} = p = N+2$.
Since $C_2 > 0 $, $C_3 > 0 $, and $D \in \{2, 3, \ldots\}$ are constants,
there exist constants $0 < L \leq U$ such that
\beno
L\; \sqrt{N \log N}
&\leq& \delta_N 
&\leq& U\; \sqrt{N \log N}.
\ee

{\bf 2. Invertibility of $ -\nabla_{\nat}^2~\ell(\nat;\, \bw)$ for all $\nat \in \mB_{\infty}(\truth,\, \epsilon^\star)$ and all $\bw \in \mH$.}
Lemma \ref{lemma.bounds.lambda} shows that $-\nabla_{\nat}^2\; \ell(\nat;\, \bw)$ is invertible for all $\nat \in \mB_{\infty}(\truth,\, \epsilon^\star)$ and all $\bw \in \mH$,
where $\mH$ is defined in \eqref{def.h}.
Lemma \ref{lemma.h} shows that there exists an integer $N_2 \in \{3, 4, \dots\}$ such that,
for all $N > N_2$,
the event $\bW \in \mW$ occurs with probability at least
\beno
1 - \upsilon(\delta_N)
&\coloneqq& 1 - \dfrac{4}{\max\{N,\, p\}^2}.
\ee

{\bf 3. Bounding $\rho_N$.}
Substituting the bounds on $\Lambda_N(\truth)$,\,
$\Psi_N$,\,
and $\spectralnorm{\mD(\truth)}$ supplied by Lemmas \ref{lemma.bounds.lambda}, 
\ref{lemma.bounds.psi}, 
and \ref{lemma.bounds.d}
into \eqref{eq:rho_n} and invoking Condition \ref{as:theta_main} reveals that
\be
\label{eq:rho_n_approx}
\rho_N 
&\coloneqq& \sqrt{96}\; \Lambda_N(\truth)\, (1+D)\, \spectralnorm{\mD(\truth)}\, \Psi_N \sqrt{\log\, \max\{N,\, p\}}\s
\\
&\leq& \sqrt{96}\;\, C_1\; \dfrac{D^7\, \chi(\truth)^9}{N}\; (2\, D)\; C_2\; D^2\; C_3\; \sqrt{N \log(N+2)}\s
\\
&\leq& \sqrt{768}\;\, C_1\; C_2\; C_3\; D^{10}\, \chi(\truth)^9\; \sqrt{\dfrac{\log N}{N}},
\ee
using $\max\{N,\, p\} = p = N+2$ and $\log(N+2) \leq \log(2\, N) \leq 2\, \log N$ ($N\geq 2$).
To bound $\chi(\truth)$,
we invoke Condition \ref{as:theta_main}:
\beno
\chi(\truth)^9
\;\coloneqq\; \exp(C\, D^2\, (|\!|\truth|\!|_\infty + \epsilon^\star))^9
\;\leq\; \exp(C\, D^2\, (A + \epsilon^\star))^9
\;=\; \exp(9\; C\, D^2\, (A + \epsilon^\star)).
\ee
Define
\beno
K
&\coloneqq& \sqrt{768}\;\, C_1\; C_2\; \; D^{10}\, \exp(9\; C\, D^2\, (A + \epsilon^\star))
&>& 0.
\ee
Since $A$, $C$, $C_1$, $C_2$, $C_3$, $D$, and $\epsilon^\star$ are independent of $N$,
so is $K$.
We conclude that
\beno
\rho_N
\lte K\; \sqrt{\dfrac{\log N}{N}} 
&\to& 0 
&\mbox{as}& N \to \infty.
\ee

{\bf Conclusion.}
Theorem \ref{thm:mple_consistency} implies that,
for all $N > N_0 \coloneqq \max\{N_1, N_2\}$,
the random set $\widehat\Nat(\delta_N)$ is non-empty and satisfies
\beno
\widehat\Nat(\delta_N) 
&\subseteq& \mB_{\infty}\left(\truth,\; K\, \sqrt{\dfrac{\log N}{N}} 
\right)	
\ee
with probability at least 
\beno
1 - \tau(\delta_N) - \upsilon(\delta_N)
&\geq& 1 - \dfrac{6}{\max\{N,\, p\}^2}
&\geq& 1 - \dfrac{6}{N^2},
\ee
using $\max\{N,\,  p\}^2 = p^2 = (N + 2)^2 \geq N^2$. 
}
\subsection{Statement and Proof of Corollary \ref{theorem.model_nonoverlapping}}
\label{sec:cor_nonoverlapping}

If subpopulations do not overlap,
$\infnormvec{\truth}$ can grow as a function $N$. 
Condition \ref{as:theta} details how fast $\infnormvec{\truth}$ can grow.

\begin{assumption} 
\label{as:theta}
{\em The parameter space is $\Nat = \mR^{N+2}$ and the data-generating parameter vector $\truth \in \mR^{N + 2}$ satisfies
\beno 
\normsup{\truth} 
&\leq& \dfrac{E  + \vartheta\, \log N}{C\, D^2} - \epsilon^\star,
\ee
where $E \geq 0$ and $\vartheta \in [0,\, 1/18)$ are constants,\,
$C > 0$ is identical to the constant $C$ in Condition \ref{as:1_main},\,
$D \in \{2, 3, \ldots\}$ is identical to the constant $D$ in \eqref{eq:def_d},\,
and $\epsilon^\star > 0 $ is identical to the constant $\epsilon^\star$ in the definition of $\Lambda_N(\truth)$ in Section \ref{sec:theory}.
}
\end{assumption}

Corollary \ref{theorem.model_nonoverlapping} replaces Condition \ref{as:theta_main} by Condition \ref{as:theta}. Resulting from this, the constant $U$ coming up in Condition \ref{as:neigh} is redefined as  $U \coloneqq (1+\exp(-D))^{-1} > 0$. 

\begin{corollary}
\label{theorem.model_nonoverlapping}
{\em
Consider a single observation of dependent responses and connections $(\bY, \bZ)$ generated by the model with parameter vector $\truth \coloneqq (\alpha_{\mZ,1}^\star,\, \dots, \alpha_{\mZ,N}^\star,\, \gamma_{\mZ,\mZ}^\star,\, \gamma_{\mX, \mY, \mZ}^\star) \in \mR^{N+2}$.
If Conditions \ref{as:1_main},
\ref{as:neigh},
and \ref{as:theta} are satisfied with $\vartheta \in [0, 1/18)$,
there exist constants $K \in (0, +\infty)$ and $0 < L \leq U < +\infty$ along with an integer $N_0 \in \{3, 4, \dots\}$ such that,
for all $N > N_0$,
the quantity $\delta_N$ satisfies 
\beno
L\, \sqrt{N \log N}
&\leq& \delta_N
&\leq& U\, \sqrt{N \log N},
\ee
and the random set $\widehat\Nat(\delta_N)$ is non-empty and satisfies
\beno
\widehat\Nat(\delta_N)
&\subseteq& \mB_{\infty}\left(\truth,\; K\,  \sqrt{\dfrac{\log N}{N^{1-18\, \vartheta}}} 
\right)
\ee
with probability at least $1 - 6\, / N^2$.
}
\end{corollary}

\textsc{Proof of Corollary \ref{theorem.model_nonoverlapping}.} 
The proof of Corollary \ref{theorem.model_nonoverlapping} resembles the proof of Corollary \ref{theorem.model},
with Condition \ref{as:theta_main} replaced by Condition \ref{as:theta}.
The proof of Corollary \ref{theorem.model} shows that 
\beno 
\rho_N
&\leq& \sqrt{768}\;\, C_1\; C_2 \, C_3\; D\; \chi(\truth)^9\; \sqrt{\dfrac{\log N}{N}}, 
\ee
where the constants $C_1 > 0 $, $C_2 > 0 $,\, $C_3\geq 1$, and $D \in \{2, 3, \ldots\}$ are defined in Lemmas  \ref{lemma.bounds.lambda}, 
\ref{lemma.bounds.psi}, 
and \ref{lemma.bounds.d}, and Equation \eqref{eq:def_d}, respectively. 
Condition \ref{as:theta} implies that
\beno 
\chi(\truth) ^9
\;\coloneqq\; \exp(C\, D^2\, (|\!|\truth|\!|_\infty + \epsilon^\star))^9
\;\leq\; \exp\left(C\, D^2\, \left(\dfrac{E  + \vartheta\, \log N}{C\, D^2}\right)\right)^9
\;=\; \exp(9\, E)\, N^{9\, \vartheta},
\ee    
which in turn implies that
\beno
\label{eq:rho_n_approx_2}
\rho_N
&\leq& \sqrt{768}\;\, C_1\; C_2 \, C_3\; D\, \exp (9\, E )\, \sqrt{\dfrac{\log N}{N^{1 - 18\, \vartheta}}}  
&=& K\; \sqrt{\dfrac{\log N}{N^{1 - 18\, \vartheta}}},
\ee
where $K \coloneqq \sqrt{768}\; C_1\; C_2 \, C_3\; D\, \exp (9\, E) > 0 $. 
The remainder of the proof of Corollary \ref{theorem.model_nonoverlapping} resembles the proof of Corollary \ref{theorem.model}.
We conclude that there exists an integer $N_0 \in \{3, 4, \dots\}$ such that,
for all $N > N_0$,
the random set $\widehat\Nat(\delta_N)$ is non-empty and satisfies
\beno
\widehat\Nat(\delta_N) 
&\subseteq& \mB_{\infty}\left(\truth,\;K\; \sqrt{\dfrac{\log N}{N^{1 - 18\, \vartheta}}}
\right)	
\ee
with probability at least $1 - 6\, / N^2$. 

\subsection{Bounding \texorpdfstring{$\Lambda_N(\truth)$}{Lambda}}

\begin{lemma}
\label{lemma.bounds.lambda}    
{\em 
Consider the model of Corollary \ref{theorem.model}.
If Conditions \ref{as:1_main} and \ref{as:neigh} are satisfied along with either Condition \ref{as:theta_main} or Condition \ref{as:theta} with $\vartheta \in [0, 1/18)$,
there exists a constant $C_1 > 0 $ along with an integer $N_0 \in \{3, 4, \dots\}$ such that,
for all $N > N_0$, 
\bi
\item $(-\nabla_{\nat}^2\; \ell(\nat;\, \bw))^{-1}$ is invertible for all $\nat \in\mB_{\infty}(\truth,\, \epsilon^\star)$ and all $\bw \in \mH$,
\item the event $\bW \in \mH$ occurs with probability at least $1 - 4\, / \max\{N,\, p\}^2$,
\item 
$\Lambda_{N}(\truth)  
~\coloneqq~ \sup\limits_{\bw\, \in\, \mH}\;\, \sup\limits_{\nat\, \in\, \mB_{\infty}(\truth,\, \epsilon^\star)}\, \infnorm{(-\nabla_{\nat}^2\; \ell(\nat;\, \bw))^{-1}}
~\leq ~C_1\,\dfrac{\chi(\truth)^9}{N}$,
\ei
where $\mH$ is defined in \eqref{def.h} and $\chi(\truth)$ is defined in \eqref{def.chi}.
}
\end{lemma}

\textsc{Proof of Lemma \ref{lemma.bounds.lambda}.}  
We first partition $-\nabla_{\nat}^2\; \ell(\nat;\, \bw)$ in accordance with $\nat \coloneqq (\nat_1,\, \nat_2)$,
given by $\nat_1 \coloneqq (\alpha_{\mZ,1}, \dots, \alpha_{\mZ,N}) \in \mR^N$ and $\nat_2 \coloneqq (\gamma_{\mZ,\mZ}, \gamma_{\mX,\mY,\mZ}) \in \mR^2$:
\be 
\label{eq:exp_hessian}
-\nabla_{\nat}^2\; \ell(\nat;\, \bw) 
&\coloneqq& 
\begin{pmatrix}
\bA(\nat,\, \bw)  & \bC(\nat,\, \bw)\\ 
\bC(\nat,\, \bw)^\top & \bB(\nat,\, \bw)
\end{pmatrix},
\ee
where the matrices $\bA(\nat,\, \bw)\in \mR^{N\times N}$ and $\bB(\nat,\, \bw)\in \mR^{2\times 2}$  define the covariance matrices of the sufficient statistics corresponding to the parameters $\nat_1$ and $\nat_2$, respectively.
Define $\bC(\nat,\, \bw) \coloneqq \left(\bc_1(\nat,\, \bw),\, \bc_2(\nat,\, \bw)\right)\in \mR^{N\times 2}$, 
where $\bc_1(\nat,\, \bw) \in \mathbb{R}^{N}$ and $\bc_2(\nat,\, \bw) \in \mathbb{R}^{N}$ are the covariances of the degree terms with the transitive connection term with weight $\gamma_{\mZ, \mZ}$ and spillover term with weight $\gamma_{\mX, \mY, \mZ}$, 
respectively.

We wish to bound the infinity norm of $\left(- \nabla_{\nat}^2\; \ell(\nat; \bw) \right)^{-1}$,
given by
\beno \label{eq:exp_inv_hessian}
(- \nabla_{\nat}^2\; \ell(\nat,\, \bw))^{-1}
&=& \begin{pmatrix}
\bA(\nat,\, \bw)  & \bC(\nat,\, \bw)\\ 
\bC(\nat,\, \bw)^\top & \bB(\nat,\, \bw)
\end{pmatrix}^{-1} \s\\ 
&=&
\begin{pmatrix}
\bA(\nat,\, \bw)^{-1}  & \bm{0}_{N,2}\\ 
\bm{0}_{2,N} & \bm{0}_{2,2}
\end{pmatrix} \\ 
&+& \begin{pmatrix}
\bA(\nat,\, \bw)^{-1}\bC(\nat,\, \bw)^\top\\
-\bm{I}_{2,2}
\end{pmatrix} \bm{V}(\nat,\, \bw)^{-1}
\begin{pmatrix}
\bA(\nat,\, \bw)^{-1}\bC(\nat,\, \bw)^\top\\
-\bm{I}_{2,2}
\end{pmatrix}^\top,
\ee
where $\bm{0}_{a,b} \coloneqq \text{diag}(0,\ldots, 0) \in \{0,1\}^{a \times b}$ and $\bm{I}_{a,b} \coloneqq \text{diag}(1,\ldots, 1) \ \in \{0,1\}^{a \times b}$ ($a, b \in \{1, 2, \ldots \}$) are diagonal matrices, 
and
\beno
\bm{V}(\nat,\, \bw) &\coloneqq&   \bB(\nat,\, \bw)  -  \bC(\nat,\, \bw)^\top \bA(\nat,\, \bw)^{-1}\, \bC(\nat,\, \bw)
\ee
is the Schur complement of $- \nabla_{\nat}^2\; \ell(\nat,\, \bw)$ with respect to the block $\bA(\nat,\, \bw)$.

The $\ell_\infty$-induced norm is submultiplicative,
so
\be
\label{hessian_break}
& & \,\infnorm{(- \nabla_{\nat}^2\; \ell(\nat,\, \bw))^{-1}}\s\\
&\leq &\, \infnorm{\bm{A}(\nat,\, \bw)^{-1}\,} \\ 
&+& \infnorm{\begin{pmatrix}
\bm{A}(\nat,\, \bw)^{-1}\,\bm{C}(\nat,\, \bw)\\
-\bm{I}_{p,p}
\end{pmatrix}}\infnorm{\bm{V}(\nat,\, \bw)^{-1}}
\infnorm{\begin{pmatrix}
\bm{A}(\nat,\, \bw)^{-1}\,\bm{C}(\nat,\, \bw)\\
-\bm{I}_{p,p}
\end{pmatrix}^\top}\s\\
&\leq& \infnorm{\bm{A}(\nat,\, \bw)^{-1}\,} \, + \, \max\{1,\; \infnorm{\bm{A}(\nat,\, \bw)^{-1}\,\bm{C}(\nat,\, \bw)}\}\\ 
&\times& \infnorm{\bm{V}(\nat,\, \bw)^{-1}}
(\infnorm{\bm{C}(\nat,\, \bw)^{\top}\bm{A}(\nat,\, \bw)^{-1}\, }+1)\\
&\leq& \infnorm{\bm{A}(\nat,\, \bw)^{-1}\,} + \max\{1,\; \infnorm{\bm{A}(\nat,\, \bw)^{-1}}\,\infnorm{\bm{C}(\nat,\, \bw)}\}\\ 
&\times&\, \infnorm{\bm{V}(\nat,\, \bw)^{-1}}\,  \left(\infnorm{\bm{C}(\nat,\, \bw)^{\top}}\, \infnorm{\bm{A}(\nat,\, \bw)^{-1}\, }+1\right).
\ee
We bound the terms $\infnorm{\bm{A}(\nat,\, \bw)^{-1}}$,\,
$\infnorm{\bm{C}(\nat,\, \bw)^\top}$,\, 
and $\infnorm{\bm{V}(\nat,\, \bw)^{-1}}$ one by one.

{\bf Bounding $\infnorm{\bA(\nat,\, \bw)^{-1}}$.} 
The proof of Lemma 9 in S25 shows that 
\be
\label{abound}
\infnorm{\bA(\nat,\, \bw)^{-1}} 
\lte \dfrac{18\, \chi(\truth)^2}{N}
\ee
for all $\nat \in \mB_{\infty}(\truth,\, \epsilon^\star)$, where $\chi(\truth)$
is an upper bound on the inverse standard deviation of connections $Z_{i,j}$ of pairs of units $\{i,j\} \subset \mP_N$ with $\mN_i\, \cap\, \mN_j \neq \emptyset$ conditional on $\bX,\, \bY,\, \bZ_{-\{i,j\}}$.
Under the model considered here,
the conditional distribution of $Z_{i,j}$ is Bernoulli,
as shown in Section \ref{subsec:GLMRepresentationOfZ}.
Therefore, $\mathbb{V}_{\mZ, i,j} (Z_{i,j})$ is given by
\beno 
\mathbb{V}_{\mZ, i,j} (Z_{i,j}) &=& \mbP(Z_{i,j} = 1\mid \bx,\, \by,\, \bz_{-\{i,j\}}) \times(1- \mbP(Z_{i,j} = 1\mid \bx,\, \by,\, \bz_{-\{i,j\}})).
\ee
Applying the bounds on $\mbP(Z_{i,j} = 1\mid \bx,\, \by,\, \bz_{-\{i,j\}})$ supplied by Lemma \ref{lemma.bounds.conditional_z} gives
\be
\label{eq:var_z}
\mathbb{V}_{\mZ, i,j}(Z_{i,j}) 
&\geq& \dfrac{1}{(\exp\left(C\, D^2\, \norm{\nat}_{\infty}\right))^2}  
&\geq& \dfrac{1}{(\exp\left(C\, D^2\, (\norm{\truth}_{\infty} + \epsilon^\star)\right))^2},
\ee
provided $D \in \{2, 3, \ldots\}$, 
where $D$ corresponds to the constant $D$ defined in \eqref{eq:def_d} and $C$ corresponds to the constant $C$ in Condition \ref{as:1_main}.
For the second inequality of \eqref{eq:var_z}, 
we use the fact that $\norm{\nat}_{\infty} \leq \norm{\truth}_{\infty} + \epsilon^\star$ for all $\nat \in \mB_{\infty}(\truth,\, \epsilon^\star)$. 
With
\beno 
\chi(\truth) 
&\coloneqq& \exp\left(C\, D^2\, (\normsup{\truth} + \epsilon^\star)\right),
\ee
we therefore deduce that $\chi(\truth)$ is an bound on the inverse standard deviation of connections $Z_{i,j}$:
\beno
\chi(\truth) 
&\geq& \dfrac{1}{ \sqrt{\mathbb{V}_{\mZ, i,j}(Z_{i,j})}}.
\ee

{\bf Bounding $\infnorm{\bC(\nat,\, \bw)^\top}$.} 
Define $\bC(\nat,\, \bw) \coloneqq  \left(\bc_1(\nat,\, \bw),\,\bc_2(\nat,\, \bw)\right)$, 
where \newline $\bc_1(\nat,\, \bw) \in \mathbb{R}^{N}$ and $\bc_2(\nat,\, \bw) \in \mathbb{R}^{N}$ are the covariance terms of the degree terms with the sufficient statistics pertaining to the transitive connection term weighted by $\gamma_{\mZ, \mZ}$  and the spillover term weighted by $\gamma_{\mX, \mY, \mZ}$, respectively.
Then
\beno
\infnorm{\bC(\nat,\, \bw)^\top} &\leq& \normsup{\bc_1(\nat,\, \bw)}\,+\,\normsup{\bc_2(\nat,\, \bw)}.
\ee
We bound the terms $\normsup{\bc_1(\nat,\, \bw)}$ and $\normsup{\bc_2(\nat,\, \bw)}$ one by one. 

By Lemma 13 of S25, 
$\normsup{\bc_1(\nat,\, \bw)} \leq 3\, D^3$. 
The term\break 
$\bC_2(\nat,\, \bw) \coloneqq (C_{2,1}(\nat,\, \bw), \ldots, C_{2,N}(\nat,\, \bw)) \in \mathbb{R}^{N}$ 
refers to the covariances between the degrees $\suff_i(\bx,\, \by,\, \bz)$ of units $i \in \{1, \dots, N\}$ and $\suff_{N+2}(\bx,\, \by,\, \bz)$.
An upper bound on $t$-th element of $\bc_{2}(\nat,\, \bw)$ can be obtained by 
\be
\label{eq:c_t}
|C_{2,t}(\nat,\, \bw)| 
&=& \left|\dsum_{i = 1}^N \, \dsum_{j  = i +1}^N \mathbb{C}_{\mZ, i,j}\left(\suff_{t}(\bx,\, \by,\, \bZ),\, \suff_{N+2}(\bx,\, \by,\, \bZ)\right)\right|\s
\\ 
&=&  \left|\dsum_{i = 1}^N \, \dsum_{j  = i +1}^N \mathbb{C}_{\mZ, i,j}\left(\dsum_{h \neq t} Z_{h,t},\;\, \sum_{h=1}^N \sum_{k=h+1}^N\, c_{h,k}\, (x_h\, y_k + x_k\, y_h)\, Z_{h,k}\right)\right|\s
\\
&=& \left|\dsum_{i \neq t:\; \mN_i \,\cap \,\mN_t \, \neq\, \emptyset} \mathbb{C}_{\mZ, i,t}\left(Z_{i,t},\; (x_i\,y_t + y_i\,x_t)\, Z_{i,t}\right)\right|\s
\\ 
&=& \left|\dsum_{i \neq t:\; \mN_i \,\cap \,\mN_t\, \neq\, \emptyset} (x_i\,y_t + y_i\,x_t)\, \mathbb{V}_{\mZ, i,t}\left(Z_{i,t}\right) \right|\s
\\ 
&\leq& \left|\dfrac{C}{2}\; \dsum_{i \neq t:\; \mN_i\,\cap\, \mN_t\, \neq\, \emptyset}^N 1\right| 
\;\;\leq\;\; C\, D^2,
\ee
where $C$ corresponds to the constant from Condition \ref{as:1_main} and $D$ is defined in \eqref{eq:def_d}.
On the third line,
note that $\suff_{t}(\bx,\, \by,\, \bz)$ only depends on connection $Z_{i,j}$ if $t\in \{i,j\}$. 
Therefore, 
the covariance of  $\suff_{t}(\bx,\, \by,\, \bz)$ with respect to any other connection is 0. 
The first inequality follows from the observation that $x_i\, y_j + x_j\, y_i \leq 2\, C$ and $\mathbb{V}_{\mZ, i,j}\left(Z_{i,j}\right) \leq 1/4$, 
which follows from $0 \leq x_i \leq C < \infty$ by Condition \ref{as:1_main} and $Y_i \in \{0, 1\}$.
{The second inequality follows from Lemma 15 in S25 bounding the pairs of units $i$ and $t$ such that $\mN_i\, \cap \mN_t\, \neq\, \emptyset$ from above by $D^2$.} 
Since the bound from \eqref{eq:c_t} holds for all $t \in \{1, \ldots, N\}$, 
we obtain $\normsup{\bC_2(\nat,\, \bw)} \leq C\, D^2$.
Taken together,
\be
\label{eq:c}
\infnorm{\bC(\nat,\, \bw)^\top} &\leq& 3\,D^3 + C\,D^2 
&\leq& \max\{3,\, C\}\,  D^3.
\ee
{\bf Bounding $\infnorm{\bm{V}(\nat,\, \bw)^{-1}}$.}
Write
\beno
\bm{B}(\nat,\, \bw) &\coloneqq& 
\begin{pmatrix}
B_{1,1}(\nat,\, \bw) & B_{1,2}(\nat,\, \bw)\\
B_{1,2}(\nat,\, \bw) & B_{2,2}(\nat,\, \bw)
\end{pmatrix} \s\\ 
\bm{V}(\nat,\, \bw) &\coloneqq&
\begin{pmatrix}
V_{1,1}(\nat,\, \bw) & V_{1,2}(\nat,\, \bw)\\
V_{1,2}(\nat,\, \bw) & V_{2,2}(\nat,\, \bw)
\end{pmatrix}.
\ee
The elements of $\bm{V}(\nat,\, \bw)$ are then given by
\beno 
V_{i,j}(\nat,\, \bw) 
&=& B_{i,j}(\nat,\, \bw) - \bc_i(\nat,\, \bw)^{\top}\bm{A}(\nat,\, \bw)^{-1}\,\bc_j(\nat,\, \bw).
\ee
The inverse of $\bm{V}(\nat,\, \bw)$ is
\beno
\bm{V}(\nat,\, \bw)^{-1} 
&=& \dfrac{1}{V_{1,1}(\nat,\, \bw)\, V_{2,2}(\nat,\, \bw)- V_{1,2}(\nat,\, \bw)^2}
\begin{pmatrix}
V_{2,2}(\nat,\, \bw) & -V_{1,2}(\nat,\, \bw)\\
-V_{1,2}(\nat,\, \bw) & V_{1,1}(\nat,\, \bw)
\end{pmatrix},
\ee
implying that
\be
\label{eq:w_inv}
\,\infnorm{(\bm{V}(\nat,\, \bw))^{-1}} 
&\leq& \dfrac{\max\left\{V_{1,1}(\nat,\, \bw), \,V_{2,2}(\nat,\, \bw)\right\} + |V_{1,2}(\nat,\, \bw)|}{|V_{1,1}(\nat,\, \bw)\, V_{2,2}(\nat,\, \bw)- V_{1,2}(\nat,\, \bw)^2|}.
\ee
Invoking the inequalities from \eqref{abound} and \eqref{eq:c},
we obtain for $i,j \in \{1,2\}$
\be
\label{eq:boundcac}
&~&|\bc_i(\nat,\, \bw)^{\top}\bm{A}(\nat,\, \bw)^{-1}\,\bc_j(\nat,\, \bw)|  \\ 
&\leq&  N\,\normsup{\bc_i(\nat,\, \bw)}\infnorm{\bm{A}(\nat,\, \bw)^{-1}\,}\normsup{\bc_i(\nat,\, \bw)}\s\\
&\leq&  N\,\infnorm{\bC(\nat,\, \bw)} \infnorm{\bm{A}(\nat,\, \bw)^{-1}\,}\infnorm{\bC(\nat,\, \bw)}\s\\
&\leq& 18\,\max\{9,\, C^2\}\,  D^6\, \chi(\truth)^2,
\ee
where $D$ corresponds to the constant $D$ defined in \eqref{eq:def_d} and $C$ corresponds to the constant  $C$ from Condition \ref{as:1_main}.

By applying Lemma \ref{lemma.bounds.v} along with \eqref{eq:boundcac}, we get for $i,j \in \{1,2\}$
\beno
|V_{i,j}(\nat,\, \bw)|\ &=&\, |B_{i,j}(\nat,\, \bw)|\, +\,|\bc_i(\nat,\, \bw)^{\top}\bm{A}(\nat,\, \bw)^{-1}\,\bc_j(\nat,\, \bw)|  \s\\ 
&\leq& \max\{1,\, C^2\} \dfrac{N \, D^5}{4} +18\,\max\{9,\, C^2\}\,  D^6\, \chi(\truth)^2\s\\
&\leq& \max\{9,\, C^2\} \,D^5 \left(\dfrac{N}{4} +18\,  D\, \chi(\truth)^2\right)
\ee 
Thus,
the numerator of \eqref{eq:w_inv} is bounded above by
\be
\label{eq:boundw}
&\max\left\{V_{1,1}(\nat,\, \bw),\, V_{2,2}(\nat,\, \bw)\right\} + |V_{1,2}(\nat,\, \bw)| \\
&\leq~  \max\{9,\, C^2\} \,D^5 \left(\dfrac{N}{2} +36\,  D\, \chi(\truth)^2\right).
\ee
The denominator of \eqref{eq:w_inv}, 
which is the determinant of $\bV(\nat,\, \bw)$, 
is
\beno 
& ~& ~ V_{1,1}(\nat,\, \bw)\, V_{2,2}(\nat,\, \bw)- V_{1,2}(\nat,\, \bw)^2\\
&=&\,(B_{1,1}(\nat,\, \bw) - \bc_1(\nat,\, \bw)^{\top}\bm{A}(\nat,\, \bw)^{-1}\,\bc_1(\nat,\, \bw))
\\
&\times& (B_{2,2}(\nat,\, \bw) - \bc_2(\nat,\, \bw)^{\top}\bm{A}(\nat,\, \bw)^{-1}\,\bc_2(\nat,\, \bw))
\\
&-& (B_{1,2}(\nat,\, \bw) - \bc_1(\nat,\, \bw)^{\top}\bm{A}(\nat,\, \bw)^{-1}\,\bc_2(\nat,\, \bw))^2
\\
&=&\,B_{1,1}(\nat,\, \bw)\,B_{2,2}(\nat,\, \bw) - B_{1,1}(\nat,\, \bw)\,\bc_2(\nat,\, \bw)^{\top}\bm{A}(\nat,\, \bw)^{-1}\,\bc_2(\nat,\, \bw)\\
&-&\, B_{2,2}(\nat,\, \bw)\,\bc_1(\nat,\, \bw)^{\top}\bm{A}(\nat,\, \bw)^{-1}\,\bc_1(\nat,\, \bw) \\ 
&+& (\bc_1(\nat,\, \bw)^{\top}\bm{A}(\nat,\, \bw)^{-1}\,\bc_1(\nat,\, \bw))\,(\bc_2(\nat,\, \bw)^{\top}\bm{A}(\nat,\, \bw)^{-1}\,\bc_2(\nat,\, \bw)) - B_{1,2}(\nat,\, \bw)^2 \\
&+&\,2\,B_{1,2}(\nat,\, \bw)\,(\bc_1(\nat,\, \bw)^{\top}\bm{A}(\nat,\, \bw)^{-1}\,\bc_2(\nat,\, \bw)) -(\bc_1(\nat,\, \bw)^{\top}\bm{A}(\nat,\, \bw)^{-1}\,\bc_2(\nat,\, \bw))^2.
\ee
Applying the property of positive semidefinite matrices $\bm{P} \in \mathbb{R}^{n \times n}$ that 
$(\bm{a}^\top \bm{P}\, \bm{a})\, (\bm{b}^\top \bm{P}\, \bm{b}) \geq (\bm{a}^\top \bm{P}\, \bm{b})^2$
is true for all vectors $\bm{a} \in \mathbb{R}^n$ and $\bm{b} \in \mathbb{R}^n$ ($n \geq 1$), 
we obtain
\beno 
& ~& ~ V_{1,1}(\nat,\, \bw)\, V_{2,2}(\nat,\, \bw)- V_{1,2}(\nat,\, \bw)^2\\
&\geq&\,B_{1,1}(\nat,\, \bw)\,B_{2,2}(\nat,\, \bw) - B_{1,2}(\nat,\, \bw)^2 \\ 
&-& 4\,\max\limits_{i,\, j}\,|B_{i,j}(\nat,\, \bw)|\,\,\max\limits_{i,\, j}\,|\bc_i(\nat,\, \bw)^{\top}\bm{A}(\nat,\, \bw)^{-1}\,\bc_j(\nat,\, \bw)| \\
&\geq&B_{1,1}(\nat,\, \bw)\,B_{2,2}(\nat,\, \bw) - B_{1,2}(\nat,\, \bw)^2-18\,  \max\{81,\, C^4\}\, N\, D^{11} \, \chi(\truth)^2\\ 
&=& U(\nat,\, \bw)- 18\,  \max\{81,\, C^4\}\, N\, D^{11} \, \chi(\truth)^2, 
\ee 
where
\be
\label{eq:def_u}
U(\nat,\, \bw) 
&\coloneqq& B_{1,1}(\nat,\, \bw)\,B_{2,2}(\nat,\, \bw) - B_{1,2}(\nat,\, \bw)^2.
\ee
The final inequality follows from invoking \eqref{eq:boundcac} along with Lemma \ref{lemma.bounds.v}.

For \eqref{eq:def_u}, 
we obtain
\beno
U(\nat,\, \bw )& =&\,B_{1,1}(\nat,\, \bw)\,B_{2,2}(\nat,\, \bw) - B_{1,2}(\nat,\, \bw)^2\s\\
&=&\left(\dsum_{i=1}^N\, \dsum_{j  = i +1}^N \mathbb{V}_{\mZ, i,j}\left( \suff_{N+1}(\bx,\, \by,\, \bZ)\right)\right)\s\\ 
& \times&    \left(\dsum_{i=1}^N\, \dsum_{j  = i +1}^N \mathbb{V}_{\mY, i}\left( \suff_{N+2}(\bx,\, \bY,\, \bz)\right) + \,\dsum_{i=1}^N\, \dsum_{j  = i +1}^N \mathbb{V}_{\mZ, i,j}\left( \suff_{N+2}(\bx,\, \by,\, \bZ)\right)\right) \s\\
&- &\left(\dsum_{i=1}^N\, \dsum_{j  = i +1}^N \mathbb{C}_{\mZ, i,j}\left(\suff_{N+1}(\bx,\, \by,\, \bZ),\; \suff_{N+2}(\bx,\, \by,\, \bZ)\right)\right)^2\s\\
&=&\left(\dsum_{i=1}^N\, \dsum_{j  = i +1}^N \mathbb{V}_{\mZ, i,j}\left( \suff_{N+1}(\bx,\, \by,\, \bZ)\right)\right)\left(\dsum_{i=1}^N \mathbb{V}_{\mY, i}\left( \suff_{N+2}(\bx,\, \bY,\, \bz)\right)\right)  \s\\
&+&\left(\dsum_{i=1}^N\, \dsum_{j  = i +1}^N \mathbb{V}_{\mZ, i,j}\left( \suff_{N+1}(\bx,\, \by,\, \bZ)\right)\right)\left(\dsum_{i=1}^N\, \dsum_{j  = i +1}^N \mathbb{V}_{\mZ, i,j}\left( \suff_{N+2}(\bx,\, \by,\, \bZ)\right)\right) \s \\
&-&\left(\dsum_{i=1}^N\, \dsum_{j  = i +1}^N \mathbb{C}_{\mZ, i,j}\left( \suff_{N+1}(\bx,\, \by,\, \bZ),\; \suff_{N+2}(\bx,\, \by,\, \bZ)\right)\right)^2.
\ee
Next, 
we show that the third term 
\beno 
\left(\dsum_{i=1}^N\, \dsum_{j  = i +1}^N \mathbb{C}_{\mZ, i,j}\left( \suff_{N+1}(\bx,\, \by,\, \bZ),\suff_{N+2}(\bx,\, \by,\, \bZ)\right)\right)^2
\ee
is smaller than the second term
\beno 
\left(\dsum_{i=1}^N\, \dsum_{j  = i +1}^N \mathbb{V}_{\mZ, i,j}\left( \suff_{N+1}(\bx,\, \by,\, \bZ)\right)\right)\left(\dsum_{i=1}^N\, \dsum_{j  = i +1}^N \mathbb{V}_{\mZ, i,j}\left( \suff_{N+2}(\bx,\, \by,\, \bZ)\right)\right).
\ee
Define
\beno 
u_{1,i,j}\coloneqq \sqrt{\mathbb{V}_{\mZ, i,j}\left( \suff_{N+1}(\bx,\, \by,\, \bZ)\right)}\ \ \ \text{and}\ \ \     
u_{2,i,j} \coloneqq \sqrt{\mathbb{V}_{\mZ, i,j}\left( \suff_{N+2}(\bx,\, \by,\, \bZ)\right)},
\ \ \ 
i=1,\ldots,N. 
\ee
Then the second term can be restated as follows:
\beno 
&\left(\dsum_{i=1}^N\, \dsum_{j  = i +1}^N \mathbb{V}_{\mZ, i,j}\left( \suff_{N+1}(\bx,\, \by,\, \bZ)\right)\right)\left(\dsum_{i=1}^N\, \dsum_{j  = i +1}^N \mathbb{V}_{\mZ, i,j}\left( \suff_{N+2}(\bx,\, \by,\, \bZ)\right)\right) \\ 
&=\left(\dsum_{i=1}^N\, \dsum_{j  = i +1}^N u_{1,i,j}^2\right)\left(\dsum_{i=1}^N\, \dsum_{j  = i +1}^N u_{2,i,j}^2\right),
\ee
while the third term is 
\beno 
&~&\left(\dsum_{i=1}^N\, \dsum_{j  = i +1}^N\, \mathbb{C}_{\mZ, i,j}\left(\suff_{N+1}(\bx,\, \by,\, \bZ),\suff_{N+2}(\bx,\, \by,\, \bZ)\right)\right)^2\s
\\
&\leq& \left(\dsum_{i=1}^N\, \dsum_{j  = i +1}^N\, |\mathbb{C}_{\mZ, i,j}\left(\suff_{N+1}(\bx,\, \by,\, \bZ),\suff_{N+2}(\bx,\, \by,\, \bZ)\right)|\right)^2\s
\\
&\leq& \left(\dsum_{i=1}^N\, \dsum_{j  = i +1}^N\, \sqrt{\mathbb{V}_{\mZ, i,j}\left( \suff_{N+1}(\bx,\, \by,\, \bZ)\right)\,\mathbb{V}_{\mZ, i,j}\left(\suff_{N+2}(\bx,\, \by,\, \bZ)\right)}\right)^2\s
\\
&=&\left(\dsum_{i=1}^N\, \dsum_{j  = i +1}^N u_{1,i,j}\, u_{2,i,j}\right)^2\s
\\ 
&\leq& \left(\dsum_{i=1}^N\, \dsum_{j  = i +1}^N u_{1,i,j}^2\right)\left(\dsum_{i=1}^N\, \dsum_{j  = i +1}^N u_{2,i,j}^2\right),
\ee
where the Cauchy-Schwarz inequality is invoked on the third and last line. 
This translates to the following lower bound on $U(\nat,\, \bw)$: 
\beno 
U(\nat,\, \bw )  &=&\, B_{1,1}(\nat,\, \bw)\,B_{2,2}(\nat,\, \bw) - B_{1,2}(\nat,\, \bw)^2\s\\
&\geq&\,\left(\dsum_{i=1}^N\, \dsum_{j  = i +1}^N \mathbb{V}_{\mZ, i,j}\left( \suff_{N+1}(\bx,\, \by,\, \bZ)\right)\right)\left(\dsum_{i=1}^N \mathbb{V}_{\mY, i}\left( \suff_{N+2}(\bx,\, \bY,\, \bz)\right)\right)\s\\
&\geq&\,\left(\dsum_{i = 1}^N \dfrac{ \infnormvec{\bH_{i,1}(\bw)}}{(1+\chi(\truth))^2}\right)\,  \dsum_{i=1}^N\left(\dsum_{j \neq i }^N c_{i,j}\, x_j\,z_{i,j}\right)^2\; \mathbb{V}_{\mY, i}\left(Y_i\right), 
\ee
where $\bH_{i,1}(\bw)$ is defined in \eqref{def.h} and the function $c_{i,j}$ is defined in \eqref{eq:cstar_sm}.
For the second inequality, 
we use the result from the proof of Lemma 13 in S25, 
which implies that
\be
\label{eq:ineq_h} 
\dsum_{i=1}^N\, \dsum_{j  = i +1}^N \mathbb{V}_{\mZ, i,j}\left(\suff_{N+1}(\by,\, \bZ)\right) 
&\geq& \dsum_{i=1}^N\, \dsum_{j  = i +1}^N c_{i,j} \, d_{i,j}(\bz)\, \mathbb{V}_{\mZ, i,j}\left(Z_{i,j}\right) \\ 
\ee
where the function $d_{i,j}(\bZ)$ is defined in \eqref{eq:cstar_sm}.
By Lemma \ref{lemma.bounds.conditional_z}, we get 
\beno 
\mathbb{V}_{\mZ, i,j} (Z_{i,j}) = \mbP(Z_{i,j}\mid \bx,\, \by_{-i},\, \bz)\times (1-\mbP(Z_{i,j}\mid \bx,\, \by_{-i},\, \bz)) 
&\geq& \dfrac{1}{(1 + \chi(\truth))^2},
\ee
where $\chi(\truth)$ is defined in \eqref{def.chi}.
When combined with \eqref{eq:ineq_h}, this results in 
\beno 
\dsum_{i=1}^N\, \dsum_{j  = i +1}^N \mathbb{V}_{\mZ, i,j}\left(\suff_{N+1}(\by,\, \bZ)\right)  &\geq& 
\dfrac{ \dsum_{i=1}^N\, \dsum_{j  = i +1}^N c_{i,j} \, d_{i,j}(\bz)}{(1+\chi(\truth))^2} \\
&\geq&  
\dsum_{i=1}^N \dfrac{\infnormvec{\bH_{i,1}(\bw)} }{2\, (1+\chi(\truth))^2},
\ee  
where the second inequality is again from the proof of Lemma 13 in S25. 

By applying Lemma \ref{lemma.bounds.conditional_y} and expanding the quadratic term, 
we obtain
\beno 
U(\nat,\, \bw) 
&\geq&\left(\dsum_{i = 1}^N \dfrac{ \infnormvec{\bH_{i,1}(\bw)}}{2\, (1 + \chi(\truth))^2}\right)\,  \dsum_{i=1}^N\, \mathbb{V}_{\mY, i}\left(Y_i\right)\, \s\\ 
&\times&\Bigg(\dsum_{j  = i +1}^N x_j^2\,c_{i,j}\,z_{i,j} +\dsum_{h= 1}^N\dsum_{k \neq h}^N \,c_{i,h}\,c_{i,k}\,   x_h \, x_k\,z_{i,h} \,z_{i,k}\Bigg)\s
\\
&\geq&\left(\dsum_{i = 1}^N \dfrac{ \infnormvec{\bH_{i,1}(\bw)}}{2\, (1 + \chi(\truth))^2}\right)\left(\dsum_{i=1}^N\, \mathbb{V}_{\mY, i}\left(Y_i\right) \, \dsum_{j  = i +1}^N x_j^2\;c_{i,j} \, z_{i,j}\right)\s
\\
&\geq& \dfrac{\left(\dsum_{i = 1}^N  \infnormvec{\bH_{i,1}(\bw)}\right)\, \left(\dsum_{i = 1}^N  \infnormvec{\bH_{i,2}(\bw)}\right)}{2\, (1+\chi(\truth))^4}
, 
\ee
where $\bH_{i,2}(\bw)$ is defined in \eqref{def.h} and $C$ corresponds to the constant from Condition \ref{as:1_main}.
The second inequality follows from the assumption $x_i \in [0,\, C]$ by Condition \ref{as:1_main} along with $c_{i,j} \in \{0, 1\}$ and $z_{i,j} \in \{0, 1\}$.
Lemma \ref{lemma.h} shows that 
\beno 
\mbP(\bW \in \mH) &\geq& 1 - \dfrac{4}{\max\{N,\, p\}^2},
\ee
where $\mH$ is defined in \eqref{def.h}.
For all $\bw \in \mH$, 
we obtain by definition 
\beno 
U(\nat,\, \bw ) 
&\geq& \dfrac{ \left(\dsum_{i = 1}^N  \infnormvec{\bH_{i,1}(\bw)}\right)\, \left(\dsum_{i = 1}^N  \infnormvec{\bH_{i,2}(\bw)}\right)}{2\, (1+\chi(\truth))^4}
&\geq& \dfrac{c^2\,N^2}{8\, (1+\chi(\truth))^7},
\ee
which results in the following bound for the denominator of \eqref{eq:w_inv}: 
\beno
V_{1,1}(\nat,\, \bw)\, V_{2,2}(\nat,\, \bw)- V_{1,2}(\nat,\, \bw)^2 &\geq& U(\nat,\, \bw )- 18\,  \max\{81,\, C^4\}\, N\, D^{11} \, \chi(\truth)^2\\ 
&\geq& \dfrac{c^2\, N^2 }{8\, (1+\chi(\truth))^7} - 18\,  \max\{81,\, C^4\}\, N\, D^{11} \, \chi(\truth)^2\s
\\
\hide{
&=& \dfrac{c^2\, N^2}{8\, (1+\chi(\truth))^7}\, \left(1 - \dfrac{72\, (1+\chi(\truth))^7\, \max\{81,\, C^4\}\, N\, D^{11} \, \chi(\truth)^2}{c^2\, N^2}\right)\s
\\
}
&>& 
\hide{
\dfrac{c^2\, N^2}{8\, (1+\chi(\truth))^7}\, \left(1 - \dfrac{72\, (2\, \chi(\truth))^7\, \max\{81,\, C^4\}\, N\, D^{11} \, \chi(\truth)^2}{c^2\, N^2}\right)\s
\\
&=& 
}
\dfrac{c^2\,N^2 }{8\, (1+\chi(\truth))^7}\, \left(1- \dfrac{18432\, \max\{81,\, C^4\}\, D^{11}\, \chi(\truth)^9}{c^2\, N}\right), 
\ee
using the fact that $C > 0$,\, $D \geq 2$, and $\epsilon^\star > 0$,
which implies that
\beno
\chi(\truth)
&\coloneqq& \exp(C\, D^2\, (|\!|\truth|\!|_\infty + \epsilon^\star))
&>& 1.
\ee
Under Conditions \ref{as:theta_main} and \ref{as:theta} with $\vartheta\in [0,\, 1/18)$,
we have,
for all $\bw \in \mH$, 
\beno 
\dfrac{18432\, \max\{81,\, C^4\}\, D^{11}\, \chi(\truth)^9}{c^2\, N}
&\to& 0 
&\mbox{as}& N \to \infty.
\ee
Thus,
there exists a real number $\epsilon> 0 $ along with an integer $N_3 \in \{3,4, \ldots\}$ such that 
\beno 
\dfrac{18432\, \max\{81,\, C^4\}\, D^{11}\, \chi(\truth)^9}{c^2\, N}
&\leq& \epsilon 
\ee
for all $N>N_3$, 
which implies that
\be
\label{eq:denom}
&& V_{1,1}(\nat,\, \bw)\, V_{2,2}(\nat,\, \bw)- V_{1,2}(\nat,\, \bw)^2
\hide{
\s
\\
&\geq& \dfrac{c^2\,N^2 }{4\, (1+\chi(\truth))^7}\, \left(1-  \dfrac{18432\,  \max\{81,\, C^4\}\, D^{11}\, \chi(\truth)^9}{c^2\, N}\right)\s\s
\\
}
&\geq& \dfrac{c^2\, N^2}{8\, (1+\chi(\truth))^7}\, (1- \epsilon).
\ee
Observe that \eqref{eq:denom} provides a positive lower bound on the determinant of $\bm{V}(\nat,\, \bw)$ for $\bw \in \mH$, demonstrating that 
\be
\label{eq:abs_determinant} 
|V_{1,1}(\nat,\, \bw)\, V_{2,2}(\nat,\, \bw)- V_{1,2}(\nat,\, \bw)^2 | &=& V_{1,1}(\nat,\, \bw)\, V_{2,2}(\nat,\, \bw)- V_{1,2}(\nat,\, \bw)^2.
\ee
Combining \eqref{eq:boundw}, \eqref{eq:denom}, and \eqref{eq:abs_determinant} shows that,
for all $\bw \in \mH$,
\be
\label{wbound}
\infnorm{ \bm{V}(\nat,\, \bw)^{-1}} &\leq&\, \dfrac{\max\left\{V_{2,2}(\nat,\, \bw), \,V_{1,1}(\nat,\, \bw)\right\}  + V_{1,2}(\nat,\, \bw) }{ V_{1,1}(\nat,\, \bw)\, V_{2,2}(\nat,\, \bw)- V_{1,2}(\nat,\, \bw)^2}\s\s\\
&\leq&\, \max\{9,\, C^2\} \,D^5 \left(\dfrac{N}{2} +32\,  D\, \chi(\truth)^2\right)\, \dfrac{8\, (1+\chi(\truth))^7}{c^2\,N^2\, (1-\epsilon)}\s\s\\
&\leq&\, K_1\, \dfrac{D^5\, \chi(\truth)^7}{N}\; \max\left\{1,\; \dfrac{D\, \chi(\truth)^2}{N}\right\},
\ee
where $K_1 > 0$ is a constant.

{\bf Conclusion.}
We show in two steps that $-\nabla_{\nat}^2\; \ell(\nat,\, \bw)$ is invertible for all $\nat\in\mB_{\infty}(\truth,\, \epsilon^\star)$ and all $\bw \in \mH$.
First, 
by Lemma 9 in S25, 
the matrix $\bA(\nat,\, \bw)$ is invertible for all $\nat\in\mB_{\infty}(\truth,\, \epsilon^\star)$ and all $\bw \in \mH$.
Second, 
\eqref{eq:abs_determinant} demonstrates that the determinant of $\bm{V}(\nat,\, \bw)$ is bounded away from $0$ for all $\nat\in\mB_{\infty}(\truth,\, \epsilon^\star)$ and all $\bw \in \mH$.
Thus,
$\bm{V}(\nat,\, \bw)$ is nonsingular for all $\nat\in\mB_{\infty}(\truth,\, \epsilon^\star)$ and all $\bw \in \mH$,
and so is $- \nabla_{\nat}^2\; \ell(\nat,\, \bw)$ by Theorem 8.5.11 of \citetsupp[][p.~99]{harville_matrix_1997}.
Combining \eqref{hessian_break}, \eqref{abound}, \eqref{eq:c}, and \eqref{wbound} shows that,
for all $\nat\in\mB_{\infty}(\truth,\, \epsilon^\star)$ and all $\bw \in \mH$,
\beno
\infnorm{(- \nabla_{\nat}^2\; \ell(\nat,\, \bw))^{-1}}
&\leq& \infnorm{\bm{A}(\nat,\, \bw)^{-1}\,}\s
\\
&+& \max\{1,\; \infnorm{\bm{A}(\nat,\, \bw)^{-1}\,}\infnorm{\bm{C}(\nat,\, \bw)}\}\; \infnorm{\bm{V}(\nat,\, \bw)^{-1}}\s\\ 
&\times& \left(N\,\infnorm{\bm{C}(\nat,\, \bw)}\infnorm{\bm{A}(\nat,\, \bw)^{-1}\, }+1\right)\s\s
\\
&\leq& \dfrac{18\, \chi(\truth)^2}{N}\s\s
\\
&+& \max\left\{1,\; \max\{3,\, C\}\,  D^3\,  \dfrac{18\, \chi(\truth)^2}{N}\right\}\, K_1\; \dfrac{D^5\, \chi(\truth)^7}{N}\, \s\\ 
&\times& \max\left\{1,\; \dfrac{D^2\, \chi(\truth)^2}{N}\right\}\; \left(\max\{3,\, C\}\,  D^3\, 18\, \chi(\truth)^2 + 1\right).
\ee
Conditions \ref{as:theta_main} and \ref{as:theta} with $\vartheta \in [0,\, 1/18)$ imply that
\beno 
\dfrac{\chi(\truth)^2}{N} ~<~\dfrac{\chi(\truth)^9}{N} \; \rightarrow\; 0 \mbox{ as } N \to \infty.
\ee
Thus,
there exists an integer $N_0 \in \{3, 4, \dots\}$ such that the two maxima in the upper bound on $\infnorm{(- \nabla_{\nat}^2\; \ell(\nat,\, \bw))^{-1}}$ are equal to $1$ for all $N > N_0$,
so that
\be
\label{last.inequality}
&& \infnorm{(- \nabla_{\nat}^2\; \ell(\nat,\, \bw))^{-1}}\s
\\
&\leq& \,\dfrac{18\, \chi(\truth)^2}{N} +  K_1\,\dfrac{D^5\, \chi(\truth)^7}{N}\left(\max\{3,\, C\}\,  D^3\,18\, \chi(\truth)^2 + 1\right)\s\s
\\
&\leq& \,\dfrac{18\, \chi(\truth)^2}{N} +  K_2\, \dfrac{D^8\, \chi(\truth)^9}{N}\s\s
\\
&\leq& C_1\,\dfrac{\chi(\truth)^9}{N},
\ee
where $K_2 > 0$ and $C_1 > 0$ are constants.
Substituting \eqref{last.inequality} into the definition of $\Lambda_{N}(\truth)$ concludes the proof of Lemma \ref{lemma.bounds.lambda}:
\beno
\Lambda_{N}(\truth)  
&\coloneqq& \sup\limits_{\bw\, \in\, \mH}\;\, \sup\limits_{\nat\, \in\, \mB_{\infty}(\truth,\, \epsilon^\star)}\, \infnorm{(-\nabla_{\nat}^2\; \ell(\nat;\, \bw))^{-1}}
&\leq& C_1\,\dfrac{\chi(\truth)^9}{N}.
\ee

\subsection{Bounding \texorpdfstring{$\Psi_N$}{Psi}}

\begin{lemma}
\label{lemma.bounds.psi}    
{\em 
Consider the model of Corollary \ref{theorem.model}.
Then $\sqrt{N/2} ~ \leq ~ \Psi_N  ~\leq ~ C_2\, \sqrt{N}$,
where $C_2 > 0 $ is a constant.
}
\end{lemma}

\textsc{Proof of Lemma \ref{lemma.bounds.psi}.}  
The term $\Psi_N$ is defined in 
\beno 
\Psi_N 
&\coloneqq& \underset{1 \leq a \leq N +2 }{\max}\; \norm{\bm{\Xi}_a}_2,
\ee
where 
\beno 
\bm{\Xi}_a 
&\coloneqq& (\Xi_{\{1\},a}, \ldots,  \Xi_{\{N\},a}, \Xi_{\{1,2\},a}, \ldots, \Xi_{\{N,N-1\},a}) = (\bm{\Xi}_{\mY,a}, \bm{\Xi}_{\mZ,a}), ~\text{$a \in \{1, \ldots, N+2\}$}.
\ee
The sensitivity of the sufficient statistic vector $\suff_a(\bx,\, \by,\, \bz)$ with respect to changes of responses is quantified by the vector $\bm{\Xi}_{\mY,a} \in \mathbb{R}^{N}$:
\beno 
\bm{\Xi}_{\mY,a} &\coloneqq (\Xi_{\{1\},a}, \ldots, \Xi_{\{N\},a}),
\ee 
where
\beno 
\Xi_{\{i\},a} &\coloneqq& \max\limits_{(\bw, \bw')\, \in\, \mW\, \times\, \mW:\;  y_{k} = y_{k}' \text{ for all } k \neq i,\,   \bz = \bz'}  |  \suff_a(\bx,\, \by,\, \bz) - \suff_a(\bx,\, \by', \bz')|.
\ee
The sensitivity of the sufficient statistic vector $\suff_a(\bx,\, \by,\, \bz)$ with respect to changes of connections is quantified by the vector $\bm{\Xi}_{\mZ,a} \in \mathbb{R}^{N}$:
\beno 
\bm{\Xi}_{\mZ,a} 
&\coloneqq& (\Xi_{\{1,2\},a}, \ldots, \Xi_{\{N,N-1\},a}),
\ee
where 
\beno 
\Xi_{\{i,j\},a} &\coloneqq& \max\limits_{(\bw, \bw')\, \in\, \mW\, \times\, \mW:\;  \by = \by',\, z_{k,l} = z_{k,l}' \text{ for all }\,\{k,\, l\}\, \neq\, \{i,\, j\}} |  \suff_a(\bx,\, \by,\, \bz) - \suff_a(\bx,\, \by', \bz')| .
\ee
Define
\be
\label{eq:psi}
\Psi_N 
&=& \max\limits_{1 \leq a \leq N+2}\, \sqrt{ \dsum_{i = 1}^N |\Xi_{\{i\},a}|^2 + \dsum_{i = 1}^N \dsum_{j=i+1}^N |\Xi_{\{i,j\},a}|^2}.
\ee
\begin{itemize}
\item For $a = 1, \ldots, N$, the statistic $\suff_a(\bx,\, \by,\, \bz)$ refers to the degree effects of unit $a$:
\beno 
\suff_a(\bx,\, \by,\, \bz) &=& \dsum_{j = 1; \, j\neq a}^N z_{a,j}.
\ee
The term $\Xi_{\{i,j\},a}$ is $1$ if $a \in \{i,\,j\}$ and is $0$ otherwise. 
Since the statistic is unaffected by the response, $\Xi_{\{i\},a} = 0$ for all $i = 1, \ldots, N$. 
For the sum in \eqref{eq:psi} over all $i<j$,\,
where $\mathbb{I} (a \in \{i,j\}) = 1$ holds $N$ times,\,
yielding $\norm{\bm{\Xi}_a}_2 \, \leq \,\sqrt{N}$ for all $a = 1, \ldots, N$. 
\item The statistic $\suff_{N +1}(\bx,\, \by,\, \bz)$ refers to the transitive connections effect given by
\beno 
\suff_{N+1}(\bx,\, \by,\, \bz) &=& \dsum_{i=1}^N\, \dsum_{j  = i +1}^N d_{i,j}(\bz)\, z_{i,j},
\ee
where the function $d_{i,j}(\bZ)$ is defined in \eqref{eq:cstar_sm}.
Since this statistic is not affected by $\by$,\,
$\Xi_{\{i\},a} = 0$ for all $i = 1, \ldots, N$.
Following Lemma 18 in S25, 
\beno 
\norm{\bm{\Xi}_{N+1}}_2  &\leq& \sqrt{N\, D^2 \,(1+D)^2} &\leq&\sqrt{4\, N \, D^4} &=& 2 \, D^2\, \sqrt{N},
\ee
where $D$ corresponds to the constant defined in \ref{eq:def_d}.
\item 
The statistic $\suff_{N +2}(\bx,\, \by,\, \bz)$ refers to the spillover effect given by
\beno 
\suff_{N+2}(\bx,\, \by,\, \bz) &=& \dsum_{i=1}^N\, \dsum_{j  = i +1}^N c_{i,j} \, (x_i\, y_j + y_i\, x_j)\, z_{i,j},
\ee
where the function $c_{i,j}$ is defined in \eqref{eq:cstar_sm}.
For $\{i, j\} \subset \mathscr{P}_N$,
the terms $\Xi_{\{i,j\},N+2}$ are $(y_i\, x_j + y_j\, x_i) \leq 2\, C$ if $\mN_i\, \cap\, \mN_j \neq \emptyset$ and $0$ otherwise.  
For all $i \in \mathscr{P}_N$,
\beno 
\Xi_{\{i\},N+2}    
&=& \dsum_{j:\,  \mN_i \, \cap \, \mN_j} x_j\, z_{i,j} 
&\leq& \dsum_{j: \mN_i \, \cap \, \mN_j} C 
&\leq& C\, D^2,
\ee
because according to Lemma 15 in S25 there are at most $D^2$ units such that $ \mN_i\, \cap\, \mN_j \neq \emptyset$ and $x_i\leq C$ for $i = 1, \ldots, N$ according to Condition \ref{as:1_main}. 
Combining $\Xi_{\{i,j\},N+2}$ and $\Xi_{\{i\},N+2}$ gives
\beno 
\norm{\bm{\Xi}_{N+2}}_2 & \leq& \sqrt{2\, N\, C\, D^2 + N\,C\,  D^2}  & \leq& D\, \sqrt{3\, N\, C} & \leq&  2 \, D\, \sqrt{N\, C},
\ee
where $C$ corresponds to the constant $C$ in Condition \ref{as:1_main}.
\end{itemize}

Combining the results for $\norm{\bm{\Xi}_{a}}_2$ for $a = 1, \ldots, N+2$ gives
\beno 
\sqrt{N/2} 
&\leq& \Psi_N  
&\leq& 2\, D\, \sqrt{N\, C} 
\\ 
\sqrt{N/2} 
&\leq& \Psi_N  
&\leq& C_2\, \sqrt{N},
\ee
where $C_2 \coloneqq 2\, D\, \sqrt{C} > 0 $ is a constant.

\subsection{Bounding \texorpdfstring{$\spectralnorm{\mD_N(\truth)}$}{D}}

\begin{lemma} 
\label{lemma.bounds.d}    
{\em 
Consider the model of Corollary \ref{theorem.model}.
If Conditions \ref{as:1_main},
\ref{as:theta_main},
and \ref{as:neigh} are satisfied with $\vartheta \in [0, 1/18)$,
there exists a constant $C_3\geq 1$ such that $\spectralnorm{\mD(\truth)} \leq C_3$ for all $N \geq 2$.
If the population $\mP_N$ consists of non-overlapping subpopulations with dependence restricted to subpopulations,
the same result holds when Condition \ref{as:theta_main} is replaced by Condition \ref{as:theta}.
}
\end{lemma}

\textsc{Proof of Lemma \ref{lemma.bounds.d}.}  
To bound $\spectralnorm{\mD(\truth)}$ from above,
we use the Hölder's inequality
\be
\label{eq:hoelder} 
\spectralnorm{\mD(\truth)} &\leq& \sqrt{\onenorm{\mD(\truth)} \, \infnorm{\mD(\truth)}},
\ee
where
\beno 
\onenorm{\mD(\truth)}&\coloneqq&\max\limits_{1\leq j \leq M} \dsum_{i =1}^M |\mD_{i,j}(\truth)|\s
\\
\infnorm{\mD(\truth)}&\coloneqq&\max\limits_{1\leq i \leq M} \dsum_{j =1}^M |\mD_{i,j}(\truth)|.
\ee
We can therefore bound $\spectralnorm{\mD(\truth)}$ by bounding the elements of the upper triangular coupling matrix $\mD(\truth) \in \mathbb{R}^{M \times M}$ which are
\beno 
\mD_{i,j}(\truth) 
&\coloneqq& 
\begin{cases}
0& \mbox{if } i<j\\ 
1& \mbox{if } i=j\\ 
\max\limits_{\bw_{1:(i-1)}\, \in\, \mW_{1:i-1}} \mbQ_{\truth, i, \bw_{1:(i-1)}}(W_{j}^{\star} \neq W_{j}^{\star\star})& \mbox{if } i>j.\\ 
\end{cases}
\ee

Next, 
we define a symmetrized version of the coupling matrix denoted by $\mT(\truth) \in \mathbb{R}^{M \times M}$ with elements
\beno 
\mT_{i,j}(\truth) 
&\coloneqq& 
\begin{cases}
\mD_{j,i}(\truth) & \mbox{if } i<j\\ 
\mD_{i,i}(\truth) & \mbox{if } i=j\\ 
\mD_{i,j}(\truth) & \mbox{if } i>j.\\ 
\end{cases}
\ee
The symmetry of $\mT(\truth) $ yields the following upper bound for \eqref{eq:hoelder}: 
\be
\label{eq:bound_d} 
\spectralnorm{\mD(\truth)} &\leq& \sqrt{\onenorm{\mT(\truth)} \, \infnorm{\mT(\truth)}} ~=~ \infnorm{\mT(\truth)} \s \\ 
&=& 1+ \underset{1\leq i \leq M}{\max} \dsum_{j = 1:\, j\neq i}^M \mbQ_{\truth, i, \bw_{1:(i-1)}}(W_j^\star \neq W_j^{\star\star}) \s\\ 
\ee
where the constant 1 in the second line stems from the diagonal elements of $\mT(\truth)$. 

Consider any $(i, j) \in \{1, \ldots , M\} \times \{1, \ldots, M\}$ such that $i \neq j$ and define the event $W_i \dpath W_j$  as the event that there exists a path of disagreement between vertices $W_i$ and $W_j$ in $\mG$. 
A path of disagreement between vertices $W_i$ and $W_j$ in $\mG$ is a path from $W_i$ to $W_j$ in $\mG$
such that the coupling $(W_{(i+1):M}^\star,\, W_{(i+1):M}^{\star\star})$ with joint probability mass function $\mbQ_{\truth, i, \bw_{1:(i-1)}}$ disagrees at each vertex on the path,
in the sense that $W^\star \neq W^{\star\star}$ holds for all vertices $W$ on the path.
Theorem 1 of \citetsupp[p.~753]{BeMa94} shows that
\be
\label{eq:coupling_trick} 
\mbQ_{\truth, i, \bw_{1:(i-1)}}(W_j^\star \neq W_j^{\star\star}) &\leq& \mathbb{B}_{\bm{\pi}}(W_i \dpath W_j\text{ in } \mG), 
\ee
where $\mathbb{B}_{\bm{\pi}}$ is a Bernoulli product measure based on $M$ independent Bernoulli experiments with success probabilities $\bm{\pi} \coloneqq (\pi_1, \dots, \pi_M) \in [0,1]^M$.
With $v\in \{1,\ldots, M\}$, the success probabilities $\pi_v$ are 
\beno 
\pi_v &\coloneqq& 
\begin{cases}
0 & \text{if $v \in \{1, \ldots, i-1\}$} \\
1 & \text{if $v = i$} \\
\underset{(\bw_{-v}, \bw_{-v}') \in \mW_{-v} \times \mW_{-v}}{\max}        \pi_{v,\bw_{-v}, \bw_{-v}'}& \text{if $v \in \{i+1, \ldots, M\},$}
\end{cases}
\ee
where
\be
\label{eq:pi_tv}
\pi_{v,\bw_{-v}, \bw_{-v}'} 
&\coloneqq& \TV{\mbP_{\nat}(\,\cdot\mid \bw_{-v} )-\mbP_{\nat}(\,\cdot\mid \bw_{-v}')}.
\ee

Lemma \ref{lemma.tv} provides the following upper bound:
\be
\label{eq:u}
\pi_{v,\bw_{-v}, \bw_{-v}'} 
&\leq&  \dfrac{1}{1+ \exp(-C\,D^2\, \normsup{\truth})} ,
\ee
where $C$ corresponds to the positive constant from Condition \ref{as:1_main} and $D$ is defined in \eqref{eq:def_d}.
Combining \eqref{eq:u} with Condition \ref{as:theta} shows that
\be
\label{def:U_N}
\dfrac{1}{1+ \exp(-C\,D^2\, \normsup{\truth})} 
&\leq& \dfrac{1}{1+\exp(-E- \vartheta \log N)} 
&\eqqcolon& U_{N}.
\ee
The constant $U_{N} = U$ coincides with the constant $U$ considered in Condition \ref{as:neigh}. 

With \eqref{def:U_N}, 
we define the vector $\bm{\xi} \in [0,1]^M$ with elements 
\beno 
\xi_v &\coloneqq& 
\begin{cases}
0 & \text{if $v \in \{1, \ldots, i-1\}$} \\
1 & \text{if $v = i$} \\
U_{N} & \text{if $v \in \{i+1, \ldots, M\}$},
\end{cases}
\ee
and obtain
\be
\label{eq:bound_pi_xi} 
\mathbb{B}_{\bm{\pi}}(W_i \dpath W_j\text{ in } \mG) 
&\leq&  \mathbb{B}_{\bm{\xi}}(W_i \dpath W_j\text{ in } \mG), 
\ee
because $\pi_v \leq \xi_v $ for all $v = 1, \ldots, M$. 

Next, 
we construct the set 
\beno 
\mM_{a,b} 
&\coloneqq& \{\{c,d\}:\; c\, \in\, \mN_a\, \cup\, \mN_b,\; d\, \in\, \mN_a\, \cup\, \mN_b \setminus\{c\}\}\; \cup\; \{\{c\}:\; c\, \in\, \mN_a\, \cup\, \mN_b\}
\ee
and two additional graphs with the same set of vertices as $\mG$:
\begin{enumerate}
\item $\mG_1 \coloneqq (\mV, \mE_1)$: 
\begin{itemize}
\item Vertex $W \in \mV_{\mZ}$ relating to connection $Z_{i,j}$ has edges to vertices that relate to all connections $Z_{h,k}$ and responses $Y_h$ with $\{h,k\}, \{h\} \in \mM_{i,j}$.
\item Vertex $W \in \mV_{\mY}$ relating to attribute $Y_{i}$ has edges  to vertices that relate to all connections $Z_{h,k}$ and responses $Y_h$ with $\{h,k\}, \{h\} \in \mM_{i,N+1}$ for a fictional unit $N+1$ with $\mN_{N+1} = \emptyset$.
\end{itemize}
\item $\mG_2 \coloneqq (\mV,\, \mE_1 \cup \mE_2)$: 
The set $\mE_2$ includes edges of all vertices $W_i \in \mV$ with $i \in \{1, \ldots, M\}$ to vertices in $\mS_{\mG_1, i, 2}$.
\end{enumerate}
The graph $\mG_2$ is a covering of $\mG$, 
so
\be
\label{eq:bound_pi_xi_cover} 
\mathbb{B}_{\bm{\xi}}(W_i \dpath W_j \text{ in } \mG) 
&\leq&  \mathbb{B}_{\bm{\xi}}(W_i \dpath W_j\text{ in } \mG_2).
\ee

Combining the previous results gives
\be
\label{eq:bound_spectral}
\spectralnorm{\mD(\truth)} &\leq& 1+ \underset{1\leq i \leq M}{\max} \dsum_{j = 1:\, j\neq i}^M \mbQ_{\truth, i, \bw_{1:(i-1)}}(W_j^\star \neq W_j^{\star\star}) \\
&\leq& 1+ \max\limits_{1\leq i \leq M} \dsum_{j = 1:\, j\neq i}^M \mathbb{B}_{\bm{\pi}}(W_i \dpath W_j\text{ in } \mG) \\
&\leq& 1+ \max\limits_{1\leq i \leq M} \dsum_{j = 1:\, j\neq i}^M \mathbb{B}_{\bm{\xi}}(W_i \dpath W_j\text{ in } \mG)\\
&\leq& 1+ \max\limits_{1\leq i \leq M} \dsum_{j = 1:\, j\neq i}^M \mathbb{B}_{\bm{\xi}}(W_i \dpath W_j\text{ in } \mG_2),
\ee
using \eqref{eq:bound_d}, \eqref{eq:coupling_trick}, \eqref{eq:bound_pi_xi}, \eqref{eq:bound_pi_xi_cover}.
Sorting the vertices without $W_i$ by the geodesic distance to $W_i$ (i.e.,
by the length of the shortest path to $W_i$), 
we obtain 
\be
\label{eq:bound_sum_b}
\dsum_{j = 1:\, j\neq i}^M \mathbb{B}_{\bm{\xi}}(W_i \dpath W_j\text{ in } \mG_2) 
&\leq& |\mS_{\mG_2, i,1}| \left(\max\limits_{W_j \in \mS_{\mG_2, i,1}} \mathbb{B}_{\bm{\xi}}(W_i \dpath W_j\text{ in } \mG_2)\right) \\ 
&+& \dsum_{k = 2}^\infty |\mS_{\mG_2, i,k}|\,\left(\max\limits_{W_j \in \mS_{\mG_2, i,k}} \mathbb{B}_{\bm{\xi}}(W_i \dpath W_j\text{ in } \mG_2)\right) \\
&\leq& |\mS_{\mG_2, i,1}| \\
&+& \dsum_{k = 2}^\infty |\mS_{\mG_2, i,k}|\,\max\limits_{W_j \in \mS_{\mG_2, i,k}} \mathbb{B}_{\bm{\xi}}(W_i \dpath W_j\text{ in } \mG_2).
\ee

For the event $W_i \dpath W_j$ in $\mG_2$ with $W_j \in \mS_{\mG_2, i, k}$ and $k\,\geq\,2$ to occur,
there must exist at least one vertex in each set $\mS_{\mG_2, i, 1}, \ldots, \mS_{\mG_2,i, k-1}$ at which the coupling disagrees. 
Therefore, we next derive bounds on $|\mS_{\mG_2, i, k}|$ to obtain an upper bound on $\mathbb{B}_{\bm{\xi}}(W_i \dpath W_j \text{ in } \mG_2)$. 
Following Lemma \ref{lemma.d}, 
Condition \ref{as:neigh} implies that for $i \in \{1, \ldots, M\}$ and $k \in \{2, 3, \ldots\}$ 
\be
\label{eq:bounds}
|\mS_{\mG_2, i, k}| &\leq& K_1 + K_2\, \log k
\ee 
and 
$|\mS_{\mG_2, i, 1}| \, \leq \, K_3$,
with constants $K_1 \, \geq\, 0, \,K_2 \, \geq\, 0$, and $K_3  \, >\, 0$ being functions of the constants $\omega_1\, \geq\, 0$ and $\omega_2\, \geq\, 0$ defined in Condition \ref{as:neigh} and the constant $D \in \{2, 3, \ldots\}$ defined in \eqref{eq:def_d}. 
The probability of event $W_i \dpath W_j$ in $\mG_2$ can then be bounded as follows:
\beno 
\mathbb{B}_{\bm{\xi}}(W_i \dpath W_j\text{ in } \mG_2) 
&\leq& U_{N}\, (1-(1-U_{N})^{K_3}) \, \dprod_{l = 2}^{k-1} \left[1-(1-U_{N})^{K_1+K_2\, \log l}\right]
\\ 
&\leq& \dprod_{l = 2}^{k-1} \left[1-(1-U_{N})^{K_1+K_2\, \log l}\right] 
\\ 
&\leq& \left[1-(1-U_{N})^{K_1+K_2\, \log(k-1)}\right]^{k-2}, 
\ee
The first inequality follows from 
\beno 
U_{N}\, (1-(1-U_{N})^{K_3}) 
&\leq& 1,
\ee
because $U_{N} \in [0, 1]$ and $K_3>0$.
Defining $K_{N} \coloneqq \exp(-K_1\, |\log(1-U_{N})|)$,
we obtain for $W_j \in \mS_{\mG, i,k}$
\be
\label{eq:boundd}
\mathbb{B}_{\bm{\xi}}(W_i \dpath W_j\text{ in } \mG_2) &\leq&  \left[1-(1-U_{N})^{K_1+K_2\, \log(k-1)}\right]^{k-2} \\ 
&\leq&  \exp(-K_{N}\, (k-1)^{1-K_2\, |\log(1-U_{N})|})
\ee
with the inequality $1 - a \,\leq\, \exp(-a)$ for all $a \in (0,1)$.

Plugging \eqref{eq:bounds} and \eqref{eq:boundd} in \eqref{eq:bound_sum_b}, 
we obtain:
\be
\label{eq:bound_b}
\dsum_{j = 1:\, j\neq i}^M &\mathbb{B}_{\bm{\xi}}&(W_i \dpath W_j \text{ in }\mG_2)  \\
&\leq& K_3  + \dsum_{k = 2}^\infty (K_1 + K_2\, \log k) \\ 
&\times&\exp\left(-K_{N}\, (k-1)^{1-K_2\, |\log(1-U_N)|}\right)\s \\ 
&=&  K_3  + K_1\,\dsum_{k = 2}^\infty  \exp\left(-K_{N}\, (k-1)^{1-K_2\, |\log(1-U_{N})|}\right) \\ 
&+& K_2\, \dsum_{k = 2}^\infty \log k \, \exp\left(-K_{N}\, (k-1)^{1-K_2\, |\log(1-U_{N})|}\right), \\ 
\ee
resulting in two series that we bound one by one.
With $\lceil \cdot \rceil: [0,\infty) \mapsto \{1, 2, \ldots\}$ being the function giving the upper ceiling of a positive real number and $u_{N}\coloneqq \lceil2/(1-K_2 \, |\log (1 - U_{N})| )\rceil$, the first series can be bounded as follows:
\beno
&~&\dsum_{k = 2}^\infty  \exp\left(-K_{N}\, (k-1)^{1-K_2\, |\log(1-U_{N})|}\right)\s
\\ 
&=& \dsum_{k = 1}^\infty  \exp\left(-K_{N}\, k^{1-K_2\, |\log(1-U_{N})|}\right)\s
\\
&\leq& \dfrac{u_{N}!}{{(K_{N})}^{u_{N}}}\dsum_{k = 1}^\infty  \dfrac{1}{k^2}
\;\;=\;\; \dfrac{u_{N}!\, \pi^2}{{(K_{N})}^{u_{N}}\, 6}.
\ee 
The above bound is based on a Taylor expansion of $\exp(z)$,
which establishes the inequality $\exp(z) > z^u \,/\, u!$ implying  for any $z > 0 $ and any $u \in \{1, 2, \dots\}$.
This,
in turn, 
implies the inequality $\exp(-z) < u! \,/\, z^u$ for any $z > 0 $ and any $u \in \{1, 2, \dots\}$.
With $v_{N}\coloneqq \lceil3/(1-K_2 \, |\log (1 - U_{N})| )\rceil$, we apply the same inequality to the second series: 
\beno
&~& \dsum_{k = 2}^\infty \log(k)\, \exp\left(-{K_{N}}\, (k-1)^{1-K_2\, |\log(1-U_{N})|}\right) \s\\
&=& \dsum_{k = 1}^\infty \log(k + 1)\, \exp\left(-{K_{N}}\, k^{1-K_2\, |\log(1-U_{N})|}\right) \s\\
&\leq& \dfrac{v_{N}!}{{(K_{N})}^{v_{N}}}\dsum_{k = 1}^\infty  \dfrac{\log (k+1)}{k^3} \s\\ 
&\leq& \dfrac{v_{N}!}{{(K_{N})}^{v_{N}}}\dsum_{k = 1}^\infty  \dfrac{k}{k^3} ~=~ \dfrac{v_{N}!}{{(K_{N})}^{v_{N}}}\dsum_{k = 1}^\infty  \dfrac{1}{k^2}  
~=~ \dfrac{v_{N}!\, \pi^2}{{(K_{N})}^{v_{N}}\, 6}.
\ee
Plugging these results into \eqref{eq:bound_b} gives 
\be
\label{eq:final_prob_bound} 
\dsum_{j = 1:\, j\neq i}^M \mathbb{B}_{\bm{\xi}}(W_i \dpath W_j \text{ in } \mG_2) &\leq&  K_3  + \dfrac{\pi^2}{6} \left( 
K_1 \, \dfrac{u_{N}!}{{(K_{N})}^{u_{N}}}+ K_2 \, \dfrac{v_{N}!}{{(K_{N})}^{v_{N}}}
\right).\\ 
\ee

Last but not least, combining \eqref{eq:final_prob_bound} with \eqref{eq:bound_spectral} yields 
\beno 
\spectralnorm{\mD(\truth)} &\leq& 1+ K_3  + \dfrac{\pi^2}{6} \left( 
K_1 \, \dfrac{u_{N}!}{{(K_{N})}^{u_{N}}}+ K_2 \, \dfrac{v_{N}!}{{(K_{N})}^{v_{N}}}
\right).
\ee
Under Condition \ref{as:theta_main}, $\vartheta = 0$ holds, hence $U_{\vartheta, N}, K_{N}, u_{\vartheta, N}$, and $v_{\vartheta, N}$ in \eqref{def:U_N} reduce to 
\beno 
U_{N} &=&  \dfrac{1}{1+\exp(-E)}  &\eqqcolon& U \s\\
K_{N} &=& \exp(-K_1\, |\log(1-U)|) &\eqqcolon& K_4 \s\\
u_{N} &=&  \left\lceil\dfrac{2}{1-K_2 \, |\log U|}\right\rceil  &\eqqcolon& u \s\\
v_{N} &=&  \left\lceil\dfrac{3}{1-K_2 \, |\log U|}\right\rceil  &\eqqcolon& v, 
\ee
which are constants independent of $\vartheta$ and $N$.
The constant $U$ corresponds to the constant $U$ from Condition \ref{as:neigh}.
This translates to
\beno 
\spectralnorm{\mD(\truth)} &\leq& C_3,
\ee
with $C_3 \coloneqq 1+ K_3  + (\pi^2/\,6)\,(K_1\, u!/K_4^u + K_2\, v!/K_4^v) \, \geq\, 1$. 
For non-overlapping subpopulations,  
we have $K_1 = K_2 = 0$ and
\beno 
\spectralnorm{\mD(\truth)} 
&\leq& 1 + K_3   &=&  C_3.
\ee

\subsection{Auxiliary Results} 
\label{sec:auxiliary_res}

\begin{lemma}
\label{lemma.d}    
{\em 
Consider the model of Corollary \ref{theorem.model}.
Condition \ref{as:neigh} implies that there exist constants $K_1 \geq 0$, $K_2 \geq 0$, $K_3 >0$ such that,
for all $k \in \{2, 3, \ldots\}$ and all $i \in \{1, \ldots, M\}$,
\beno
|\mS_{\mG_2, i, k}|   &\leq& K_1 + K_2\, \log k\s
\\
|\mS_{\mG_2, i, 1}|  &\leq& K_3.
\ee
}
\end{lemma}
\textsc{Proof of Lemma \ref{lemma.d}.}  
With the set 
\beno 
\mM_{a,b} 
&\coloneqq& \{\{c,d\}:\; c\, \in\, \mN_a\, \cup\, \mN_b,\; d\, \in\, \mN_a\, \cup\, \mN_b \setminus\{c\}\}\; \cup\; \{\{c\}:\; c\, \in\, \mN_a\, \cup\, \mN_b\},
\ee
we constructed from $\mG$ two additional graphs $\mG_1$ and $\mG_2 $ as follows:
\begin{enumerate}
\item $\mG_1 \coloneqq (\mV, \mE_1)$: 
\begin{itemize}
\item Vertex $W \in \mV_{\mZ}$ relating to connection $Z_{i,j}$ has edges to vertices that relate to all connections $Z_{h,k}$ and responses $Y_h$ with $\{h,k\}, \{h\} \in \mM_{i,j}$.
\item Vertex $W \in \mV_{\mY}$ relating to attribute $Y_{i}$ has edges  to vertices that relate to all connections $Z_{h,k}$ and responses $Y_h$ with $\{h,k\}, \{h\} \in \mM_{i,N+1}$ for a fictional unit $N+1$ with $\mN_{N+1} = \emptyset$.
\end{itemize}
\item $\mG_2 \coloneqq (\mV,\, \mE_1 \,\cup\, \mE_2)$: 
The set $\mE_2$ includes edges of all vertices $W_i \in \mV$ with $i \in \{1, \ldots, M\}$ to vertices in $\mS_{\mG_1, i, 2}$.
\end{enumerate}

The graph $\mG_1$ is equivalent to the graph cover $\mG^\star$ defined in Lemma 16 of S25. 
Therefore, we are able to use results from the proof of Lemma 16 in S25 demonstrating that Condition \ref{as:neigh} implies the following bound for $\mS_{\mG_1, i, k}$:
\beno 
|\mS_{\mG_1, i, k}| &\leq& (\omega_1+1)(2\, D^3\, \omega_1+\omega_2\, \log (k-1)), & & k \in \{2, 3, \ldots\},
\ee
where $D$ corresponds to the constant defined in \eqref{eq:def_d} and the constants $\omega_1\, \geq \, 0$ and $0 \, \leq \, \omega_2\, \leq \, \min\limits\{\omega_1, \, 1/((\omega_1+1)\, |\log(1-U)|)\}$
with $U \coloneqq (1+\exp(-A))^{-1} > 0 $ correspond to the constant from Condition \ref{as:neigh}.
Defining $K_5 \coloneqq 2 \, \omega_1 \, (\omega_1+1)\, D^3 \, \geq\, 0$ and $K_6 \coloneqq \omega_2 \, (\omega_1+1)\, \geq\, 0$, 
this bound is:
\beno 
|\mS_{\mG_1, i, k}| &\leq& K_5 + K_6\, \log (k-1), & & k \in \{2, 3, \ldots\}.
\ee
The bound for $\mS_{\mG_1, i, 1} \leq 4\,D^2 + D$  differs to the result from S25 since for our definition of $\mM_{i,j}$ there are additional $|\mN_i \cup \mN_j| \, \leq\,  D$ responses in $\mM_{i,j}$. 

Adding edges $\mE_2$, defined as the edges from vertices to other vertices with a geodesic distance of two in $\mG_1$, to $\mG_2$ reduces the geodesic distance between all vertices from $k\in \{1,2,\ldots\}$ in $\mG_1$ to $\lceil k/2 \rceil$ in $\mG_2$. 
Therefore, $|\mS_{\mG_2, i, k}| = |\mS_{\mG_1, i,2\,k}| + |\mS_{\mG_1, i,2\,k-1}|$ holds for $k \in \{1,2,\ldots\}$ and $i\in \{1, \ldots, M\}$.
This allows us to relate the bounds for $|\mS_{\mG_1, i, k}| $ to bounds for $|\mS_{\mG_2, i, k}| $ with $k = 2, 3, \ldots$ and $i \in \{1, \ldots, M\}$:
\beno
|\mS_{\mG_2, i, k}| &=& |\mS_{\mG_1, i, 2\,k }| + |\mS_{\mG_1, i, 2\,k-1}| \\
&\leq& 2\,K_5 + K_6\, (\log (2\,k)+ \log (2\,k -1))\\
&\leq& 2\,K_5 + 2\,K_6\, \log (2\,k)\\
&=& K_1 + K_2\, \log k
\ee
and 
\beno 
|\mS_{\mG_2, i, 1}|  &\leq& 4\,D^2 + D + K_1  \eqqcolon K_3,
\ee
with $K_1\coloneqq  2\,K_5 + 2\,K_6\log 2$ and $K_2\coloneqq 2\,K_6$.
This proves the statement with $K_1 \geq 0, K_2\geq0,$ and $K_3 > 0$. 

\begin{lemma}
\label{lemma.bounds.conditional_z}    
{\em 
Consider the model of Corollary \ref{theorem.model}.
Then,
for any pair of units $\{i, j\} \subset \mP_N$ such that $\mN_i\, \cap\, \mN_j \neq \emptyset$, 
\beno 
\dfrac{1}{1+\exp\left(C\, D^2\, \norm{\nat}_{\infty}\right)}~\leq~ \mbP_{\nat}(Z_{i,j} = 1\mid \bx,\, \by,\, \bz_{-\{i,j\}}) ~ \leq ~\dfrac{1}{1+\exp\left(-
C\, D^2\, \norm{\nat}_{\infty}\right) }.
\ee
}
\end{lemma}
\textsc{Proof of Lemma \ref{lemma.bounds.conditional_z}.}  
For all $\{i, j\} \subset \mP_N$ such that $\mN_i \cap \mN_j \neq \emptyset$, 
the conditional probability of $Z_{i,j}$ given $(\bX, \bY, \bZ_{-\{i,j\}}) = (\bx, \by, \bz_{-\{i,j\}})$ is 
\beno 
&\mbP_{\nat}(Z_{i,j} = z_{i,j}\mid \bx,\, \by,\, \bz_{-\{i,j\}}) \s\\ 
= &\dfrac{\exp\left(\nat^\top\, \bsuff(\bx,\, \by,\, \bz_{-\{i,j\}}, z_{i,j})\right)}{\exp\left(\nat^\top\, \bsuff(\bx,\, \by,\, \bz_{-\{i,j\}}, 1)\right) +\exp\left(\nat^\top\, \bsuff(\bx,\, \by,\, \bz_{-\{i,j\}}, 0)\right)} \s\\ 
= &\dfrac{1}{1 + \bg(1- z_{i,j};  \bz_{-\{i,j\}},z_{i,j}, \nat)}, 
\ee
with 
\beno  
\bg(z;\, \bz_{-\{i,j\}}, z_{i,j}, \nat) &=&  \exp\left(\nat^\top \left(\bsuff(\bx,\, \by,\, \bz_{-\{i,j\}}, z)-\bsuff(\bx,\, \by,\, \bz_{-\{i,j\}}, z_{i,j})\right)\right).
\ee
Note that 
\beno 
\underset{z_{-\{i,j\}} \in \mZ_{-\{i,j\}}}{\max} |\suff_a(\bx,\, \by,\,  \bz_{-\{i,j\}}, 0) - \suff_a(\bx,\, \by,\,  \bz_{-\{i,j\}}, 1)| \s\\ 
\leq \begin{cases}
0 & \text{if $a \in \{1, \ldots, N\} \setminus \{i,j\}$} \\
1 & \text{if $a \in \{i,j\}$} \\
1+D & \text{if $a = N+1$} \\
2\, C & \text{if $a = N+2$} \\
\end{cases},
\ee
where $\mZ_{-\{i,j\}}  \coloneqq \Cross_{(k,h) \neq (i,j)}^N\, \mZ_{k,h}$ is the domain of $\bZ$ excluding $Z_{i,j}$,
$C$ corresponds to the constant from Condition \ref{as:1_main}, and $D$ matches the constant defined in \eqref{eq:def_d}.
.  
The bounds for $a = 1, \ldots, N$ follow from the observation, that the degree statistic of unit $a$ can, first, only affected by connections $z_{i,j}$ with $a \in \{i,j\}$ and, second, be at most 1 if this is the case. 
For $a = N + 1$, 
the bound follows from Lemma 18 of S25. 
For $a = N+2$, 
the sufficient statistic counts the number of connections with overlapping neighborhoods and either $Y_i\, x_j > 0$ or $Y_j\, x_i > 0$. 
For $\mN_i\, \cap\, \mN_j \neq \emptyset$, 
the maximal change in the statistic is $2\, C$ since $y_i \in \{0,1\}$ and $x_i\leq C$ for $i \in \mP_N$, 
otherwise the maximal change is 0. 

Upon applying the triangle inequality,
\beno 
|\nat^\top\,\bsuff(\bx,\, \by,\, \bz_{-\{i,j\}}, z)-\nat^\top\,\bsuff(\bx,\, \by,\, \bz_{-\{i,j\}}, z_{i,j})| ~\leq~ (2+2\,C+D)\; \norm{\nat}_{\infty}, 
\ee
we obtain for $\mN_i \cap \mN_j \neq \emptyset$
\beno 
\exp\left(-(2+2\,C+D)\, \norm{\nat}_{\infty}\right) ~\leq~ g(1-z_{i,j};  \bz_{-\{i,j\}}, z_{i,j}, \nat)  ~\leq~ \exp\left((2+2\,C+D)\, \norm{\nat}_{\infty}\right).
\ee
Upon collecting terms,
we obtain the final result:
\beno 
\dfrac{1}{1+\exp\left(C_6\, \norm{\nat}_{\infty}\right)}&\leq& \mbP_{\nat}(Z_{i,j} = 1\mid \bx,\, \by,\, \bz_{-\{i,j\}})& \leq&\dfrac{1}{1+\exp\left(-C_6\, \norm{\nat}_{\infty}\right)} \s\\ 
\dfrac{1}{1+ \exp\left(C\, D^2\, \norm{\nat}_{\infty}\right)}&\leq& \mbP_{\nat}(Z_{i,j} = 1\mid \bx,\, \by,\, \bz_{-\{i,j\}}) &\leq& \dfrac{1}{1+\exp\left(-C\,D^2\, \norm{\nat}_{\infty}\right)}
\ee
where $D \in \{2, 3, \ldots\}$ and $C_6 \coloneqq 2+2\, C+D > 0 $ are constants.

\begin{lemma}
\label{lemma.bounds.conditional_y}    
{\em 
Consider the model of Corollary \ref{theorem.model}.
Then, for any $i \in \{1, \ldots, M\}$
\beno 
\dfrac{1}{1+\exp\left(C\, D^2\, \norm{\nat}_{\infty}\right)}~\leq~ \mbP_{\nat}(Y_{i} = 1\mid \bx,\, \by_{-i},\, \bz) ~ \leq ~\dfrac{1}{1+\exp\left(-C\, D^2\, \norm{\nat}_{\infty}\right) }.
\ee
}
\end{lemma}
\textsc{Proof of Lemma \ref{lemma.bounds.conditional_y}.}  
The conditional probability of $Y_{i}$ given $(\bX,\, \bY_{-i}, \bZ) = (\by_{-i},\, \bz)$ is
\beno 
\mbP_{\nat}(Y_{i} = y_{i}\mid \bx,\, \by_{-i},\, \bz)
&=& \dfrac{\exp\left(\nat^\top\, \bsuff(\bx,\, \by_{-i},\, y_i, \bz)\right)}{\exp\left(\nat^\top\, \bsuff(\bx,\, \by_{-i},0, \bz)\right) + \exp\left(\nat^\top\, \bsuff(\bx,\, \by_{-i},1, \bz)\right)}\s
\\
&=&\dfrac{1}{\bg(0; \by_{-i},\, y_i,\nat) +\bg(1; \by_{-i},\, y_i,\nat) },
\ee
where
\beno  
g(y;\, \by_{-i},\, y_i,\, \nat) 
&=&  \exp\left(\nat^\top \left(\bsuff(\bx,\, \by_{-i},\, y,\, \bz) - \bsuff(\bx,\,\by_{-i},\, y_i,\, \bz)\right)\right).
\ee
Note that 
\beno 
\underset{y_{-i} \in \mY_{-i}}{\max}\; |\suff_a(\bx,\, \by_{-i},\, 0,  \bz) - \suff_a(\bx,\, \by_{-i},1,   \bz)| 
\;\leq\; \begin{cases}
0 & \text{if $a \in \{1, \ldots, N+1\}$} \\
C\,D^2 & \text{if $a = N+2$,} \\
\end{cases}
\ee
where $\mY_{-i}  \coloneqq \Cross_{j \neq i}^N\, \mY_{j}$ is the domain of $\bY$ without $Y_i$, 
$C$ corresponds to the constant from Condition \ref{as:1_main}, and $D$ matches the constant defined in \eqref{eq:def_d}. 
The bounds for $a = 1, \ldots, N+1$ are 0 as the corresponding statistics are not affected by changes 
in $\by$. 
For $a = N+2$, 
the maximal change is bounded by the number of units $j$ such that $\mN_i\, \cap\, \mN_j \neq \emptyset$,
which is $D^2$, 
times the maximal value $C$ of the predictors. 
The remainder of the proof of Lemma \ref{lemma.bounds.conditional_y} resembles the proof of Lemma \ref{lemma.bounds.conditional_z}.

\begin{lemma}
\label{lemma.bounds.v}    
{\em 
Consider the model of Corollary \ref{theorem.model}.
If Conditions \ref{as:1_main},
\ref{as:neigh},
and \ref{as:theta} are satisfied with $\vartheta \in [0, 1/18)$,
we obtain the following bounds for all elements of $\bB(\nat,\, \bw)$, 
being the covariance matrix of the sufficient statistics $\suff_{N+1}(\bx,\,\by,\, \bz)$ and $\suff_{N+2}(\bx,\,\by,\, \bz)$ defined in Section \ref{sec:setup}, 
for all $\nat \in \Nat$ and all $\bw \in \mW$:
\beno 
B_{1,1}(\nat,\, \bw) &\leq&\, \dfrac{ND^5}{4}\s
\\ 
|B_{1,2}(\nat,\, \bw)| &\leq&\, \dfrac{N\, C^2\, D^5}{4}\s
\\  
B_{2,2}(\nat,\, \bw) &\leq&\ \dfrac{N\, C^2\, D^5}{4}.
\ee    
}
\end{lemma}
\textsc{Proof of Lemma \ref{lemma.bounds.v}.}   
We first bound $B_{1,1}(\nat,\, \bw)$ from above as follows:
\be
\label{eq:b11}
B_{1,1}(\nat,\, \bw) 
&=& \dsum_{i=1}^N\, \dsum_{j  = i +1}^N\mathbb{V}_{\mZ, i,j}\left( s_{N+1}(\bx,\, \by,\, \bZ)\right)\s \\ 
&=& \dsum_{i=1}^N\, \dsum_{j  = i +1}^N  \mathbb{V}_{\mZ, i,j}\left( \dsum_{a= 1}^N\dsum_{b = a + 1}^N  d_{a,b}(\bZ) \, Z_{a,b}\right)\s\\
& =& \dsum_{i=1}^N\, \dsum_{j  = i +1}^N c_{i,j}\, \mathbb{V}_{\mZ, i,j}\,  \left(\dsum_{a= 1}^N\dsum_{b = a + 1}^N Z_{a,b} \, d_{a,b}(\bZ)\right) \s\\
&\leq& \dsum_{i=1}^N\, \dsum_{j  = i +1}^N c_{i,j}\,D^2\left(\dsum_{a= 1}^N\dsum_{b = a + 1}^N\mathbb{V}_{\mZ, i,j}\, \left( Z_{a,b}\,d_{a,b}(\bZ)\right)\right)
\ee
where $D$ matches the constant defined in \eqref{eq:def_d} and the function $d_{a,b}(\bZ)$ is defined in \eqref{eq:cstar_sm}.
On the second line of \eqref{eq:b11},
we use that the fact that $\mathcal{N}_i\, \cap\, \mathcal{N}_j\, =\, \emptyset$ implies that $d_{i,j}(\bZ) \, Z_{i,j} = 0$ and $d_{a,b}(\bZ) \, Z_{a,b}$ does not depend on $Z_{i,j}$ for any $\{a, b\} \neq \{i, j\}$. 
For the inequality in the last line of \eqref{eq:b11}, we use the fact that the number of pairs $(a,b)$ for which $d_{a,b}(\bZ) \, Z_{a,b}$ is a function of $Z_{i,j}$ is bounded above by $D$ (see proof of Lemma 19 in S25). 
Invoking Lemma 15 of S25 together with applying  
\beno 
\dsum_{a = 1}^N\dsum_{b =a +1}^N\mathbb{V}_{\mZ, i,j}\, \left( d_{a,b}(\bZ)\, Z_{a,b}\right) &\leq& \dfrac{D}{4}
\ee
gives: 
\beno
B_{1,1}(\nat,\, \bw) &\leq& \dsum_{i=1}^N\, \dsum_{j  = i +1}^N c_{i,j}\,D^2\left(\dsum_{a= 1}^N\dsum_{b = a + 1}^N\mathbb{V}_{\mZ, i,j}\, \left( Z_{a,b}\,d_{a,b}(\bZ)\right)\right)
\;\;\leq\;\; \dfrac{N\, D^5}{4}
\ee

We proceed with bounding $B_{2,2}(\nat,\, \bw)$:
\beno 
B_{2,2}(\nat,\, \bw) &=&  \dsum_{i = 1}^N \mathbb{V}_{\mY, i}\left( s_{N+2}(\bx,\bY,\, \bz)\right) + \dsum_{i=1}^N\, \dsum_{j  = i +1}^N  \mathbb{V}_{\mZ, i,j}\left( s_{N+2}(\bx,\, \by,\, \bZ)\right)\\
&=& \dsum_{i = 1}^N \mathbb{V}_{\mY, i}\left(\left(\dsum_{j = 1}^N c_{i,j}\, x_j\,z_{i,j}\right)\,Y_i\right) +\dsum_{i=1}^N\, \dsum_{j  = i +1}^N  \mathbb{V}_{\mZ, i,j}\left( c_{i,j}\, (x_i\,y_j + x_j\,y_i)\,Z_{i,j}\right)\\
&=& \dsum_{i = 1}^N \left(\dsum_{j = 1}^N c_{i,j}\, x_j\,z_{i,j}\right)^2 \mathbb{V}_{\mY, i}\left(Y_i\right)+ \dsum_{i=1}^N\, \dsum_{j  = i +1}^N  c_{i,j}\,(x_i\,y_j + x_j\,y_i)^2\,\mathbb{V}_{\mZ, i,j}\left( Z_{i,j}\right)\s\\
&\leq&\,\dfrac{5\,N\, C^2\,D^4}{4}  ~\leq~ \dfrac{N\, C^2\, D^5}{4},
\ee
because $|x_j|\, \leq C$ according to Condition \ref{as:1_main}. 
For the first inequality, we also use that
\beno 
\dsum_{j = 1}^N c_{i,j} &\leq& D^2
\ee
by Lemma 15 in S25. 
We obtain
\beno 
\max\{B_{1,1}(\nat,\, \bw),\, B_{2,2}(\nat,\, \bw)\}
&\leq&  \dfrac{N\, C^2\, D^5}{4},
\ee
which provides an upper bound on $|B_{1,2}(\nat,\, \bw)|$ by the Cauchy-Schwarz inequality:
\beno 
|B_{1,2}(\nat,\, \bw)|&\leq& \sqrt{B_{1,1}(\nat,\, \bw)} \, \sqrt{B_{2,2}(\nat,\, \bw)}\ \leq\,   \dfrac{N\, C^2\, D^5}{4}.
\ee

\begin{lemma}
\label{lemma.tv}    
{\em 
Consider the model of Corollary \ref{theorem.model}.
Define
\beno 
\pi_{v,\bw_{-v}, \bw_{-v}'} 
&\coloneqq& \TV{\mbP_{\nat}(\,\cdot\mid \bw_{-v} )-\mbP_{\nat}(\,\cdot\mid \bw_{-v}')}
\ee
\beno 
\pi^\star 
&\coloneqq& \underset{1\,\leq \, v \,\leq \,M}{\max}\; \underset{(\bw_{-v},\, \bw_{-v}')\, \in\, \mW_{-v} \times \mW_{-v}}{\max}\; \pi_{v,\bw_{-v}, \bw_{-v}'}.
\ee        
Let $D \in \{2, 3, \ldots\}$ be the maximum degree of vertices $Z_{i,j}$ in $\mG$. 
Then
\beno
\pi^\star \lte \dfrac{1}{1+ \exp(-C\,D^2\, \norm{\truth}_{\infty})}.
\ee        
}
\end{lemma}
\textsc{Proof of Lemma \ref{lemma.tv}.}  
The proof of Lemma \ref{lemma.tv} resembles the proof of Lemma 21 in S25,
adapted to the bounds on the conditional probabilities derived in Lemmas \ref{lemma.bounds.conditional_z} and \ref{lemma.bounds.conditional_y}. 
We distinguish four cases, where $W_v$ with $v \in \{1, \ldots, M\}$ relates to:
\begin{enumerate}
\item Connection $Z_{i,j}$ of a pair of nodes $\{i,\, j\} \subset \mP_N$ with $\mN_i \cap \mN_j = \emptyset$.
\item Attribute $Y_i$ with $i \in \mP_N$ and $\{j \in \mP_N:\; \mN_i\, \cap \,\mN_j\, \neq \,\emptyset\} = \emptyset$.
\item Connection $Z_{i,j}$ of a pair of nodes $\{i,\, j\} \subset \mP_N$ with $\mN_i \cap \mN_j \neq \emptyset$.
\item Attribute $Y_i$ with $i \in \mP_N$ and $\{j \in \mP_N:\; \mN_i\, \cap \,\mN_j\, \neq \,\emptyset\} \neq \emptyset$.
\end{enumerate}
In cases 1 and 2,
$W_v$ is independent of $\bW_{-v}$,
so that $\pi_{v,\bw_{-v}, \bw_{-v}'} = 0$;
note that case 2 cannot occur,
because Condition \ref{as:1_main} ensures that there are no units $i \in \mP_N$ with $\{j \in \mP_N:\; \mN_i\, \cap \,\mN_j\, \neq \,\emptyset\} = \emptyset$. 
In cases 3 and 4, 
$W_v$ depends on a non-empty subset of other vertices in $\mG$.  
Consider any $v \in \{1, \ldots, M\}$ such that $\pi_{v,\bw_{-v}, \bw_{-v}'} >0$ for some $(\bw_{-v}, \bw_{-v}') \in \mW_{-v} \times \mW_{-v}$ and define
\beno 
a_{0, v} &\coloneqq& \mbP_{\nat} (W_v = 0 \mid \bW_{-v} = \bw_{-v}) ~\text{  and  }~a_{1, v} &\coloneqq& \mbP_{\nat} (W_v = 1 \mid \bW_{-v} = \bw_{-v}) \\ 
b_{0, v} &\coloneqq& \mbP_{\nat} (W_v = 0 \mid \bW_{-v} = \bw_{-v}') ~\text{  and  }~b_{1, v} &\coloneqq& \mbP_{\nat} (W_v = 1 \mid \bW_{-v} = \bw_{-v}').
\ee
Lemma 21 in S25 shows that 
\beno
\pi_{v,\bw_{-v}, \bw_{-v}'} 
&\leq& \min\{\max\{a_{0,v},\, b_{0,v}\},\; \max\{a_{1,v},\, b_{1,v}\}\}.
\ee
Plugging in the bounds on the conditional probabilities in Lemmas \ref{lemma.bounds.conditional_z} and \ref{lemma.bounds.conditional_y},
we obtain
\beno
\pi_{v,\bw_{-v}, \bw_{-v}'} &\leq& \dfrac{1}{1+\exp\left(-C\, D^2\, \norm{\truth}_{\infty}\right)},
& v \in \mathscr{V}_Z 
\ee
and
\beno
\pi_{v,\bw_{-v}, \bw_{-v}'} &\leq& \dfrac{1}{1+\exp\left(-C\, D^2\, \norm{\truth}_{\infty}\right)},
& v \in \mathscr{V}_Y. 
\ee
Since $D \in \{2, 3, \ldots\}$,  
we obtain
\beno
\pi^\star 
&\coloneqq& \underset{1\,\leq \, v \,\leq \,M}{\max}\;\; \underset{(\bw_{-v},\, \bw_{-v}')\, \in\, \mW_{-v} \times \mW_{-v}}{\max}\; \pi_{v,\bw_{-v}, \bw_{-v}'} \lte \dfrac{1}{1+ \exp(-C\,D^2\, \norm{\truth}_{\infty})}.
\ee    

\begin{lemma}
\label{lemma.h}    
{\em 
Consider the model of Corollary \ref{theorem.model}.
If Conditions \ref{as:1_main} and \ref{as:neigh} are satisfied along with either Condition \ref{as:theta_main} or Condition \ref{as:theta} with $\vartheta \in [0, 1/18)$,
there exists an integer $N_0 \in \{3, 4, \ldots\}$ such that,
for all $N > N_0$,
\beno 
\mbP(\bW \not\in \mH) 
&\leq& \dfrac{4}{\max\{N,\, p\}^2},
\ee 
where $\mH$ is defined in \eqref{def.h}.
}
\end{lemma}

\textsc{Proof of Lemma \ref{lemma.h}.}   
We prove Lemma \ref{lemma.h} by showing that
\be
\label{eq:bound_H_1}
\mbP\left(\dsum_{i=1}^N \infnormvec{\bH_{i,1}(\bW)}~<~\dfrac{N}{2\, (1 + \chi(\truth))^2}\right) 
\lte \dfrac{2}{\max\{N,\, p\}^2} \s\s \\ 
\mbP\left(\dsum_{i=1}^N \infnormvec{\bH_{i,2}(\bW)}~<~\dfrac{c^2\,N}{2\, (1 + \chi(\truth))}\right) 
\lte \dfrac{2}{\max\{N,\, p\}^2}.
\ee
To prove the first line of \eqref{eq:bound_H_1}, 
we first bound $(1/2) \sum_{i=1}^N \infnormvec{\bH_{i,1}(\bW)}$ from below. 
We then use Theorem 1 of \citetsupp[][p.\ 207]{Chetal07} to concentrate $\sum_{i=1}^N \infnormvec{\bH_{i,1}(\bW)}$.
Last,
but not least,
we show that there exists an integer $N_0 \in \{3, 4, \ldots\}$ such that the obtained lower bound for $(1/2) \sum_{i=1}^N \infnormvec{\bH_{i,1}(\bW)}$ is,
with high probability, 
greater than the deviation of $\sum_{i=1}^N \infnormvec{\bH_{i,1}(\bW)}$ from its mean.
The first line of \eqref{eq:bound_H_1} follows from combining these steps. 
The second line of \eqref{eq:bound_H_1} can be established along the same lines.
A union bound then establishes the desired result:
\beno 
\mbP(\bW \not\in \mH)
\lte 
\mbP\left(\dsum_{i=1}^N \infnormvec{\bH_{i,1}(\bW)}~<~\dfrac{ N}{2\, (1 + \chi(\truth))^2}\right)\s
\\
&+&
\mbP\left(\dsum_{i=1}^N \infnormvec{\bH_{i,2}(\bW)}~<~\dfrac{c^2\,N}{2\, (1 + \chi(\truth))}\right)\s
\\
&\leq& \dfrac{4}{\max\{N,\, p\}^2}.
\ee 

\s

\hide{
{\bf Remarks by Michael: 
\bi
\item First, we need to make sure that the lemma numbers cited are aligned with the lemma numbers in S25.
\item Second, Subhankar detected small issues in Lemma 11, which I fixed.
We need to check the entire poof of Lemma 11. 
\ei
} 
}

{\bf Step 1:}
Condition \ref{as:1_main} implies that,
for each unit $i \in \mP_N$,
there exists a unit $j \in \mP_N \setminus \{i\}$ such that $\mN_i \cap \mN_j \neq \emptyset$ and $x_j \in [c,\, C]$.
Thus,
by Lemma \ref{lemma.bounds.conditional_z}, 
Lemma 17 of S25, 
and Conditions \ref{as:1_main} and \ref{as:theta_main},
we obtain 
\beno 
\dfrac{1}{2}\, \dsum_{i = 1}^N \infnormvec{\bH_{i,1}(\bw)} 
&\geq& \dfrac{1}{2}\, \dsum_{i = 1}^N \mbE\, d_{i,j}(\bZ) 
&\geq& \dfrac{N}{2\, (1 + \chi(\truth))^2}
&\geq& \dfrac{N}{4\, \chi(\truth)^2}\s 
&\geq& \dfrac{N^{1 - 2\, \vartheta}}{4\, \exp(2\, E)}   .
\ee
Theorem 1 of \citetsupp[][p.\ 207]{Chetal07} implies
\beno 
\mbP\left(\left| \dsum_{i = 1}^N \infnormvec{\bH_{i,1}(\bw)}  - \mathbb{E} \dsum_{i = 1}^N \infnormvec{\bH_{i,1}(\bW)} \right| < t\right) &\geq& 1- 2 \exp \left(- \dfrac{2\, t^2}{\Psi_N^2 \, \spectralnormsqrt{\mD_N(\truth)} }\right).
\ee
Choosing
\beno 
t 
&\coloneqq& \sqrt{\log \max\{N,\, p\}}\,\Psi_N \,\spectralnorm{\mD_N(\truth)}
\ee
gives
\beno 
&~&\mbP\left(\left| \dsum_{i = 1}^N \infnormvec{\bH_{i,1}(\bW)}  - \mathbb{E} \dsum_{i = 1}^N \infnormvec{\bH_{i,1}(\bW)} \right| <\sqrt{\log \max\{N,\, p\}}\,\Psi_N \,\spectralnorm{\mD_N(\truth)} \right)\s
\\
&\geq& 1 - \dfrac{2}{\max\{N,\, p\}^2}.
\ee

Next, 
we demonstrate that there exists an integer $N_1 \in \{3, 4, \dots\}$ such that,
for all $N > N_1$, 
\beno 
\sqrt{\log \max\{N,\, p\}}\, \Psi_N \,\spectralnorm{\mD_N(\truth)}&\leq&  \dfrac{N^{1-2\, \vartheta}}{4\, \exp(2\, E)}.
\ee
To do so, 
we bound the three terms one by one.
Using $\max\{N,\, p\} = N + 2$,
the first term, $\sqrt{\log \max\{N,\, p\}}$, is bounded above by $\sqrt{\log \max\{N,\, p\}}\, \leq\, 2\, \sqrt{\log N}$ provided $N \geq 2$.
The second term is bounded above by $\Psi_N ~\leq~ D\, \sqrt{N}$ as shown in the proof of Lemma 14 in S25. 
The third term is bounded above by $\spectralnorm{\mD_N(\truth)}<C_3$ by Lemma \ref{lemma.bounds.d},
where $C_3 > 0 $ is a constant.

Combining these results gives 
\beno 
2\,\sqrt{N\, \log N}\, C_3, D &\leq&  \dfrac{N^{1-\vartheta}}{4\, \exp(E)}\s
\\ 
8\, C_3\, D\, \exp(E) &\leq&  \sqrt{\dfrac{N^{1-2\, \vartheta}}{\log N}}.
\ee
Similar to the proof of Lemma 14 in S25, 
this implies
\beno 
\mbP\left(\dsum_{i=1}^N \infnormvec{\bH_{i,1}(\bW)}  
~\geq~ \dfrac{N}{2\, (1 + \chi(\truth))^2}\right) 
&\geq& 1 - \dfrac{2}{\max\{N,\, p\}^2}.
\ee

{\bf Step 2:} 
Conditions \ref{as:1_main} and \ref{as:theta_main} along with Lemma \ref{lemma.bounds.conditional_z} establish 
\beno 
\dfrac{1}{2}\, \dsum_{i = 1}^N \infnormvec{\bH_{i,2}(\bw)} 
&\geq& \dfrac{c^2}{2}\, \dsum_{i = 1}^N \mbE\, Z_{i,j} 
&\geq& \dfrac{c^2\, N}{2\, (1 + \chi(\truth))}
&\geq& \dfrac{c^2\, N}{4\, \chi(\truth)} 
&\geq& \dfrac{c^2\, N^{1-\vartheta}}{4\, \exp(E)}. 
\ee
Once more,
we invoke Theorem 1 of \citetsupp[][p.\ 207]{Chetal07} to obtain 
\beno 
&& \mbP\left(\left|\dsum_{i = 1}^N \infnormvec{\bH_{i,2}(\bW)} - \mathbb{E} \dsum_{i = 1}^N \infnormvec{\bH_{i,2}(\bW)}\right|\, <\, \sqrt{\log \max\{N,\, p\}}\, \Psi_N\, \spectralnorm{\mD_N(\truth)}\right)\s
\\
&\geq& 1 - \dfrac{2}{\max\{N,\, p\}^2}.
\ee
We proceed by showing that there exists an integer $N_2 \in \{3, 4, \dots\}$ such that,
for all $N > N_2$,
\be
\label{eqeq}
\sqrt{\log \max\{N,\, p\}}\, \Psi_N\, \spectralnorm{\mD_N(\truth)}
&\leq& \dfrac{c^2\, N^{1-\vartheta}}{4\, \exp(E)}.
\ee
We bound the three terms on the left-hand side of \eqref{eqeq} one by one.
The bounds on the first term, $\sqrt{\log \max\{N,\, p\}}$, and third term, $\spectralnorm{\mD_N(\truth)}$, are the same as in the first step.
With regard to the second term, 
we obtain $\Psi_N ~\leq~ C_2\, \sqrt{N}$ by the proof of Lemma \ref{lemma.bounds.psi} with $C_2 > 0 $.

Combining these bounds gives
\beno 
2\, \sqrt{N\, \log N}\, C_3
&\leq&  \dfrac{c^2\, N^{1-\vartheta}}{4\, \exp(E)}\s
\\ 
\dfrac{8}{c^2}\, C_3\,  \exp(E) 
&\leq& \sqrt{\dfrac{N^{1-2\, \vartheta}}{\log N}},
\ee
which vanishes as $N \rightarrow \infty$ under Conditions \ref{as:theta_main} and \ref{as:theta} with $\vartheta\in [0,\, 1/18)$.
Thus,
for all $N > N_2$,
\beno 
\mbP\left(\dsum_{i=1}^N \infnormvec{\bH_{i,2}(\bW)}
~\geq~ \dfrac{c^2\, N}{2\, (1 + \chi(\truth))}\right) 
&\geq& 1 - \dfrac{2}{\max\{N,\, p\}^2}.
\ee

\s

\section{Quasi-Newton Acceleration}
\label{sec:quasi.newton}

The two-step algorithm described in Section \ref{sec:two.stage} iterates two steps:
\bi
\item[] {\bf Step 1:}
Update $\nat_1^{(t)}$ given $\nat_2^{(t-1)}$ using a MM algorithm with a linear convergence rate (\citealpsupp{bohning_monotonicity_1988}, Theorem 4.1).
\item[] {\bf Step 2:}
Update $\nat_2^{(t)}$ given $\nat_1^{(t)}$ using a Newton-Raphson update with a quadratic convergence rate.
\ei
To accelerate Step 1,
we use quasi-Newton methods (\citealpsupp{lange_optimization_2000}):
We approximate the difference between  $(\bA^\star)^{-1}$ and $[\bA(\nat^{(t)})]^{-1}$,
defined in Lemma~\ref{lemma.minorizer} and Equation~(\ref{A}), 
respectively, 
by rank-one updates. 

A first-order Taylor approximation of $\nabla_{\nat_1}\, \ell(\nat_1, \nat_2^{(t)})$ around $\nat_1^{(t)}$ shows that 
\begin{align}
\label{eq:taylor_approx2}
-\bA(\nat^{(t)})\; \bm{k}(\nat_1^{(t)},\, \nat_1;\, \nat_2^{(t)})\; \approx\; \nat_1^{(t)} - \nat_1,
\end{align}
where
\be
\label{eq:s_matrix}
\bm{k}(\nat_1^{(t)},\, \nat_1;\, \nat_2^{(t)}) &\coloneqq& \nabla_{\nat_1}\, \ell(\nat_1, \nat_2^{(t)})\Big|_{\nat_1=\nat_1^{(t)}}-\nabla_{\nat_1}\, \ell(\nat_1, \nat_2^{(t)}) .
\ee
Since a standard Newton-Raphson algorithm corresponds to \eqref{eq:taylor_approx2} with $\nat_1 = \nat_1^{(t-1)}$, 
the change in consecutive estimates carries information on $[\bA(\nat^{(t)})]^{-1}$, 
which we want to approximate. 
More specifically, 
we approximate the difference between $\left(\bA^\star \right)^{-1}$ and $[\bA(\nat^{(t)})]^{-1}$. 
Thus we write 
$\left(\bA^\star \right)^{-1} - \left[\bA(\nat^{(t)})\right]^{-1}  \eqqcolon \bm{M}^{(t)} $ and set $\nat_1 = \nat_1^{(t-1)}$, so that \eqref{eq:taylor_approx2} becomes
\be
\bm{M}^{(t)}\, \bm{k}(\nat^{(t)}_1, \nat_1^{(t-1)}; \nat_2^{(t)})&=& (\nat_1^{(t)} - \nat_1^{(t-1)}) + \left(\bA^\star \right)^{-1} \, \bm{k}(\nat^{(t)}_1, \nat_1^{(t-1)}; \nat_2^{(t)}) \eqqcolon \bm{r}^{(t)}, 
\label{eq:M_update}
\ee
which is called the inverse secant condition for updating  $\bm{M}^{(t)}$. 
Given that \eqref{eq:M_update} relates $[\bA(\nat^{(t)})]^{-1} $ to the score functions through the definition of $\bm{k}(\nat_1^{(t)},\, \nat_1^{(t-1)};\, \nat_2^{(t)})$ in \eqref{eq:s_matrix} and estimates $\nat_1^{(t)}$ and $\nat_1^{(t-1)}$, the updates of $\bm{M}^{(t+1)}$ will be based on \eqref{eq:M_update}. 
We employ the parsimonious symmetric, rank-one update of \citetsupp{davidon_variable_1991} to satisfy \eqref{eq:M_update} by updating $\bm{M}^{(t)}$ as follows: 
\be
\label{eq:update_acc}
\bm{M}^{(t)} &=& \bm{M}^{(t-1)} + \bm{q}^{(t)}\left(\bm{q}^{(t)}\right)^\top \left[c^{(t)}\right]^{-1},
\ee
with $ \bm{q}^{(t)} \coloneqq \bm{r}^{(t)} -  \bm{M}^{(t-1)}\,  \bm{k}(\nat^{(t)}, \nat^{(t-1)}; \, \nat_2^{(t)})$ and $c^{(t)} \coloneqq    (\bm{q}^{(t)})^\top\,  \bm{k}(\nat^{(t)}, \nat^{(t-1)}; \, \nat_2^{(t)})$.
We seed the algorithm with the MM update described in Section \ref{sec:two.stage} by setting $\bm{M}^{(0)} = \bm{0}$, 
the $N\times N$ null matrix.

In summary,
the quasi-Newton acceleration of the MM algorithm updates $\nat_1^{(t)}$ given $\nat_2^{(t)}$  as follows: 
\begin{enumerate}
\item[]{\bf Step 1:} Calculate $\bm{k}(\nat_1^{(t)},\, \nat_1^{(t-1)}; \,\nat_2^{(t)})$ defined in \eqref{eq:s_matrix}. 
\item[]{\bf Step 2:} Update $\bm{M}^{(t)}$ according to \eqref{eq:update_acc}.
\item[]{\bf Step 3:} Update $\nat_1^{(t+1)}$ from $\nat_1^{(t)}$: 
\begin{align}
\label{eq:updated_accelerate}
\nat_1^{(t+1)} = \nat_1^{(t)} + \left[\left(\bA^\star \right)^{-1} - \bm{M}^{(t)}\right]\left[\nabla_{\nat_1}\, \ell(\nat_1, \nat_2^{(t)})\Big|_{\nat_1=\nat_1^{(t)}}\right].
\end{align}
\end{enumerate}
Unlike the unaccelerated MM algorithm, the described quasi-Newton algorithm does not guarantee 
that $\ell(\nat_1^{(t+1)},\, \nat_2^{(t+1)})\, \ge\, \ell(\nat_1^{(t+1)},\, \nat_2^{(t)})$.
Therefore, 
$\nat_1^{(t+1)}$ is updated by either the quasi-Newton update \eqref{eq:updated_accelerate} or the MM update \eqref{Step1Update}, 
whichever gives rise to the higher pseudo-likelihood.
The resulting updates slightly increase
the computing time per iteration while potentially dramatically decreasing
the total number of iterations.

\section{MM Algorithm: Directed Connections}
\label{sec:mm_directed}

If connections are directed,
$Z_{i,j}$ may differ from $Z_{j,i}$.
In such cases,
the pseudo-loglikelihood can be written as
\beno
\ell(\nat)
&\coloneqq& \dsum_{i=1}^N \ell_i (\nat) + \dsum_{i=1}^{N-1}\, \dsum_{j=1,\, j \neq i}^N \ell_{i,j}(\nat),
\ee
where $\ell_i$ and $\ell_{i,j}$ are defined by
\beno
\ell_{i}(\nat) 
\ \coloneqq\   \log\, p_{\nat}(y_i \mid \by_{-i},\, \bz) 
\quad\mbox{and}\quad
\ell_{i,j}(\nat) 
\ \coloneqq\ 
\log\, p_{\nat}(z_{i,j} \mid \by,\, \bz_{-\{i,j\}}).
\ee
We partition the parameter vector $\nat \coloneqq (\nat_1,\, \nat_2) \in \mR^{2 N + 12}$ into 
\bi
\item the nuisance parameter vector:
$\nat_1 \coloneqq (\alpha_{\mZ,O,1}$, $\dots$, $\alpha_{\mZ,O,N}, \alpha_{\mZ,I,1}$, $\dots$, $\alpha_{\mZ,I,N-1}) \in \mR^{2\, N - 1}$;
\item the parameter vector of primary interest: 
$\nat_2 \coloneqq (\alpha_{\mY},$ $\, \beta_{\mX, \mY,1},\, \beta_{\mX, \mY,2},\, $ $\beta_{\mX, \mY,3},$ $ \, \lambda, \, \gamma_{\mZ, \mZ,1}$,
$\gamma_{\mZ, \mZ,2}$, 
$\gamma_{\mX, \mZ,1}$,
$\gamma_{\mX, \mZ,2}$,
$\gamma_{\mX, \mZ,3}$,
$\gamma_{\mX, \mZ,4}$,
$\gamma_{\mY, \mZ}$,
$\gamma_{\mX, \mY, \mZ}) \in \mR^{13}$.
\ei
As explained in Section \ref{sec:application.model}, $\alpha_{\mZ,N, I}$ is set to $0$ in order to address identifiability issues.
The negative Hessian is partitioned in accordance: 
\beno
-\nabla_{\nat}^2\; \ell(\nat) 
&\coloneqq& 
\begin{pmatrix}
\bA(\nat)  & \bB(\nat)\\ 
\bB(\nat)^\top & \bC(\nat) &
\end{pmatrix},
\ee
where $\bA(\nat) \in \mR^{(2\,N - 1) \times (2\,N - 1)}$,
$\bB(\nat) \in \mR^{(2\,N - 1) \times 13}$,
and $\bC(\nat) \in \mR^{13 \times 13}$.
Writing $\ell(\nat_1, \nat_2)$ in place of $\ell(\nat)$,
we compute at iteration $t+1$:
\bi
\item[] {\bf Step 1:}
Given $\nat_2^{(t)}$,
find $\nat_1^{(t+1)}$ satisfying $\ell(\nat_1^{(t+1)}, \nat_2^{(t)})\, \ge\, \ell(\nat_1^{(t)}, \nat_2^{(t)})$.
\item[] {\bf Step 2:}
Given $\nat_1^{(t+1)}$,
find $\nat_2^{(t+1)}$ satisfying $\ell(\nat_1^{(t+1)}, \nat_2^{(t+1)})\, \ge\, \ell(\nat_1^{(t+1)}, \nat_2^{(t)})$.
\ei

In Step 1, 
it is inconvenient to invert the high-dimensional $(2\, N-1) \times (2\, N-1)$ matrix 
\beno
\bA(\nat^{(t)})
&\coloneqq& -\dsum_{i=1}^N\, \dsum_{j=1,\, j \neq i}^N\, \nabla_{\nat_1}^2\, \ell_{i,j}(\nat_1, \nat_2^{(t)}) \Big|_{\nat_1=\nat_1^{(t)}}
&=&  \dsum_{i=1}^N\, \dsum_{j=1,\, j \neq i}^N \pi_{i,j}^{(t)}\, (1-\pi_{i,j}^{(t)})\, \bm{e}_{i,j}\, \bm{e}_{i,j}^\top.
\ee
Note that the definition of vector $\bm{e}_{i,j} \in \mathbb{R}^{2\, N-1}$ differs from the undirected case described in Section \ref{sec:two.stage}.  
For $j \neq N$, 
let $\bm{e}_{i,j}$ be the $(2\, N-1)$-vector whose $i$th and $(j+N)$th coordinates are $1$ and whose other coordinates are $0$. 
For $j = N$, 
let $\bm{e}_{i,j}$ be the $(2\, N-1)$-vector whose $i$th coordinate is $1$ and whose other coordinates are $0$.
Along the lines of the MM algorithm for undirected connections described in Section \ref{sec:two.stage}, 
we increase $\ell$ by maximizing a minorizing function of $\ell$,
replacing $\bA(\nat^{(t)})$ by a constant matrix $\bA^\star$ that is more convenient to invert. 
The constant matrix $\bA^\star$ is defined as
\beno 
\bA^\star 
&\coloneqq& 
\begin{pmatrix}
\bA_{1,1}^\star  &         \bA_{1,2}^\star\\ 
\left(\bA_{1,2}^\star\right)^\top  &         \bA_{2,2}^\star\\ 
\end{pmatrix}
\ee
where
\begin{itemize}
\item $\bA_{1,1}^\star\in \mathbb{R}^{N \times N}$ and $\bA_{2,2}^\star \in \mathbb{R}^{(N-1) \times (N-1)}$ are diagonal matrices with elements $(N-1)/4$ on the main diagonal; 
\item $\bA_{1,2}^\star \in \mathbb{R}^{N \times (N-1)}$ is a matrix with vanishing elements on its main diagonal and off-diagonal elements $1/4$.
\end{itemize}

Applying Theorem 8.5.11 in \citetsupp{harville_matrix_1997} to $\bA_{1,2}^\star$ and $\bA^\star$ shows that matrix can be inverted in $O(N)$ operations. 
With the above change in the constant matrix $\bA^\star$, 
we estimate $\nat$ along the lines of Section \ref{sec:two.stage}.

\hide{
\section{Simulation Study: Godambe Information}
\label{sec:app.add_sim}

\begin{figure}[t!]
\centering
\includegraphics[width = .48\linewidth, keepaspectratio]{plots/plot_simulation_corrected_uncertainty.pdf}
\includegraphics[width = .48\linewidth, keepaspectratio]{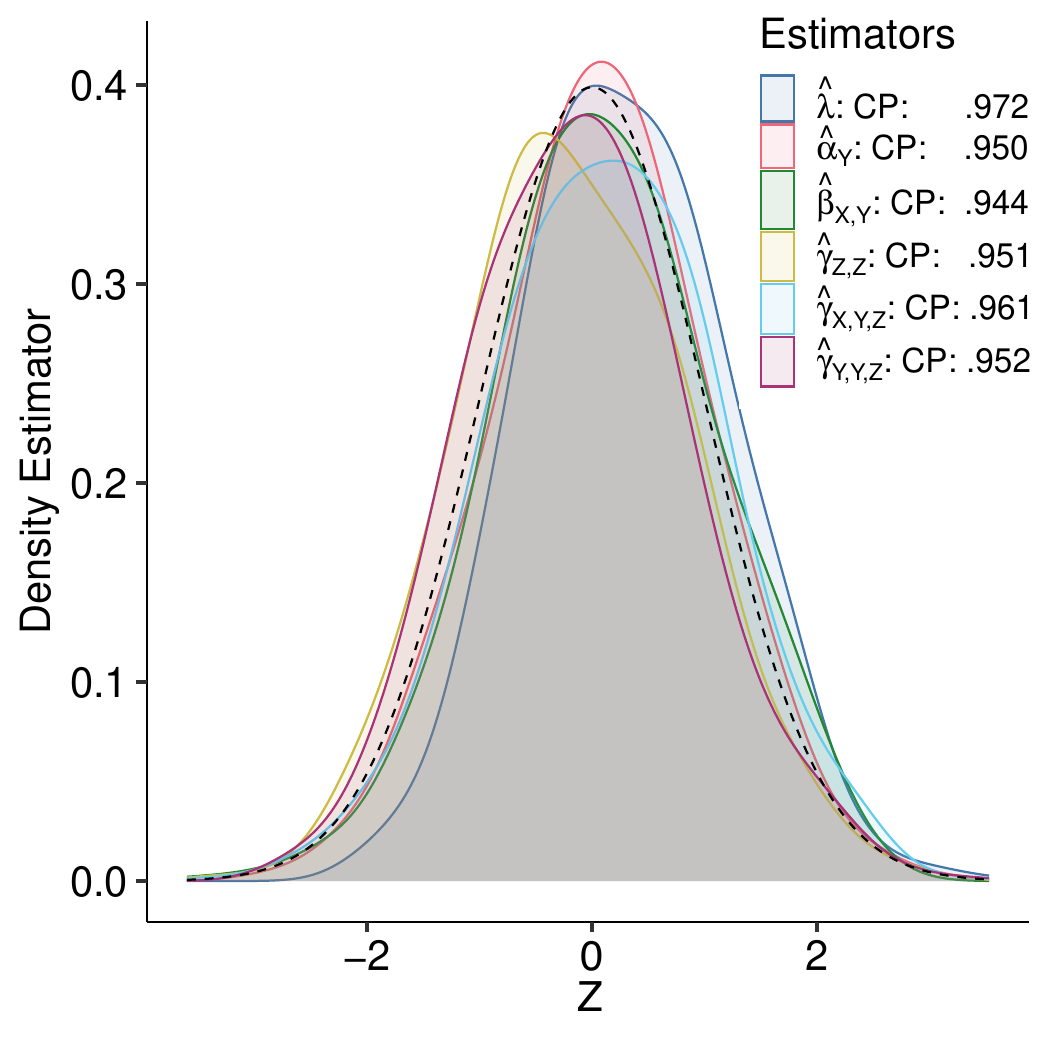}
\caption{Simulation results based on 1,000 simulated data sets.
Right:
Kernel density estimators of $Z$-scores $(\widehat\lambda - \lambda^\star)/\text{S.E.}(\widehat \lambda)$,\,
$(\widehat\alpha_{\mY} - \alpha_{\mY}^\star)/\text{S.E.}(\widehat\alpha_{\mY})$,\,
$(\widehat\beta_{\mX,\mY} - \beta_{\mX,\mY}^\star)/\text{S.E.}(\widehat\beta_{\mX,\mY})$,\,
$(\widehat\gamma_{\mZ,\mZ} - \gamma_{\mZ,\mZ}^\star)/\text{S.E.}(\widehat\gamma_{\mZ,\mZ})$,\,
$(\widehat\gamma_{\mX,\mY,\mZ} - \gamma_{\mX,\mY,\mZ}^\star)/ \text{S.E.}(\widehat\gamma_{\mX,\mY,\mZ})$,\,
and $(\widehat\gamma_{\mY,\mY,\mZ} - \gamma_{\mY,\mY,\mZ}^\star)/\text{S.E.}(\widehat\gamma_{\mY,\mY,\mZ})$ based on $N = 500$ units,
where $\text{S.E.}$ are the standard errors based on the variance approximation described in Section \ref{sec:standard_errors}. 
Right:
Kernel density estimators of $Z$-scores $(\widehat\lambda - \lambda^\star)/\widetilde{\text{S.E.}(\widehat \lambda)}$,\,
$(\widehat\alpha_{\mY} - \alpha_{\mY}^\star)/\widetilde{\text{S.E.}(\widehat\alpha_{\mY})}$,\,
$(\widehat\beta_{\mX,\mY} - \beta_{\mX,\mY}^\star)/\widetilde{\text{S.E.}(\widehat\beta_{\mX,\mY})}$,\,
$(\widehat\gamma_{\mZ,\mZ} - \gamma_{\mZ,\mZ}^\star)\widetilde{/\text{S.E.}(\widehat\gamma_{\mZ,\mZ})}$,\,
$(\widehat\gamma_{\mX,\mY,\mZ} - \gamma_{\mX,\mY,\mZ}^\star)/ \widetilde{\text{S.E.}(\widehat\gamma_{\mX,\mY,\mZ})}$,\,
and $(\widehat\gamma_{\mY,\mY,\mZ} - \gamma_{\mY,\mY,\mZ}^\star)/\widetilde{\text{S.E.}(\widehat\gamma_{\mY,\mY,\mZ})}$,
where $\widetilde{\text{S.E.}}$ are the standard errors based on the Godambe information. 
The dashed line corresponds to the standard normal density.
CP is the coverage probability of interval estimators with nominal coverage probability $.95$.
}
\label{fig:simulation_godambe}
\end{figure}

As outlined in Section \ref{sec:standard_errors}, the variance of the maximum pseudo-likelihood estimator $\widehat\nat$ can be approximated using the Godambe information \citep{schmid_computing_2023}: 
\be
\label{eq:godambe}
\mathbb{V}_{\truth}(\widehat\nat)\, \approx\, \mathbf H(\widehat\nat)^{-1}\; \mathbb{V}[\mathbf G(\widehat\nat)]\; \mathbf H(\widehat\nat)^{-1},
\ee
where $\mathbf G(\widehat\nat)$ and $\mathbf H(\widehat\nat)$ are the gradient and Hessian of \eqref{eq:comlik} evaluated at the maximum pseudo-likelihood estimator $\widehat\nat$.
If $-\nabla_{\nat}^2\,\, \ell(\nat;\, \bY, \bZ)|_{\nat=\widehat\nat}^{-1}$\, is invariant under the distribution of $(\bY, \bZ) \mid \bX$,
\eqref{approximate.covariance} coincides with \eqref{eq:godambe}.
The use of the Godambe information comes with computational advantages: 
First, it requires a single inversion of the Hessian rather than repeated inversions across simulated data sets. 
Second, simulated Hessians may be singular, making \eqref{eq:godambe} more stable.

To assess the empirical performance of this approximation, we revisit the simulation study in Section \ref{sec:simulations}, replacing the variance formula \eqref{approximate.covariance} detailed in Section \ref{sec:standard_errors} with \eqref{eq:godambe}. 
The left panel of Figure \ref{fig:simulation_godambe} presents the updated results alongside the original estimates from Figure \ref{fig:simulation_N_increasing},
demonstrating that both approaches to uncertainty quantification give rise to similar results.
}

\hide{

\begin{figure}[t]
\centering
\includegraphics[width = .48\linewidth, keepaspectratio]{plots/plot_simulation_2_new.pdf}
\includegraphics[width = .48\linewidth, keepaspectratio]{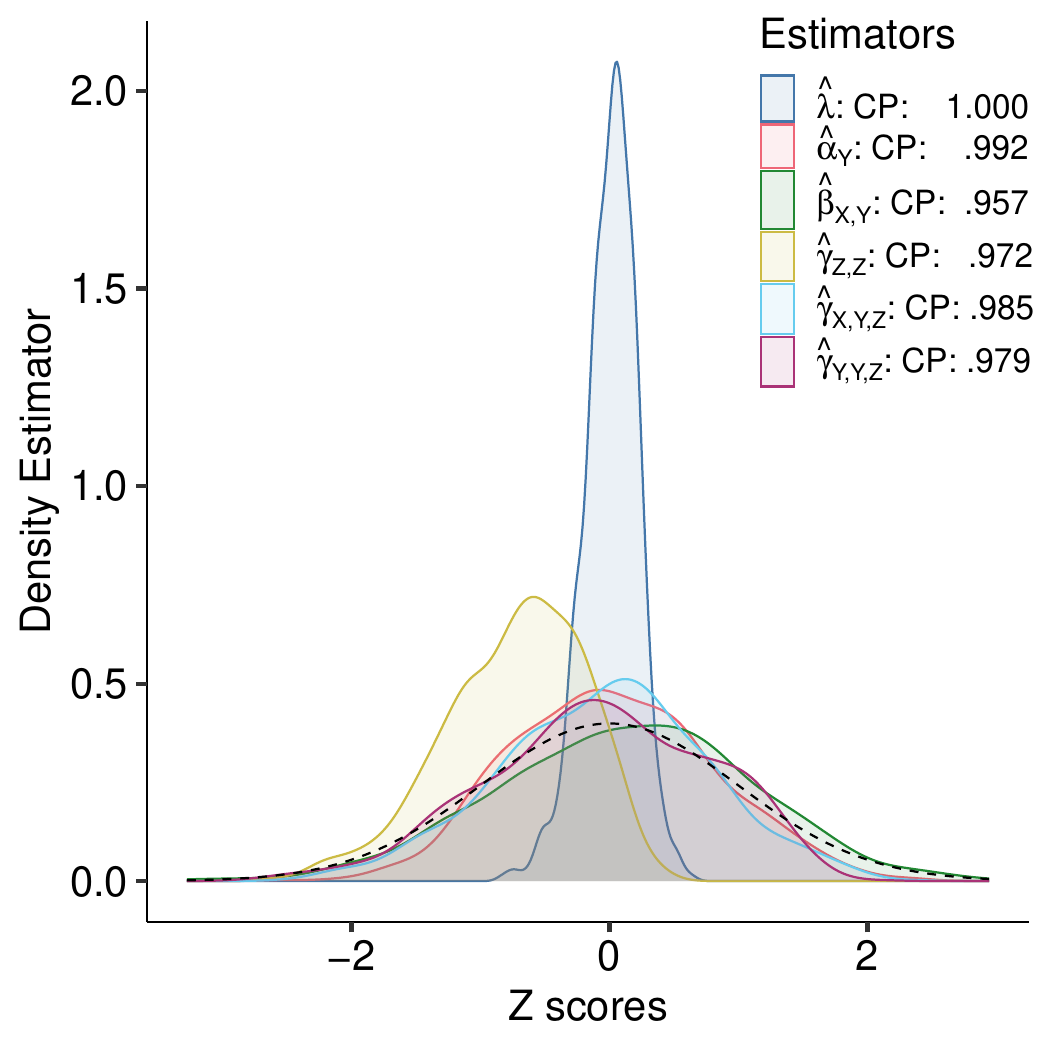}
\caption{Simulation results based on 1,000 simulated data sets.
Left:
Statistical error $|\!|\widehat\nat - \truth|\!|_\infty$ of the maximum pseudo-likelihood estimator $\widehat\nat \in \mR^{N+6}$ as a function of $N$.
Right:
Statistical errors 
$|\widehat\lambda - \lambda^\star|$,\,
$|\widehat\alpha_{\mY} - \alpha_{\mY}^\star|$,\,
$|\widehat\beta_{\mX,\mY} - \beta_{\mX,\mY}^\star|$,\,
$|\widehat\gamma_{\mZ,\mZ} - \gamma_{\mZ,\mZ}^\star|$,\,
$|\widehat\gamma_{\mX,\mY,\mZ} - \gamma_{\mX,\mY,\mZ}^\star|$,\,
and $|\widehat\gamma_{\mY,\mY,\mZ} - \gamma_{\mY,\mY,\mZ}^\star|$ in case $N = 250$.
CP denotes the coverage probability of interval estimators with a nominal coverage probability of $.95$. \textcolor{red}{The second plot still needs to be changed!}
}
\label{fig:simulation_N_increasing_suppl}
\end{figure}
\subsection{Setup with Low-Degree Heterogeneity}

In line with the simulation study reported in Section \ref{sec:simulations},  
we simulate data from the same example model specified by Equations~(\ref{eq:gi}) and~(\ref{eq:hij})
with data-generating parameter vector $\nat^\star \coloneqq (\nat_1^\star,\, \nat_2^\star) \in \mR^{N+6}$.
To demonstrate the effect if low variability, we now assume that 
$\nat_1^\star \coloneqq (\alpha_{\mZ,1}^\star$, $\dots$, $\alpha_{\mZ,N}^\star) \in \mR^N$,
are independent draws from a Gaussian with mean $-7/5$ and standard deviation $1/5$.
The parameter vector,
$\nat_2^\star \coloneqq (\lambda^\star$, 
$\alpha_{\mY}^\star$,
$\beta_{\mX,\mY}^\star$,
$\gamma_{\mZ,\mZ}^\star$,
$\gamma_{\mX,\mY,\mZ}^\star$,
$\gamma_{\mY,\mY,\mZ}^\star) \in \mR^6$,
is specified as $(3/10, -2,\, 2,\, 2/10,\, 1/10,\, 1/10)$, 
while the neighborhood structure is set along the lines of Section \ref{sec:simulations}.

To assess the properties of interval estimators with a nominal coverage probability of $.95$ based
on the Godambe information in Section~\ref{sec:standard_errors},
we generate 1,000 data sets with $N = 250$ units.
The right plot in Figure~\ref{fig:simulation_N_increasing_suppl} shows that the empirical coverage
probabilities approximately match the nominal coverage probability.
Estimators related to connections ($\widehat\gamma_{\mZ,\mZ}$,
$\widehat\gamma_{\mX,\mY,\mZ}$,
$\widehat\gamma_{\mY,\mY,\mZ}$) exhibit lower statistical errors compared to those related to responses ($\widehat\alpha_{\mY}$ and $\widehat\beta_{\mX,\mY}$),
reflecting the fact that the number of possible connections ${N \choose 2}$ exceeds the number of responses $N$.
}

\section{Hate Speech on X: Additional Information}
\label{sec:app.add}

\subsection{Data}

For the application, we use posts of $N = $ 2,191  U.S.\ state legislators on the social media platform X collected by \citetsupp{kim_attention_2022} 
in the six months leading up to and including the insurrection at the United States Capitol on January 6, 2021.  
We restrict attention to active legislators,
that is,
legislators who posted during the aforementioned period and mentioned or reposted content from other active legislators. 
Since reposts do not necessarily reflect politicians' opinions, we exclude all reposts and non-unique posts that are direct copies of other users' messages to gather information on responses. 
Employing large language models of \citetsupp{camacho-collados_tweetnlp_2022} pre-trained on these posts enables categorizing the 109,974 posts into those containing hate speech statements versus those that do not.
Accordingly, the binary attribute $Y_i$ equals 1 if the corresponding legislator sent at least one post classified as hate speech and 0 otherwise.  
The algorithm of  \citetsupp{camacho-collados_tweetnlp_2022} provides for each Tweet a continuous value between 0 and 1. We classify the respective Tweet as using hate speech if its value is larger than 0.5. 
The attribute $x_{i,1} \in \{0, 1\}$ is 1 if legislator $i$ is a Republican and 0 otherwise.  
In addition, 
we incorporate information on each legislator's gender ($x_{i,2} = 1$ if legislator $i$ is female and $0$ otherwise), race ($x_{i,3} = 1$ if legislator is white and $0$ otherwise), and state ($x_{i,4}$).  
On the social media platform X, users have the ability to either mention or repost other users' posts. 
The resulting network, denoted as $\bZ$, is based on the mentions and reposts exchanged between January 6, 2020 and January 6, 2021:
$Z_{i,j} = 1$ if legislator $i$ mentioned or reposted legislator $j$ in a post. 

\hide{

To construct the neighborhoods $\mN_i$ of the legislators $i$, we respect the choices of the legislators:  the neighborhood $\mN_i$ is defined as the set of users who follow legislator $i$. Alternatively, the neighborhood structure could be inferred from observed data using model-based clustering techniques, as in \citetsupp{latouche2011}. The specified neighborhoods encode local dependence, identifying which legislators are likely to exert influence on others. Given that users primarily view information from their followers, they are able to chose who can influence them. Consequently, we argue that the observed neighborhood --- supported by substantive reasoning --- is more appropriate than treating the neighborhoods as random or unknown. 

}

\subsection{Plots}
\label{sec:app.plots}

\begin{figure}[t!]
\centering
\includegraphics[width=0.6\linewidth]{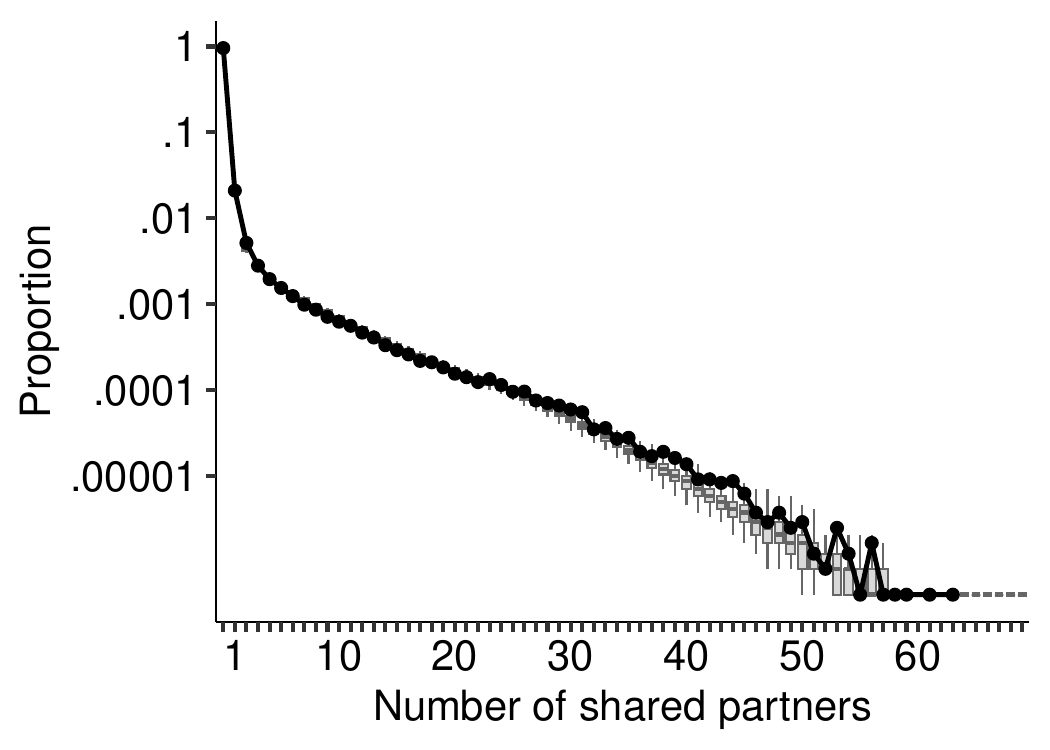}
\caption{Hate speech on X:
The red line indicates the observed shared partners distribution of the network of repost and mention interactions of U.S.\ legislators, 
while the boxplots represent the shared partners distributions of simulated networks from the estimated model.}\s
\label{fig:sharing_partnets}
\end{figure}

In addition to the goodness-of-fit checks reported in Section \ref{sec:applications}, 
we assess whether the model preserves salient characteristics of connections $\bZ$.
Figure \ref{fig:sharing_partnets} suggests that the proposed model captures the shared partner distribution,
i.e.,
the numbers of connected pairs of legislators $\{i, j\} \subset \mP_N$ with $1$, $2$, $\dots$ shared partners.

\newpage

\bibliographystylesupp{chicago} 
\bibliographysupp{base}

\end{document}